\documentclass[a4paper,fleqn,usenatbib]{mnras}
\usepackage{newtxtext,newtxmath}
\usepackage[T1]{fontenc}
\usepackage{ae,aecompl}

\usepackage{graphicx}	
\usepackage{amsmath}	
\usepackage{amssymb}	

\usepackage[dvipsnames]{xcolor}

\newcommand{\rr}[1]{#1} 

\title[Mergers moulding the CGM I]{Galaxy mergers moulding the circum-galactic medium -- I. The impact of a major merger}

\author[M. H. Hani et al.]{
  \parbox[t]{\textwidth}{
  {Maan H. Hani,$^{1}$\thanks{E-mail: mhani@uvic.ca}\thanks{Vanier Scholar}}
  {Martin Sparre,$^{2,3,4}$ }
  {Sara L. Ellison,$^{1}$ }
  {Paul Torrey,$^{4}$ }
  {Mark Vogelsberger$^{4}$}
  }
\\\\
$^{1}$Department of Physics and Astronomy, University of Victoria, Victoria, British Columbia, V8P 1A1, Canada\\
$^{2}$Institut f\"ur Physik und Astronomie, Universit\"at Potsdam, Karl-Liebknecht-Str.\,24/25, 14476 Golm, Germany\\
$^{3}$Leibniz-Institut f\"ur Astrophysik Potsdam (AIP), An der Sternwarte 16, 14482 Potsdam, Germany\\
$^{4}$Department of Physics, Kavli Institute for Astrophysics and Space Research, MIT, Cambridge, MA 02139, USA
}

\date{Accepted XXX. Received YYY; in original form ZZZ}

\pubyear{2017}

\begin{document}
\label{firstpage}
\pagerange{\pageref{firstpage}--\pageref{lastpage}}
\maketitle



\begin{abstract}
Galaxies are surrounded by sizeable gas reservoirs which host a significant amount of metals: the circum-galactic medium (CGM). The CGM acts as a mediator between the galaxy and the extra-galactic medium. However, our understanding of how galaxy mergers, a major evolutionary transformation, impact the CGM remains deficient. We present a theoretical study of the effect of galaxy mergers on the CGM. We use hydrodynamical cosmological zoom-in simulations of a major merger selected from the Illustris project such that the $z=0$ descendant has a halo mass and stellar mass comparable to the Milky Way. To study the CGM we then re-simulated this system at a 40 times better mass resolution, and included detailed post-processing ionization modelling. Our work demonstrates the effect the merger has on the characteristic size of the CGM, its metallicity, and the predicted covering fraction of various commonly observed gas-phase species, such as \ion{H}{i}, \ion{C}{iv}, and \ion{O}{vi}. We show that merger-induced outflows can increase the CGM metallicity by $0.2-0.3$ dex within 0.5 Gyr post-merger. These effects last up to 6 Gyr post-merger. While the merger increases the total metal covering fractions by factors of $2-3$, the covering fractions of commonly observed UV ions decrease due to the hard ionizing radiation from the active galactic nucleus, which we model explicitly. Our study of the single simulated major merger presented in this work demonstrates the significant impact that a galaxy interaction can have on the size, metallicity, and observed column densities of the CGM.
\end{abstract}


\section{Introduction}
\label{sec:intro}
\noindent
%
Galaxies are surrounded by vast gaseous haloes which extend well beyond the hosts' stellar components: Early observations of quasar sight lines attributed the presence of absorption at multiple intermittent redshifts to gaseous haloes of intervening galaxies \citep[e.g., ][]{Bergeron1986, Bergeron_et_Boiss1991, Lanzetta1995, Tripp2000, Chen2001}. In the past decade, owing to the rise of large spectroscopic surveys of galaxies with well-determined physical properties (e.g., SDSS), all sky UV surveys (e.g., \textit{GALEX}), and improved sensitivity of UV spectrographs (e.g., COS), studies of the gaseous haloes of galaxies could systematically connect gas absorption properties to galaxy properties in statistically meaningful samples \citep[e.g., ][]{Cooksey2010, Prochaska2011, Tumlinson2013, Liang-et-Chen2014, Lehner2015}. The aforementioned gaseous haloes are commonly referred to as the circum-galactic medium (CGM) and are ubiquitous in galaxies regardless of mass or star formation activity: even sub$-L^*$ galaxies \citep{Bordoloi2014}, and passive galaxies host a CGM \citep{Thom2012}. The current model of the CGM suggests the presence of a clumpy multiphase medium which extends beyond the virial radius of the host galaxy, with a declining radial density profile, containing a substantial amount of gas and metals \citep[e.g., ][]{COS-Halos_metals, Werk2014, Liang-et-Chen2014, Lehner2014, Lehner2015, Prochaska2017}. Observational studies targeting the CGM of $L^*$ galaxies showed that the CGM gas content is comparable to the mass of the interstellar medium \citep[ISM; e.g., ][]{Chen2010, Tumlinson2011, Werk2014, Prochaska2017} and correlates positively with ISM properties \citep{Borthakur2015}. Additionally, CGM observations infer a significant amount of metals \citep[e.g., ][]{COS-Halos_metals, Peeples2014} where CGM metallicities can extend to supersolar metallicities \citep{Prochaska2017}. The clumpy multiphase CGM consist of a warm gas $T\sim 10^{4-5}$ K (clumpy in nature) embedded within a hot diffuse $T\sim 10^6$ K medium \citep[e.g., ][]{Heitsch2009, bordoloi_ionization, Armillotta2017}. The multiphase structure of the CGM is corroborated by the variety of observed ionic species which survive at a vast range of temperatures: While the warm gas hosts the low ionization species (e.g., \ion{H}{i}, \ion{Si}{ii}, \ion{Si}{iii}, \ion{C}{ii}, \ion{C}{iv}), the hot medium is home for the most highly ionized species (e.g., \ion{O}{vi}, \ion{O}{vii}). Additionally, the spectral line profiles of absorbers in the CGM can be reproduced by invoking a patchy medium \citep[e.g., ][]{Stern2016, Werk2016}: i.e. multiple high density gas clouds contribute to the optical depth along the line of sight thus leaving their kinematic imprint on the absorption line profile. For a review of the CGM, see \cite{Putman2012} and \citet{Tumlinson2017}.

Numerical simulations have provided a controlled environment to study the CGM and test feedback models and their ability to reproduce observations \citep[e.g., ][]{Hummels2013, Bird2014, Christensen2016, Oppenheimer2016}. Cosmological hydrodynamical simulations reproduce the observed sizable, multiphase, non-pristine gas reservoir in the CGM \citep[e.g., ][]{Shen2013, Stinson2012, Ford2013, Nelson2016, Suresh2017} while emphasizing the important role of galactic outflows at populating the CGM with metals even to large impact parameters \citep[e.g., ][]{Oppenheimer_et_Dave2006, Hummels2013, Suresh2015b, Bird2016}. Additionally, numerical simulations facilitate studying the redshift evolution of the CGM's structure \citep[e.g., filaments, outflows, streams; e.g., ][]{Shen2012, Corlies2016, Ford2014, Muratov2015, Christensen2016, vandeVoort2012}.

%
Under the current paradigm for structure formation, galaxy major mergers are thought to play a key role in the assembly and evolution of the observed structure \citep[e.g., ][]{LC1993}. Numerical simulations have demonstrated that by driving non-axisymmetric gravitational torques galaxy mergers can create gravitational instabilities in the galaxies' ISM which compress the gas and cause it to lose angular momentum thus being funnelled into the central regions \citep[e.g., ][]{H89, BH91, Mihos-et-Hernquist1996, Hopkins2008i, Hopkins2008ii}. The resultant increase in the central gas densities dilutes the central metallicity \citep[e.g., ][]{Montouri2010, Torrey2012}, fuels excess star formation \citep[e.g., ][]{merger_properties, Cox2006, Cox2008, Moreno2015}, and possibly feeds a central supermassive black hole (BH) causing a significant increase in the BH's mass \citep[e.g., ][]{DiMatteo2005, Hopkins_et_Quataert2010}. Additionally, the morphological signatures of mergers manifest in the form of shells, ripples, tidal tails, and tidal plumes \citep[e.g., ][]{DiMatteo2007, Lotz2008, Lotz2010massratio, Lotz2010gasfraction, Pop2017}.

The framework in which galaxy mergers alter galaxy morphologies and induce star formation and active galactic nuclei (AGN) is borne by various observational studies. Merging galaxies exhibit enhanced star formation rates \citep[SFRs; e.g., ][]{Lambas2003, Dasyra2006, Patton2011, Ellison2013, Scott_et_Kaviraj2014, Knapen2015} and AGN fractions \citep[e.g., ][]{Ellison2011, Satyapal2014, Weston2017} in contrast to their non-interacting counterparts. Additionally, the observed central metallicity dilution in galaxy pairs is a strong observational signature of gas being funnelled to the central regions during a merger \citep[e.g., ][]{Kewley2006, Ellison08, Scudder2012}. Observed merging galaxies also show enhanced morphological asymmetries \citep[e.g., ][]{Hernquist-et-Quinn1987, HernandezToledo2005, Casteels2014, Patton2016}. Galaxy mergers have therefore been both theoretically predicted and observationally proven to provide an efficient way of triggering and accelerating gas evolution (i.e. enrichment/depletion via star formation, accretion, outflows, and/or heating due to AGN feedback) and thus represent a major driver of galaxy evolution.

%
Given their potentially profound impact on both the stellar and gas properties, galaxy mergers are often invoked to explain the presence of metals and low ionization species in the CGM at large impact parameters \citep[e.g., ][]{Farina2013, Farina2014, Johnson2015QPQ}. Indeed, merger-induced tidal torques can give rise to observed stellar tidal features extending to many tens of kpc \citep[e.g., ][]{HernandezToledo2006, Patton2011, Patton2013, Casteels2014}. Although some gas asymmetries coexist with asymmetries in the stellar profiles of interacting galaxies, tidally induced asymmetries can persist even longer in \ion{H}{i} gas \citep[e.g., ][]{HIasymm, Lelli2014i, Lelli2014ii}. Such tidal debris can potentially diffuse into the CGM contributing a significant gas and metal mass. In addition to morphological disturbances, galaxy mergers can trigger vigorous galactic outflows associated with feedback from the enhanced star formation \citep[e.g., ][]{Martin2005, Rupke2005_starburst_outflow, Strickland2009, Hayward2017} and AGN activity \citep[e.g., ][]{Rupke2005_agn_outflow, Veilleux2013, Zschaechner2016, JongHak_outflows} which can populate the CGM with metals while giving rise to multiphase absorbers \citep[e.g., ][]{Borthakur2013, Bird2015, bordoloi_ionization, COS-Burst_old2016, COS-Burst}. Besides, merger-induced shocks and feedback can increase the CGM's internal energy; numerical simulations of galaxy mergers show that CGM gas can be significantly heated (to X-ray emitting temperatures) through shocks and feedback processes \citep[e.g., ][]{Cox2004, Cox2006_agn,  Sinha2009}. Furthermore, studies of galaxies in dense environments (higher merger probability) show different CGM properties when compared to a matched isolated galaxy sample: The CGM of galaxies in groups show distinct kinematics \citep[][]{Pointon2017} and ionic covering fractions \citep{Johnson2015, Burchett2016}. Despite the possibly significant influence of galaxy mergers on the CGM, the details of the interplay between galaxy mergers and the CGM remain relatively unexplored and we are currently lacking clear and quantitative predictions of how the CGM will be affected during the merger process. Current observations of the impact of galaxy mergers on the CGM are limited to a few case studies \citep[e.g., ][]{Keeney2011, Johnson2014}. A systematic survey targeting the CGM of kinematic galaxy pairs (i.e. COS-Pairs: Bordoloi et al. in preparation) is needed to place observational constraints on the effect of mergers on the CGM.

%
In this work, we present a study of the effect of galaxy--galaxy major mergers on the CGM. We showcase a zoom-in cosmological hydrodynamical simulations of a major merger from \citet{Sparre2016a} and demonstrate the effects of the merger on the CGM size, metal content, and ionization. This paper is structured as follows: In Section \ref{sec:methods} we describe the halo selection method and the galaxy formation model. Additionally, we discuss the ionization analysis and the radiative model implemented in post-processing. We then present the results of the simulations in Section \ref{sec:results} focusing on the effect of the merger on the covering fractions of oxygen and hydrogen, the size of the CGM, and the CGM ionization. In Section \ref{sec:discussion} we discuss the role of merger-induced outflows in populating the CGM with metal rich gas, the thermal state of the CGM, and assess the effect of the simulation resolution on the results. Lastly, Section \ref{sec:conclusions} summarizes the conclusions of this work.

\section{Methods}
\label{sec:methods}
\noindent
The primary goal of this work is to study the impact of galaxy--galaxy major mergers on the physical and observable properties of the CGM. We use cosmological zoom-in simulations to study the evolution of the CGM throughout the merger and during the postmerger epoch. The use of cosmological zoom-in simulations allows for self-consistent modelling of the CGM in a cosmological framework while offering superior spatial and mass resolution compared to full-volume cosmological simulations which is required when studying the CGM. In cosmological zoom-in simulations, a higher resolution mesh and a smaller target mass are used in regions of interest (zoom-in regions), while the resolution in the rest of the box is degraded, thus permitting faster computational times, and higher resolution than large-scale cosmological box simulations. Unlike isolated galaxy merger simulations, the cosmological nature of the simulations offers a proper treatment of relevant external processes, i.e. cosmological tidal fields and mass accretion. Therefore, studying the CGM in cosmological simulations avoids many assumptions about CGM properties (i.e. structure, physical properties, chemical properties). The initial conditions used in this paper are the same used in \citet{Sparre2016a}, which are drawn from the Illustris project and specifically target galaxy major mergers.

This section briefly describes the simulations; specifically focusing on the physics implementation of the galaxy formation model. Additionally, the post-processing techniques used to analyse the simulations are described here.

%
\subsection{Simulation code}
\label{sec:methods/code}
\noindent
We use the moving-mesh magneto-hydrodynamical code \textsc{arepo} \citep{Arepo, Arepo_revised} to solve the Euler equations. \textsc{arepo} handles the hydrodynamical evolution of gas by discretizing the Euler equations over a dynamic Voronoi mesh where the mesh-generating points trace the local fluid flow. The nature of the code reduces advection errors compared to the adaptive static mesh approach as the (de)refinement of the mesh is less frequent due to the adaptive, dynamic mesh. 

%
\subsection{Simulations and major merger selection}
\label{sec:methods/merger_selection}
\noindent
In this study, we showcase one\footnote{Note that the choice of merger does not affect the results of this study. All three other mergers yield qualitatively similar results.} of the four mergers introduced in \citet{Sparre2016a} (merger: 1605-3) to demonstrate the effect of galaxy mergers on the CGM metal budget and ionization. The galaxy merger was selected from the Illustris-1 box of the Illustris project \citep{IntroIllustris_G1, IntroIllustris_V1, IntroIllustris_V2,  IntroIllustris_S1}, a large scale hydrodynamical cosmological simulation, and re-simulated at a 40 times higher mass resolution\footnote{Note that the lower resolution re-runs of the same merger introduced in \citet{Sparre2016a} (i.e. 1605-1, 1605-2) were only used for a convergence test in Section \ref{sec:discussion/resolution_effect}} (see Table \ref{tab:sim-resolution}). The merger was selected such that the remnant at $z = 0$ is Milky Way-like with stellar mass $M_\star = 10^{10.89}$ M$_\odot$ residing in a dark matter halo with total mass $M_{200} = 10^{12.00}$ M$_\odot$. The merger occurs at $z = 0.66$ which allows the study of the CGM well beyond the merger event. The progenitors undergoing the merger have stellar masses of $10^{10.27}$ and $10^{10.21}$ M$_\odot$, hence a mass ratio of 1:1.16. The merger chosen for this study is characterized by the presence of a sizeable ($10^{7.14}$ M$_\odot$) very dense star-forming gas reservoir prior to the merger and therefore exhibits a significant starburst during the merger with SFRs of $70-100$ M$_\odot /$yr \citep[$3.5 - 5 \times$ higher than the pre-merger SFR; see][]{Sparre2016a}. We note that this merger is gas-rich, and that the merger remnant remains star forming. Such a behaviour is often seen in simulated mergers in this mass range \citep{SH2005, Governato2009, Rodriguez-Gomez2017}. Additionally, the progenitors' recent merger histories, as well as the descendant's merger history, are relatively quiet which makes it possible to highlight the effects of the major merger, i.e. no other major mergers, with $\text{mass ratio} >  \text{1:3}$, occur post-merger (only one minor merger with $\text{1:10} < \text{mass ratio} < \text{1:3}$). For the exact selection criteria, see \citet{Sparre2017}.

We track the most massive galaxy participating in the merger by identifying the galaxy's stellar population at $z=0.93$, well before the merger, as a reference population. The associated progenitors and descendants are defined to be the galaxies with the most common stellar particles with the reference population. Each galaxy is identified using the \textsc{subfind} algorithm \citep{Subfind}.

\begin{table}
  \centering
  \caption{An overview of the three resolution re-runs of the merger simulation showcased in this work \citep{Sparre2016a}. The results of this paper use the highest resolution simulation: 1605-3. We follow the same naming convention as \citet{Sparre2016a} where the simulations are named AAAA-B, where AAAA is the z = 0 friends-of-friends group in Illustris, and B is the `zoom-factor'. $\epsilon_\mathrm{dm}$ is the maximum softening length for the dark matter resolution elements in physical units. $m_\mathrm{b}$ and $m_\mathrm{dm}$ are the masses for the baryon and dark matter resolution elements, respectively.}
\label{tab:sim-resolution}
\begin{tabular}{c|cccc}
  \hline
Sim-name & $\epsilon_\mathrm{dm}$ [kpc] & $m_\mathrm{b}$ [M$_\odot$] & $m_\mathrm{dm}$ [M$_\odot$] \\
  \hline \hline
1605-1    &  0.64                      & $8.90\times 10^5$    & $4.42\times 10^6$     \\
1605-2    &  0.32                      & $1.11\times 10^5$    & $5.53\times 10^5$     \\
1605-3    &  0.21                      & $3.30\times 10^4$    & $1.64\times 10^5$     \\
\hline
\end{tabular}
\end{table}

%
\subsection{Galaxy formation model}
\label{sec:methods/GFM}
\noindent
The galaxy formation model implemented in the simulations was presented in detail in the Auriga simulation paper \citep{Auriga}. The model is based on the original physical model in the Illustris project \citep{Illustris} and the modifications of \citet{Marinacci2014}. 

The ISM is modelled as a multiphase gas governed by an effective equation of state according to the sub-resolution model of \citet{ISM}. In this model, star formation occurs stochastically in gas cells denser than the star formation threshold, $n_\mathrm{sfr}=0.13$ cm$^{-3}$, with a star formation time-scale, $t_\mathrm{sfr} = 2.2$ Gyr. Each newly formed star particle is modelled as a simple stellar population. The evolution of star particles and associated mass-loss and chemical enrichment are tracked continuously in the simulation assuming a Chabrier \citep{Chabrier} initial mass function. Mass lost from asymptotic giant branch stars and supernovae (SN) is returned to the environment using a top-hat kernel enclosing the 64 nearest neighbouring gas cells. We follow the abundances of nine elements (H, He, C, N, O, Ne, Mg, Si, Fe) using the yields described in \citet{Illustris}. {\sc arepo} allows metals to advect self-consistently between gas cells therefore creating a smooth metal profile by allowing high metallicity gas to diffuse amongst neighbouring cells. Stellar feedback from SN II is implemented by probabilistically spawning wind particles from star-forming gas cells. Wind particles are decoupled from the hydrodynamics and travel isotropically at a velocity which scales with the local dark matter velocity dispersion until the surrounding gas density drops below a threshold density ($0.05 n_\mathrm{sfr}$) or a maximum travel time interval has elapsed. When wind particles re-couple to the hydrodynamics, they deposit their momentum, thermal energy, and metal content into the nearest gas cell.  

The simulations also include BHs and consequently feedback from AGNs. When haloes reach a mass of $7.1 \times 10^{10}$ M$_\odot$ they are seeded with a BH of mass $1.4 \times 10^5$ M$_\odot$ which grows by Eddington-limited Bondi--Hoyle--Lyttleton accretion. Two modes of AGN feedback are implemented: quasar mode (high accretion rate) and radio mode (low accretion rate) feedback. The quasar mode returns thermal energy to the BH's neighbouring gas cells. On the other hand, the radio mode exerts mechanical feedback where the AGN jets inflate hot bubbles which rise buoyantly in the halo. The radiative loss from the radio mode uses the scaling relations of \citet{Nulsen_et_Fabian2000}. \rr{In addition to the two feedback modes, the simulation includes radiative feedback (photoheating) from the AGN which affects the gas heating/cooling rates. The ionization balance is not explicitly calculated within the simulation, instead the cooling/heating rates are interpolated from \textsc{cloudy} tables. However, we calculate the ionization balance in post-processing whilst accounting for an AGN radiation field to be consistent with the heating/cooling rate calculations done within the simulation (see Section \ref{sec:methods/ionization} for details).} Radiative processes and gas heating/cooling are also treated within the simulation as described in \citet{Illustris}. The associated physics model is discussed in detail in Section \ref{sec:methods/ionization}. We refer the interested reader to the above references for more details about the galaxy formation model.

\rr{The physical model used in this study and close variations thereof have been used in both large-scale cosmological simulations and cosmological zoom-in simulations. The model has been successful at reproducing observable CGM properties, e.g. (i) the bimodality in \ion{O}{vi} covering fractions between passive and star forming galaxies \citep{Suresh2017}, (ii) the correlation between the circum-galactic \ion{H}{i} content and the galaxy's SFR \citep{Marinacci2017}, (iii) the presence of hot coronae around galaxies \citep{Bogdan2015}, and (iv) the number of damped Ly $\alpha$ and Ly limit systems at high redshift \citep{Suresh2015b}. Furthermore, studies employing the same physical model placed constraints on feedback models (stellar and AGN feedback) by comparing simulated CGM properties to observations. AGN feedback has been shown to suppress gas accretion on to the galaxy and launch gas into the CGM and beyond \citep{Auriga}. Additionally, thermal AGN feedback heats the CGM giving rise to hot gaseous haloes \citep{Bogdan2015}, and driving the observed \ion{O}{vi} bimodality \citep{Suresh2017}. On the other hand, \citet{Marinacci2014feedback} used cosmological zoom-in simulations to demonstrate the significant effects of stellar feedback (i.e. galactic winds) on the gas morphology and heavy element distribution in the CGM. Studies of \ion{C}{iv} in the CGM of simulated galaxies indicate that energetic stellar winds, which are superior at enriching the CGM, are favoured \citep{Bird2016, Suresh2015b}.}

%
\subsection{Tracking the CGM gas and metals}
\label{sec:methods/gas}
\begin{figure*}
  \includegraphics[width=\linewidth]{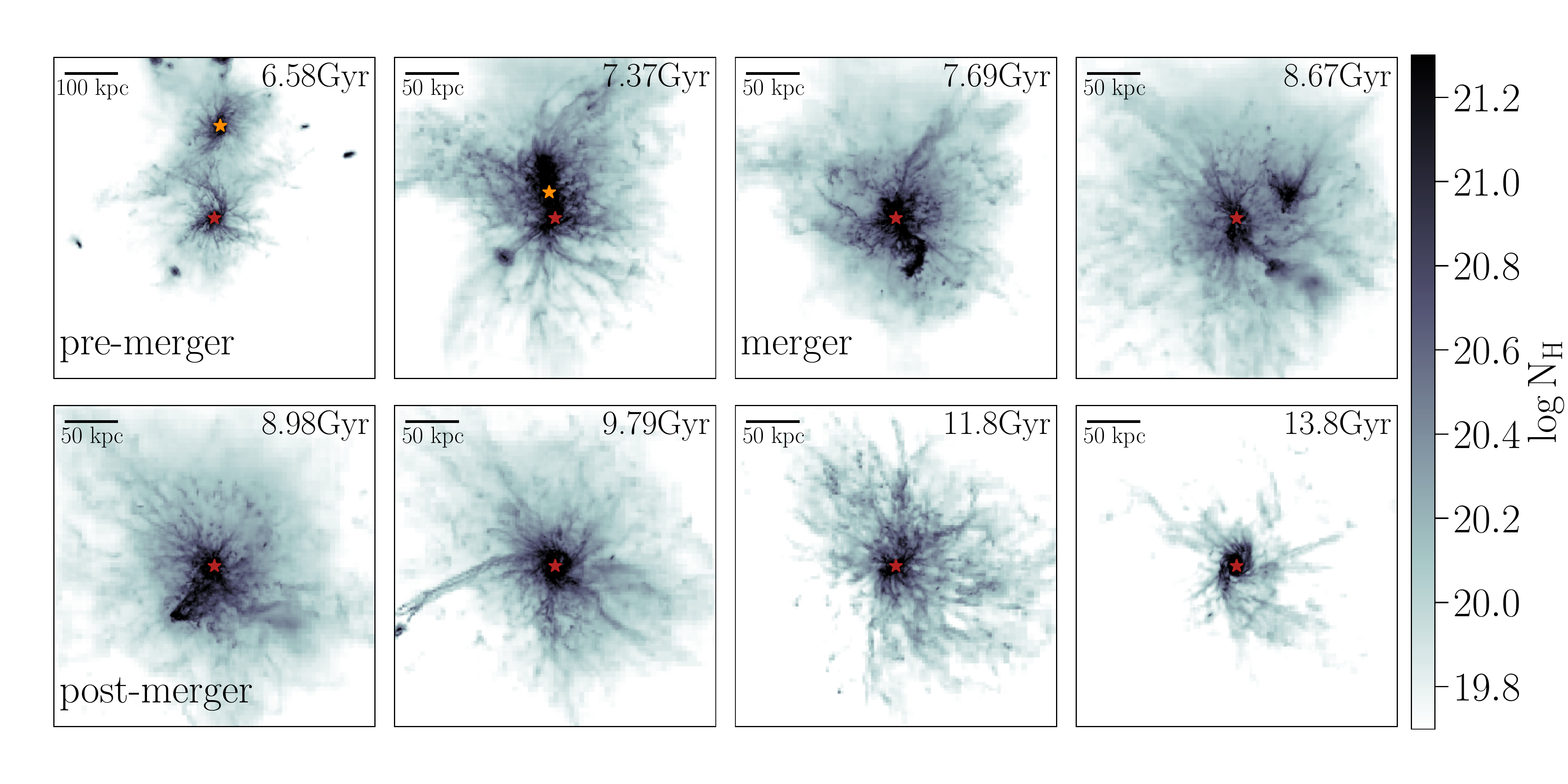}
  \includegraphics[width=\linewidth]{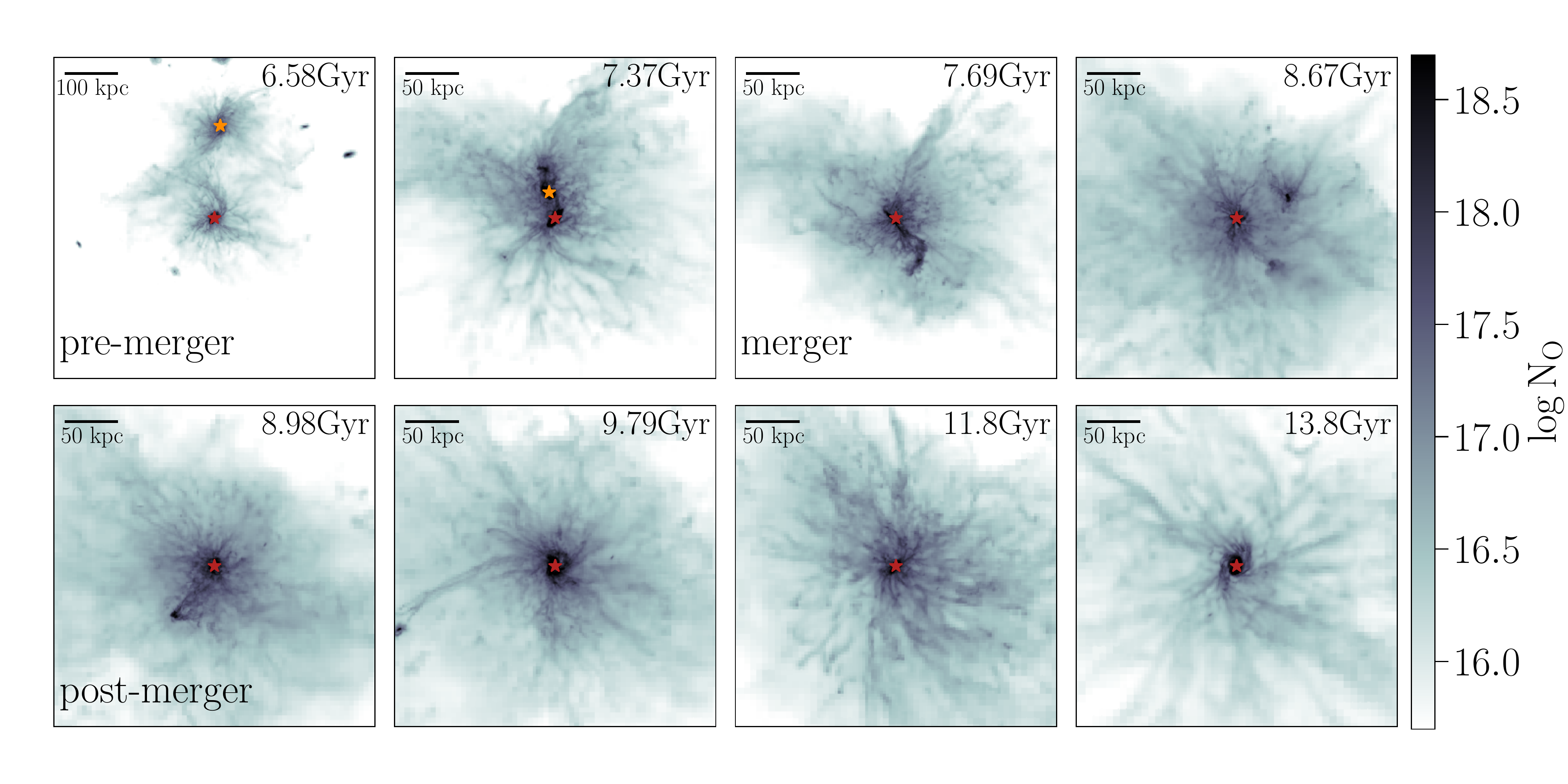}
\caption{$300 \times 300$ physical kpc$^2$ projections \rr{(except the left-most projections in the first and third rows for demonstration purposes only: $600 \times 600$ physical kpc$^2$)}, centred at the most massive progenitor, showing the evolution of the galaxy throughout the stages of the merger. The top two rows depict projections of the hydrogen gas, while the bottom two rows show the projections of the oxygen gas. The bar in the upper-left corner of each frame represents $50$ kpc for scale \rr{($100$ kpc for the left-most projections in the first and third rows)}, while the cosmic time stamp is annotated in the upper-right corner. \rr{The position of the primary (secondary before coalescence) galaxy is marked by a red (orange) star.} The frames corresponding to iconic evolutionary stages are annotated in the lower-left corner accordingly (i.e. pre-merger, merger, post-merger). The post-merger CGM contains significantly more oxygen compared to the pre-merger galaxy.}
\label{fig:projections_H-et-O}
\end{figure*}

\noindent
In order to track the CGM properties of the merging galaxies in an even-handed fashion with CGM observations, viz. calculating the column densities of various ionic species in the CGM, we project a slice of depth 1.35 physical Mpc, centred at the most massive progenitor in the merger, along a random projection axis which is then fixed for all snapshots throughout the analysis. Note that the choice of projection axis does not influence our results. The projections use a static Cartesian grid with 293 pc pixels (comparable to the spatial resolution of the simulation) and a $300 \times 300$ physical kpc$^2$ field of view (FOV) corresponding to impact parameters ($\lesssim 150$ kpc) which are in accord with current observational surveys targeting the CGM \citep[e.g., ][]{Tumlinson2011, Tumlinson2013, COS-Halos_metals,  Bordoloi2014, Peeples2014, Werk2014, Borthakur2015, COS-Gass, Werk2016, Prochaska2017}. Figure \ref{fig:projections_H-et-O} shows the gas projections in hydrogen (top two rows) and oxygen (bottom two rows). The projections were performed using the oct-tree projection routines implemented in the \texttt{yt} analysis tool-kit version 3.3.2 \citep{yt}.

\begin{figure}
\includegraphics[width=\columnwidth]{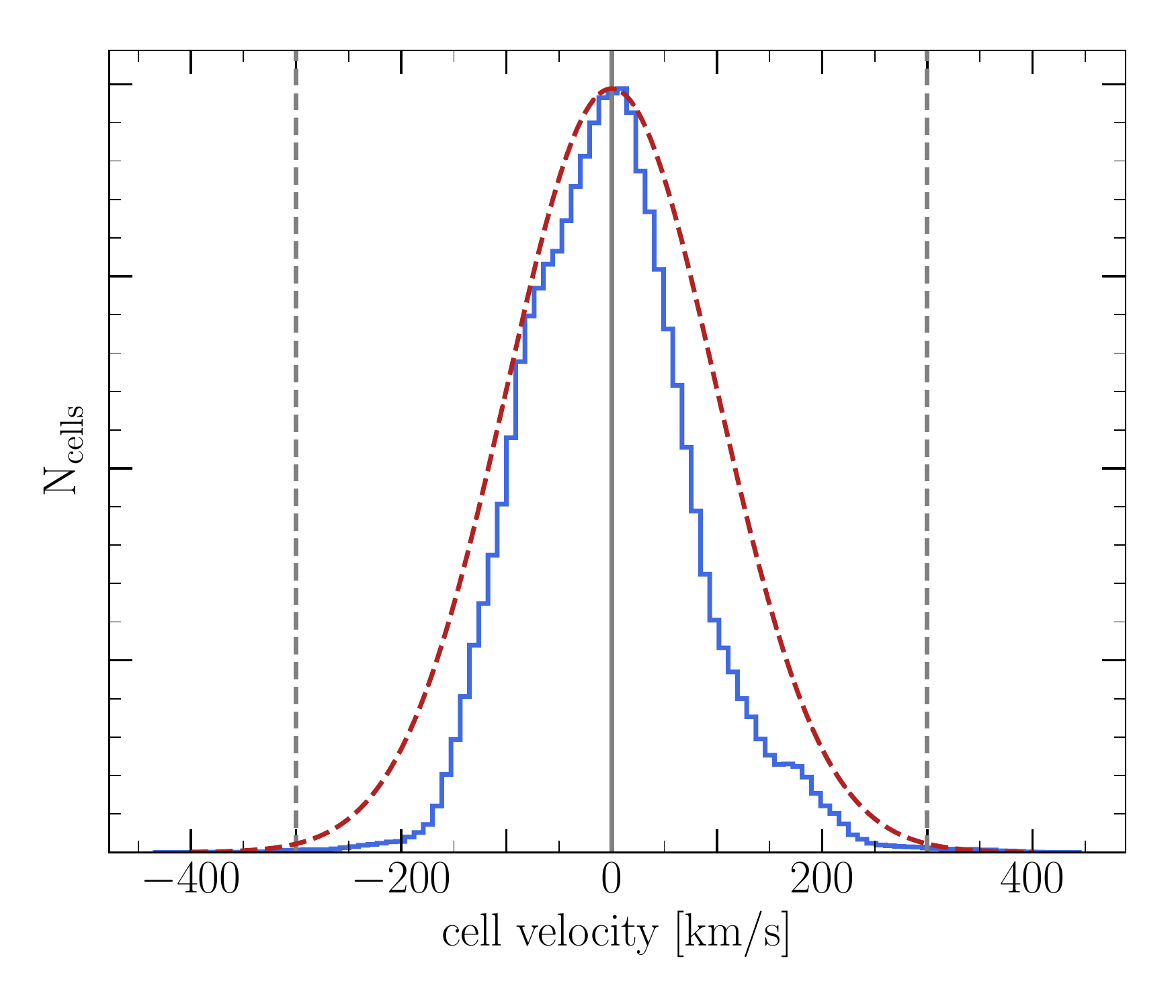}
\caption{The line-of-sight velocity distribution of gas cells within the projection slice relative to the galaxy's systemic velocity (blue histogram). The distribution of gas resolution elements is well within  $\pm 300$ km/s (grey dashed lines) which is more stringent than that nominally adopted in observational studies of the CGM ($\pm 500$ km/s), thus motivating the lack of need for further constraints on the velocities. For visual comparison, the dashed brick-red line shows a normal distribution with a characteristic $\sigma = 100$ km/s.}
\label{fig:los-vel_distribution}
\end{figure}

Unlike observational surveys targeting galaxy mergers, we do not impose a line-of-sight velocity cut since our stringent spatial projection domain implies a velocity distribution well within the $500$ km/s cut-off traditionally employed in observations \citep{Tumlinson2011, Tumlinson2013, COS-Halos_metals,  Bordoloi2014, Peeples2014, Werk2014, Borthakur2015, COS-Gass, Werk2016, Prochaska2017}. Figure \ref{fig:los-vel_distribution} shows a typical line-of-sight velocity distribution of gas elements in our projection slice which lies well within $v_\mathrm{los} \sim \pm 300$ km/s. Because element abundances are tracked continuously in the simulation (see Sections \ref{sec:methods/GFM}), our projections include projected number densities (column densities) of each element included in the simulation. Therefore, we are able to analyse the evolution of the CGM's chemical composition on an element-by-element basis.

We will frequently make reference to some key evolutionary stages that mark the evolution of the merger. We define the merger to correspond to the maximum specific BH accretion rate. The pre-merger (post-merger) time is defined to be 1 Gyr prior to (post) the merger (see Figure \ref{fig:projections_H-et-O}).

%
\subsection{CGM ionization}
\label{sec:methods/ionization}
\noindent
It is particularly interesting to study the ionization of the CGM as this includes valuable information about the observed temperature and density structure of the gas within. Moreover, predictions of the column densities of particular species facilitate a comparison with observations. In this section, we discuss the details of the ionization calculations.

We use {\sc cloudy} v13.03 \citep{Cloudy} to calculate the ionization fractions of metal species under the assumption of ionization equilibrium for a gas exposed to an external radiation field. Both photoionization and collisional ionization processes are included in our calculations. We run a grid of {\sc cloudy} simulations which we then interpolate, using multilinear interpolation, to calculate the ionization fractions of every gas cell. The grid of {\sc cloudy} simulations spans a five-dimensional space of metallicities, temperatures, densities, redshifts, and bolometric AGN intensities summarized in Table \ref{tab:ionization_grid}.

\begin{table}
  \centering
\caption{The grid of {\sc cloudy} simulations used to calculate gas ionization fractions.}
\label{tab:ionization_grid}
\begin{tabular}{lcc}
\hline
                                 & Grid range   & Grid increments\\
\hline
\hline
$\log(Z \, [Z_\odot])$                      & [$-4$, 1]        & 1.0 dex  \\
$\log(T \, [$K$])$                         & [1, 9]         & 0.1 dex  \\
$\log({n_\text{H}} \, [$cm$^{-3}])$                & [$-8$, 2]        & 0.25 dex \\
$z$                                        & [0, 1.05]      & 0.05     \\
$\log(J_{\mathrm{AGN}} \, [$erg s$^{-1}$ cm$^{-2}])$ & \{0, [$-5$, 5]\} & 1.0 dex  \\
\hline
\end{tabular}
\end{table}

We include two radiation fields in the {\sc cloudy} simulations which are consistent with the radiation fields included in the simulations. We employ the publicly available spatially-uniform redshift-dependent ultraviolet background (UVB) of \citet{UVB} which accounts for radiation from quasars and star-forming galaxies, although the radiation from quasars dominates the UVB at the redshift of the merger ($z \le 1$). AGN radiation is also included assuming a universal, time-independent AGN spectral energy distribution (SED) which is scaled with the bolometric AGN intensity ($J_{\mathrm{AGN}}$) at the position of each gas cell. We adopt the same parametrization used in the simulations \citep{Illustris}, such that the AGN SED is given as
\begin{equation}
  f^{\mathrm{AGN}}(\nu) = \nu^{\alpha_\text{UV}} \exp \left(-\frac{h\nu}{kT_\text{BB}} - \frac{0.01\mathrm{Ryd}}{h\nu} \right) + a\nu^{\alpha_\text{X}} ,
\end{equation}
\noindent
where $T_{\text{BB}}=10^6$ K, $\alpha_{\text{UV}} = -0.5$, $\alpha_\text{X} = -1$, and $a$ is defined by the ratio of the X-ray to the UV component as follows:
\begin{equation}
\frac{f^{\mathrm{AGN}}(2 \, \mathrm{keV})}{f^{\mathrm{AGN}}(2500 \, \text{\AA})} = 403.3^{\alpha_\text{0X}}
\end{equation}
\noindent
with $\alpha_\text{0X} = -1.4$.

At high densities ($n \geq 10^{-3}$ cm$^{-3}$), gas can no longer be assumed to be optically thin to ionizing radiation and it becomes self-shielding to radiation more energetic than 1 Ryd: At high densities hydrogen absorbs notable fractions of the incident radiation for photon energies $\ge 1$ Ryd, therefore attenuating the radiation field. This, in turn, affects the heating and cooling rates of the gas, hence affecting its ionization. To account for self-shielding we suppress the intensity of the incident radiation field, at energies $>1$ Ryd, by the factor given in equation A1 of \citet{Rahmati2013}. The values of the parameters in the self-shielding equation are computed by linearly interpolating table A1 of \citet{Rahmati2013}. Although self-shielding is not expected to play a significant role in the CGM where the densities are lower than the star formation threshold, it is included in our calculations for completeness especially considering that (i) we expect dense clumps to appear in the CGM due to the tidal effects of the merger and outflows, and (ii) we also calculate the ionization of the ISM where self-shielding effects become dominant.

We stress some caveats of the ionization calculations described above. First, {\sc cloudy} is run in single cell mode which assumes that the temperature and densities of the gas are constant. Therefore, we do not resolve any density or thermal substructure in the gas. This is a reasonable assumption given that the {\sc cloudy} calculations are applied below the resolution scale. We also assume a molecule and dust free gas in the {\sc cloudy} simulations because molecules and dust form in molecular clouds at scales well below the resolution limit of our simulations. Moreover, the ionization balance is calculated for a gas with relative solar abundances. Abundance deviations from solar have insignificant effects on the cooling rates of the gas especially given the large uncertainties associated with metal enrichment \cite[e.g., ][]{sim_chem_evol_3}. Therefore, calculating the ionization balance on an element-by-element basis would introduce an unreasonable computational overhead with limited enhancement in the accuracy of the ionization calculations. Additionally, our galaxy formation model does not include an on-the-fly treatment of non-equilibrium ionization \citep[e.g., ][]{Oppenheimer2016, Oppenheimer2017_OVI, Oppenheimer2017}; consequently, we assume ionization equilibrium (collisional and photoionization equilibrium) in all the {\sc cloudy} simulations which poses a limitation to the ionization modelling presented in this study. We revisit the assumption of ionization equilibrium in section \ref{sec:results/ionization}. On the other hand, knowing that star-forming gas is treated with a sub-grid model where the vast majority of the gas is in the form of cold/dense clouds embedded in an ambient hot medium \citep[see figure 1 of][]{ISM}, we consider star-forming gas to be fully neutral in our post-processing. This does not have any effect on the results of our work as we focus on the CGM, however, one should proceed with caution when extending our analysis to regions where star formation is dominant (i.e. ISM; galactic disc). Lastly, we do not include the stellar radiation component of the host galaxy's SED. This is solely due to the radiation from the stellar component not being included in the galaxy formation model used in the simulations. Therefore, including a stellar radiation component in post-processing would result in an inconsistent treatment of the state of the gas (i.e. underestimating the effect of the radiation field on gas heating). Moreover, \citet{Werk2016} showed that a stellar radiation field is only effective at altering the ionization of the CGM for extreme SFRs and low impact parameters \citep[see section 4.3 of][]{Werk2016}.

\section{Results}
\label{sec:results}
\noindent

In the course of our analysis, we will present the results for oxygen as a representative species for other metals (except when discussing the ionization of the gas); all other elements which are tracked in the simulation show similar qualitative trends. 

In keeping with observational studies of galaxy mergers \citep[e.g., ][]{Ellison08, Ellison2011, Patton2011, Scudder2012, Ellison2013,  Patton2013,  Patton2016} we look at the differential changes in the CGM as a result of the merger. Such an approach facilitates quantifying the effects of the merger. The aforementioned observational studies compare a `merger sample' to an `isolated (control) sample'. Similarly, previous numerical work studying the effects of mergers \citep[e.g., ][]{Torrey2012, Moreno2015} simulated their galaxies in isolation which allowed a comparison to  an isolated `control sample'. Alas, that option is a sacrifice which we must bare when using cosmological simulations which provide a more realistic approach to study the CGM (see Section \ref{sec:methods}). Hence, we use the pre-merger properties of the CGM as the reference point in our comparison. The pre-merger CGM properties are calculated over \rr{a $\sim 100-200$ Myr range 1 Gyr before the merger (at $t \sim 6.6$ Gyr) where the CGM of the galaxy shows remarkable stability (in the metrics we study).}

%

\begin{figure}
  \centering
  \includegraphics[width=\columnwidth]{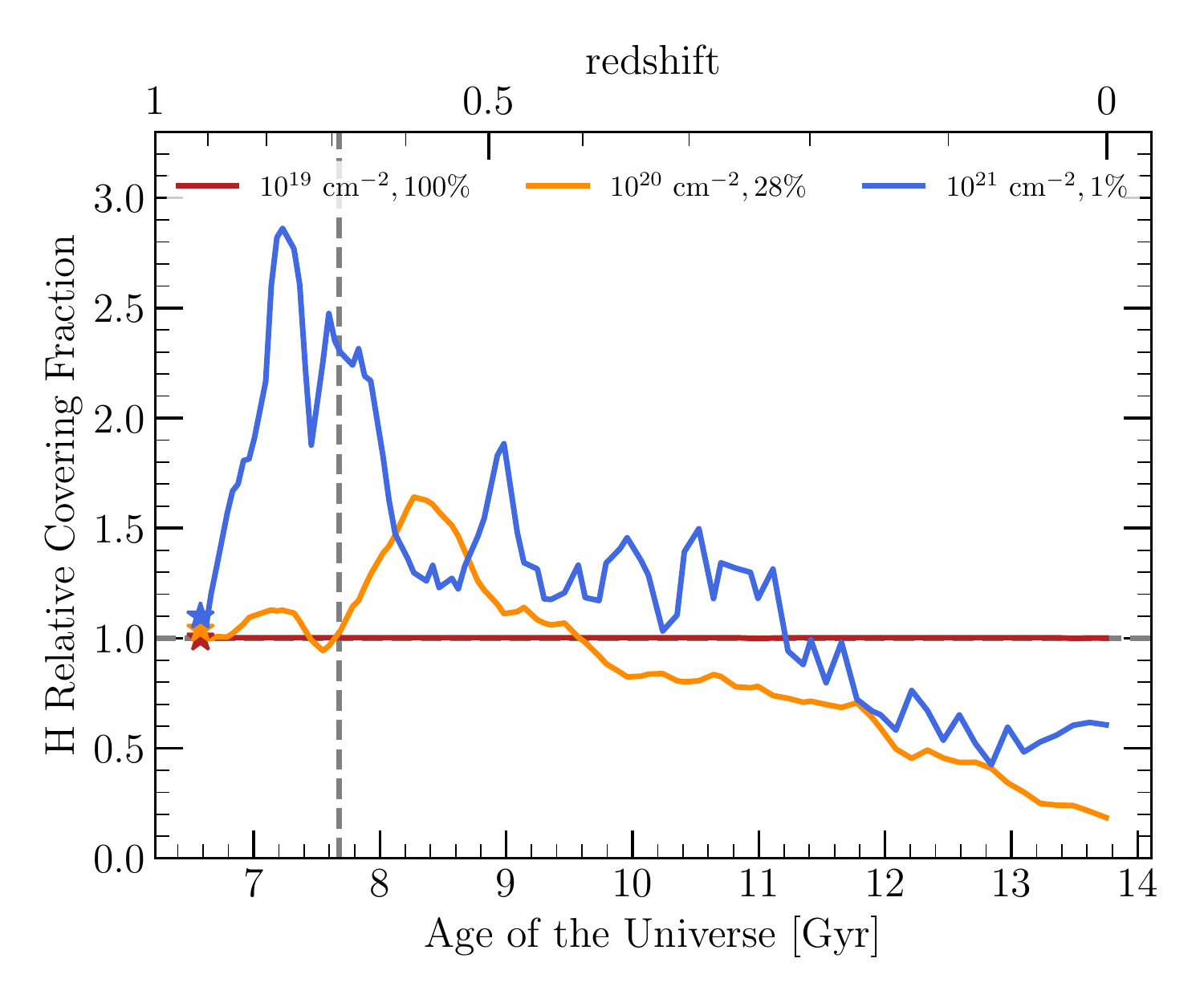}
\caption{The evolution of the covering fraction of hydrogen absorbers (normalized to the pre-merger covering fraction) during the merger. The brick-red, orange, and blue curves represent absorbers with column densities higher than $10^{19}$, $10^{20}$, and $10^{21}$ cm$^{-2}$, respectively. \rr{The covering fractions in the CGM of the secondary galaxy are marked by stars; the CGM of both galaxies show remarkable similarity.} The legend shows the column density threshold and the associated pre-merger absolute covering fraction. The vertical dashed grey line indicates the time of the merger. The covering fractions are normalized relative to the pre-merger epoch; hence the variation in the CGM covering fractions is a relative variation. As the second galaxy enters the FOV, the covering fraction of all absorbers increases. Shortly after, the covering fractions decrease during the first passage as the galaxies overlap along the line of sight. After the merger, the covering fractions of the highest column density absorbers remain enhanced (compared to the pre-merger stage) yet decrease compared to the merger as tidal torques and feedback processes redistribute the gas. This is evident in the increasing covering fractions of absorbers in the neighbouring column density bin.}
\label{fig:H_covering_fraction}
\end{figure}

\begin{figure}
  \centering
  \includegraphics[width=\columnwidth]{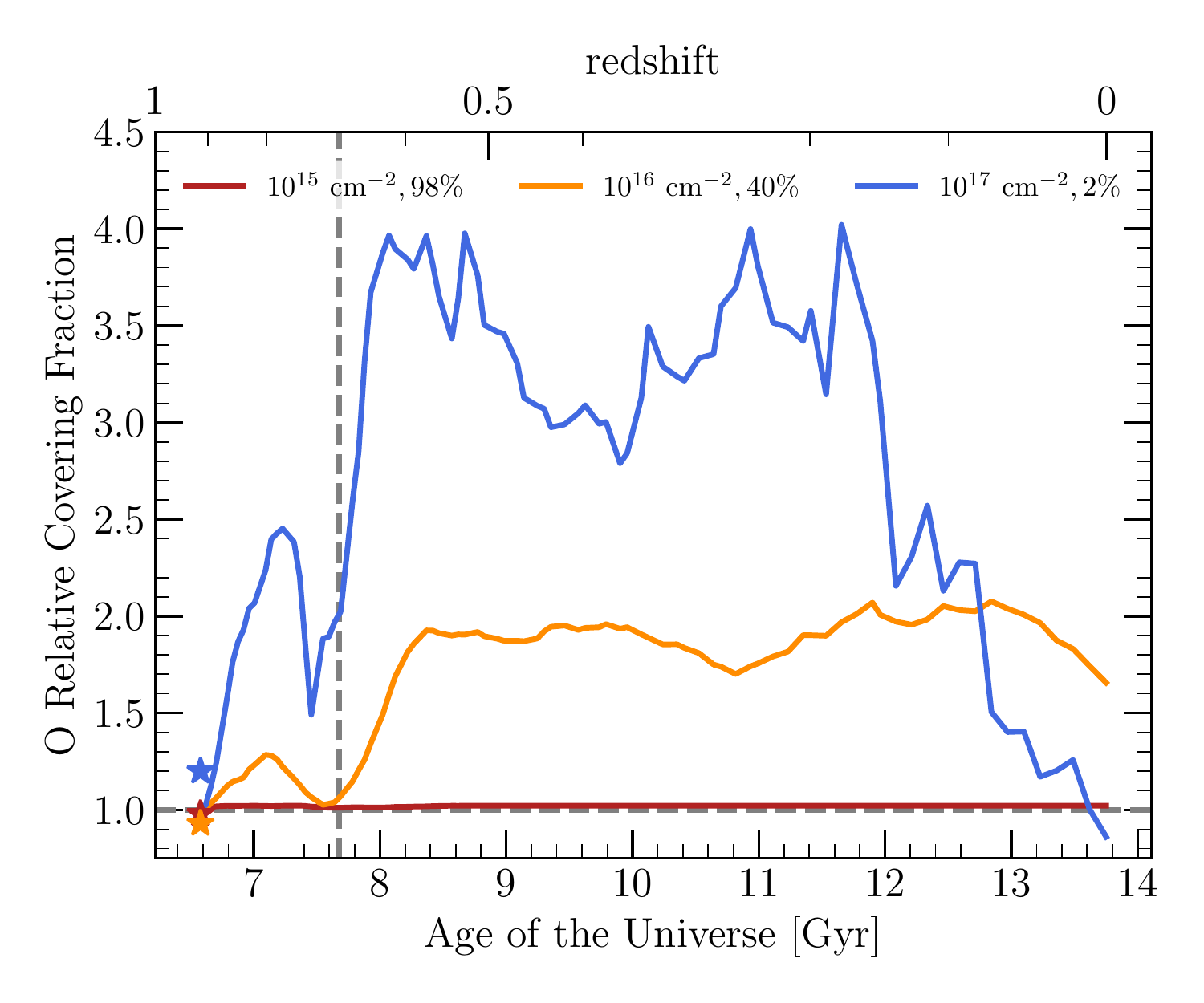}
\caption{Similar to Figure \ref{fig:H_covering_fraction}, this figure represents the evolution of the covering fraction of oxygen during the major merger (normalized relative to the pre-merger epoch). The brick-red, orange, and blue curves represent column densities higher than $10^{15}$, $10^{16}$, and $10^{17}$ cm$^{-2}$, respectively. \rr{The covering fractions in an identical FOV centred at the secondary galaxy are marked by stars.} The merger populates the CGM with metals which are transported via galactic winds. The enhanced metal covering fractions persist for many Gyrs after the merger has concluded.}
\label{fig:O_covering_fraction}
\end{figure}

\subsection{Gas and metal covering fraction}
\label{sec:results/cover}

%
\subsubsection{CGM covering fractions}
\noindent
To quantify the global effects of the merger on the CGM gas and metal content, we first examine the evolution of the covering fraction of gas throughout the merger. The covering fraction is a metric frequently used in observational studies of the CGM \citep[e.g., ][]{Tumlinson2011, Tumlinson2013, COS-Halos_metals,  Bordoloi2014, Peeples2014, Werk2014, Borthakur2015, COS-Gass, Werk2016, Prochaska2017}. The covering fraction is defined as the fraction of lines of sight intersecting an absorber denser than a chosen column density threshold. Equivalently, it can be thought of as the fraction of the FOV (`observed' area) containing absorbers above a given column density threshold. The FOV in the following analysis is defined as a square\footnote{Note that the choice of the FOV's shape does not affect the presented results (square versus circle). We use a square FOV for computational simplicity.} with a length of 300 kpc centred at the more massive merging galaxy. \rr{For comparison, all metrics used in this study are also calculated in an identical FOV centred at the secondary galaxy at the pre-merger time. Note that centring the FOV at the secondary galaxy has no effect on the conclusions of this study as both galaxies show remarkable similarities in their CGM properties.}

Figures \ref{fig:H_covering_fraction} and \ref{fig:O_covering_fraction} show the evolution of the covering fraction of gas (hydrogen and oxygen respectively) above a given column density threshold. The vertical dashed grey line represents the merger time at $t=7.6$ Gyr. For a visual projection following the merger's evolution see Figure \ref{fig:projections_H-et-O}.

Before the merger ($t \sim 6.58$ Gyr), the galaxy gas distribution exhibits a disc-like morphology extending to $\sim 100$ kpc. Figure \ref{fig:projections_H-et-O} shows the galaxy in a near face-on projection with an extended gas reservoir ($N_\mathrm{H}>5\times10^{19}$ cm$^{-2}$ out to $\sim 100$ kpc). As the merger proceeds, the second galaxy enters the FOV causing an increase in the gas covering fractions for all column densities until $t\sim 6.6$ Gyr when the galaxies begin to overlap. This increase is due to an increasing area which is covered by absorbers associated with the merging galaxy. At first passage ($t \sim 7$ Gyr), the superposition of the two galaxies along the line of sight causes a geometrically driven decrease in the gas covering fraction shortly before the merger occurs. The covering fraction of hydrogen and oxygen increases at the time of merger ($t\sim 7.6$ Gyr, indicated by the vertical dashed grey line) by factors of 2 and $2-3$ relative to the pre-merger covering fractions, respectively.

Immediately after the merger ($t\sim 8$ Gyr), the covering fraction of the highest column density hydrogen gas ($\log(N_\mathrm{H}) > 21$) declines while coinciding with an increase in the covering fraction of absorbers with column densities $\log(N_\mathrm{H}) > 20$ (see Figure \ref{fig:H_covering_fraction}). During the merger, dense gas is heated by SN and AGN feedback and tidally disrupted. The combined impact of feedback and tidal fields disrupts and diffuses the gas disc, which drives a decline in the highest hydrogen column densities. On the other hand, outflows primarily contain metal rich gas and the tidally disrupted disc is more metal rich compared to the surrounding medium. Therefore, the covering fraction of oxygen increases after the merger ($t > 7.6$ Gyr) by factors of 2 and 3 for column densities $\log(N_\mathrm{O})> 16$ and $\log(N_\mathrm{O})> 17$ respectively.

The major merger has a lasting impact on the CGM. The covering fraction of hydrogen gradually decreases for several Gyrs post-merger ($t>8$ Gyr) but remains above the pre-merger covering fraction until $t\sim 11$ Gyr. The enriched CGM persists well above its pre-merger stage (increase in the covering fraction of oxygen by factors of $2-3$) for several Gyrs post-merger (Figure \ref{fig:O_covering_fraction}). Metals remain present at significant column densities until $z=0$.

%
\subsubsection{Measuring the extent of the CGM}
\begin{figure}
  \centering
  \includegraphics[width=\columnwidth]{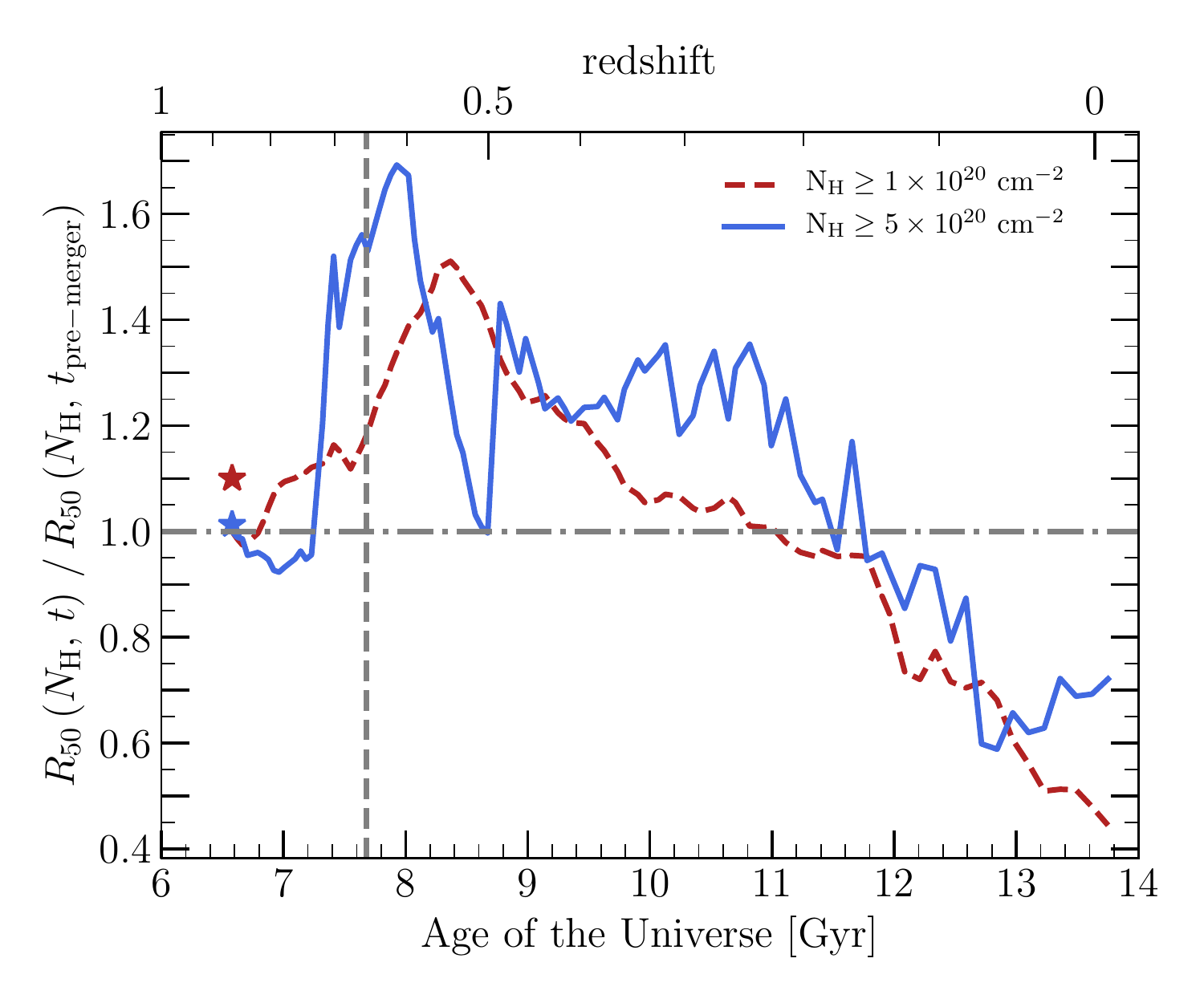}
\caption{The evolution of the $R_{50, \, \mathrm{H}} (10^{20}\, \mathrm{cm}^{-2}, \, t)$ (dashed brick-red line) and $R_{50, \, \mathrm{H}} (5\times 10^{20}\, \mathrm{cm}^{-2}, \, t)$  (blue solid line) during the merger. The plot shows $R_{50, \, \mathrm{H}}$ normalized to the pre-merger epoch ($R_{50, \, \mathrm{H}}(N_\mathrm{H}, \, 6.58\, \mathrm{Gyr}$)). \rr{The stars show the corresponding $R_{50}$ of the secondary galaxy relative to the primary galaxy's $R_{50}$.} The merger increases $R_{50, \, \mathrm{H}}$ as the merger vigorously redistributes the gas. As the merger settles, the $R_{50, \, \mathrm{H}}$ remains enhanced, albeit to a lesser extent, until the galaxy's activity perishes and the dense CGM gas starts to diffuse without being replenished.}
\label{fig:H_radius_vs_t}
\end{figure}

\begin{figure}
  \centering
  \includegraphics[width=\columnwidth]{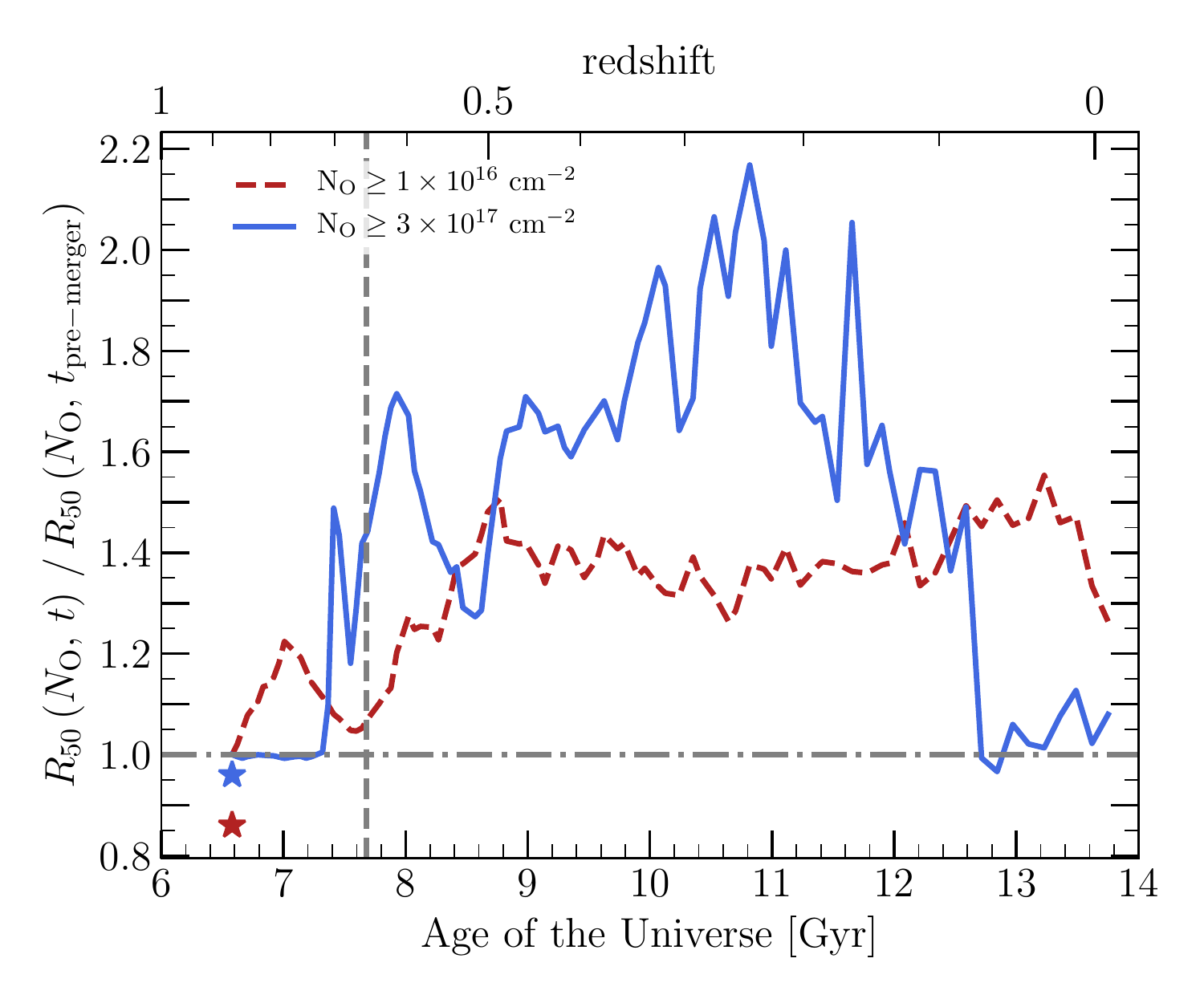}
\caption{Similar to figure \ref{fig:O_radius_vs_t}, showing the evolution of $R_{50, \, \mathrm{O}} (10^{16}\, \mathrm{cm}^{-2}, \, t)$ (dashed brick-red line) and $R_{50, \, \mathrm{O}} (3\times 10^{17}\, \mathrm{cm}^{-2}, \, t)$ (blue solid line) during the merger. Similar to hydrogen, the merger increases the covering of oxygen absorbers as the CGM is populated with metal enriched gas. The increase in $R_{50, \, \mathrm{O}}$ persists long after the merger concludes.}
\label{fig:O_radius_vs_t}
\end{figure}

To further quantify the changes in the extent of the CGM as a result of the merger, we define $R_{50, \, \mathrm{X}} =  R_{50, \, \mathrm{X}}( N_\mathrm{X}, \, t)$  as the radius where the covering fraction of lines of sight with column densities $N > N_\mathrm{X}$ is at least 50\%. The definition of $R_{50, \, \mathrm{X}}$ therefore defines a characteristic size for the CGM. Figures \ref{fig:H_radius_vs_t} and \ref{fig:O_radius_vs_t} show the evolution of $R_{50, \, \mathrm{H}}$ (hydrogen) and $R_{50, \, \mathrm{O}}$ (oxygen), respectively.

Prior to coalescence ($t<7.6$ Gyr), the area populated with dense hydrogen absorbers increases, therefore increasing $R_{50, \, \mathrm{H}}$. After the merger ($t\sim 8$ Gyr), as the tidally disrupted merger remnant settles and the feedback driven outflows push gas outwards, the total hydrogen $R_{50, \, \mathrm{H}} (5\times 10^{20} \, \mathrm{cm}^{-2}, \, 8 \,\mathrm{Gyr})$ decreases back to the pre-merger state. The highest column density hydrogen is more centrally located (the galactic disc) while the outflowing gas diffuses outwards causing an increase in $R_{50, \, \mathrm{H}}(10^{20}\, \mathrm{cm}^{-2}, \, t)$ by a maximum factor of $\sim 1.5$ which then declines as the gas continues to diffuse below the threshold. The characteristic size of the CGM as observed in hydrogen remains enhanced by a factor of $\sim 1.3$ for $\sim 3$ Gyr for $N_\mathrm{H} > 5\times 10^{20}$ cm$^{-2}$.

$R_{50, \, \mathrm{O}}$ for oxygen absorbers increases after the first passage ($t\sim 7.4$ Gyr) as two processes contribute to the observed increase: (1) the induced starburst enriching the gas, and (2) the geometric superposition of the galaxies increases the column densities. After the merger ($t>7.6$ Gyr), metal rich outflows cause an increase in $R_{50, \, \mathrm{O}} ( 10^{16} \, \mathrm{cm}^{-2}, \, t>7.6 \, \mathrm{Gyr})$ and $R_{50, \, \mathrm{O}}(3\times 10^{17} \, \mathrm{cm}^{-2}, \, t>7.6 \, \mathrm{Gyr})$ by factors of 2.2 and 1.6, respectively. The increase persists for $> 3$ Gyr post-merger.


\textbf{The effect of mergers on the CGM `size':} In summary, using two different metrics for CGM `size', the covering fraction and $R_{50, \, \mathrm{X}}$, we have demonstrated that the merger can lead to a long-lasting ($>4$ Gyr) increase in the physical extent of halo gas. The CGM growth is evident in both hydrogen and metals. It is worth noting that the aforementioned CGM growth may not translate into an increase in the observed covering fractions of the ionized species which are influenced by the underlying radiation field (see Section \ref{sec:results/ionization}).

%
\subsection{Gas migration and enrichment}
\label{sec:results/content}

%
\subsubsection{The CGM density profile}
\label{sec:results/content/density}

\begin{figure}
  \includegraphics[width=\columnwidth]{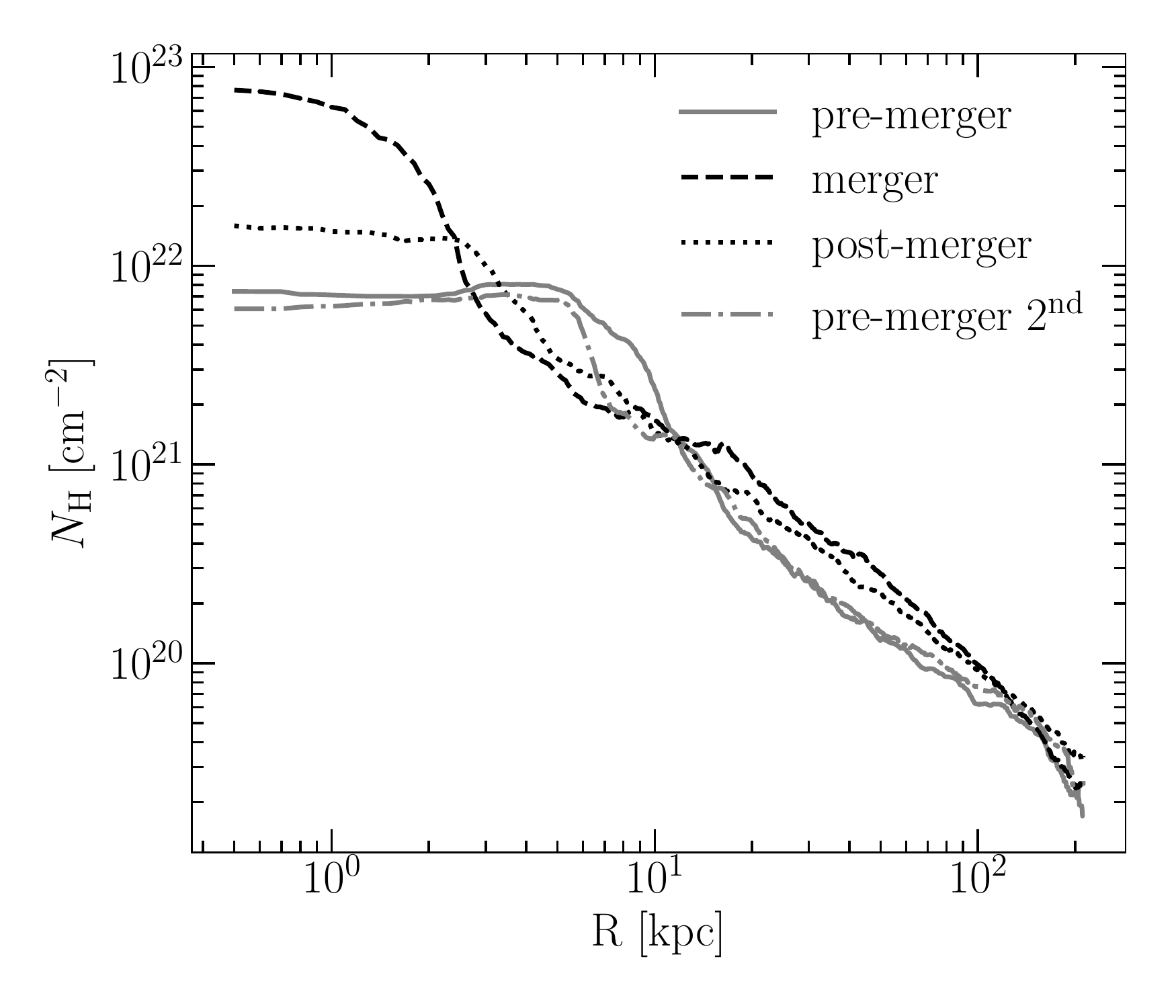}
  \includegraphics[width=\columnwidth]{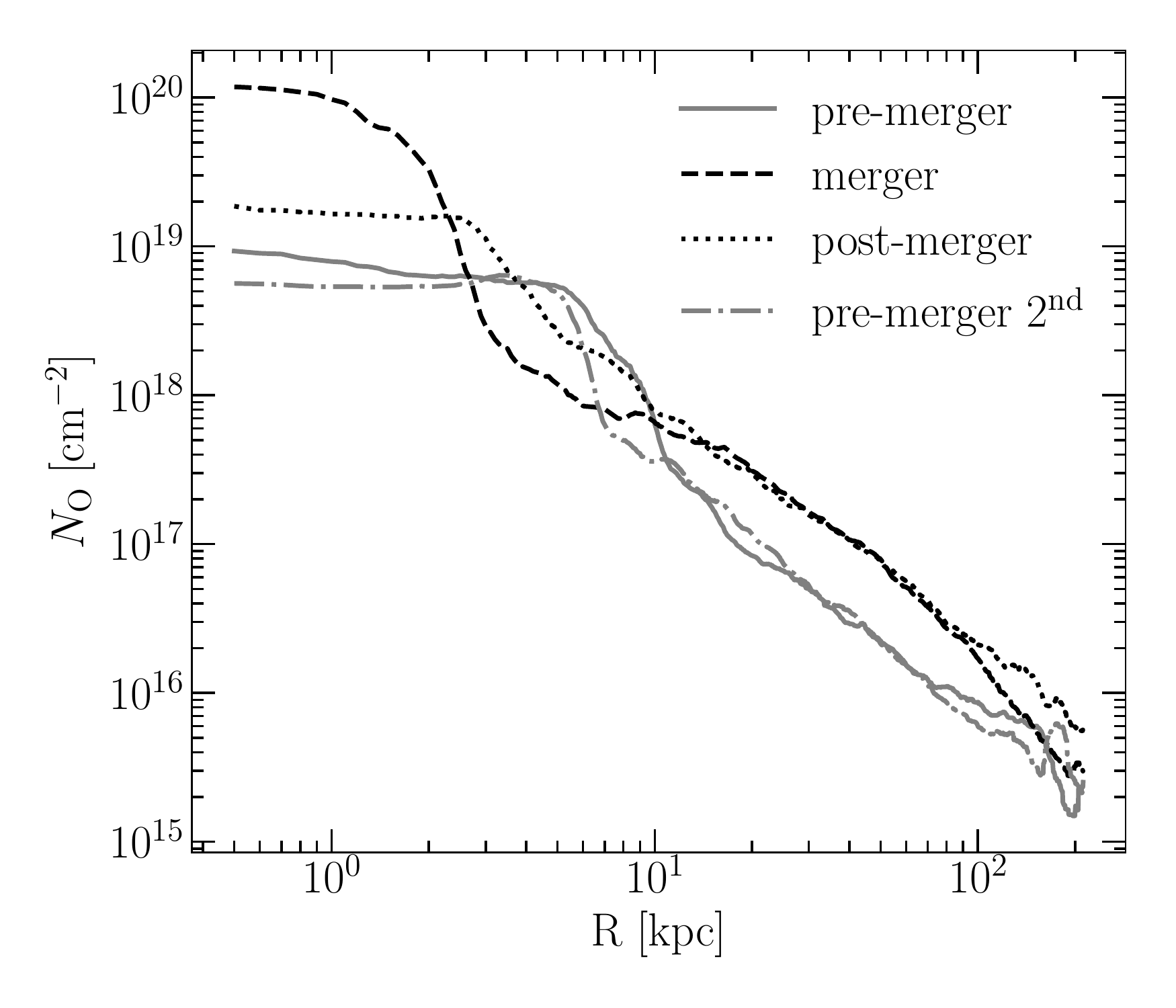}
  \caption{The median hydrogen (top) and oxygen (bottom) projected 2D density profile for three selected epochs: pre-merger (grey solid line), merger (black dashed line), and post-merger (black dotted line). \rr{The secondary galaxy's profiles are shown by the dot--dashed grey lines at the pre-merger epoch. The two galaxies show remarkable similarity in their density profiles.} The merger funnels gas into the central regions of the galaxy while populating the CGM with enriched gas from the galactic disc eventually creating a flatter surface density profile. The enhanced column densities in the CGM persist long after the merger and progressively propagate to larger radii.}
\label{fig:gas_median_vs_r}
\end{figure}

\noindent
In Section \ref{sec:results/cover} we demonstrated how the merger increases the covering fractions of hydrogen and metal absorbers in the CGM. The effects of the merger populating the CGM with metal enriched gas through feedback processes can also be seen by examining the galaxy's projected density profile. Figure \ref{fig:gas_median_vs_r} shows the projected column density profiles of hydrogen and oxygen at three epochs: pre-merger (solid grey line), merger (dashed black line), and post-merger (dotted black line). During the merger, the column densities of hydrogen and oxygen in the central regions ($<2$ kpc) increase by an order of magnitude as the merger tidally induces gas inflows due to the strong tidal torques \citep[e.g., ][]{BH91, BH96, Kewley2006, Ellison08,  Scudder2012, Torrey2012}. Consequently, the gas column densities in the outer disc region ($R \in [2,10]$ kpc) decline. Additionally, outflows induced by star formation and AGN activity after the first peri-centre passage populate the CGM with gas thus increasing the density profile between 10 and 100 kpc by a factor of $\sim 3$. After the merger, the central gas reservoir is further depleted being converted to stars, feeding the central BH, or being launched through outflows causing a decline in the central densities. Nonetheless, the post-merger central density profile remains enhanced when compared to its pre-merger counterpart. The outflows responsible for the increased density profile in the CGM continue to propagate outwards causing an increase at radii $>100$ kpc post-merger compared to the pre-merger and merger profiles.

\textbf{The effects of the merger on the CGM density profile:} Succinctly, Figure \ref{fig:gas_median_vs_r} shows that the merger produces a shallower projected density profile by increasing the column densities at larger radii ($>10$ kpc). After the merger, the central densities decline as the gas is converted into stars or launched in outflows. The CGM remains enriched compared to the pre-merger stage at radii $>10$ kpc until $z=0$; the enrichment progressively propagates outwards consistent with the gas being transported outwards by outflows.

%
\subsubsection{The CGM metallicity}
\label{sec:results/content/metallicity}
\begin{figure}
  \centering
  \includegraphics[width=\columnwidth]{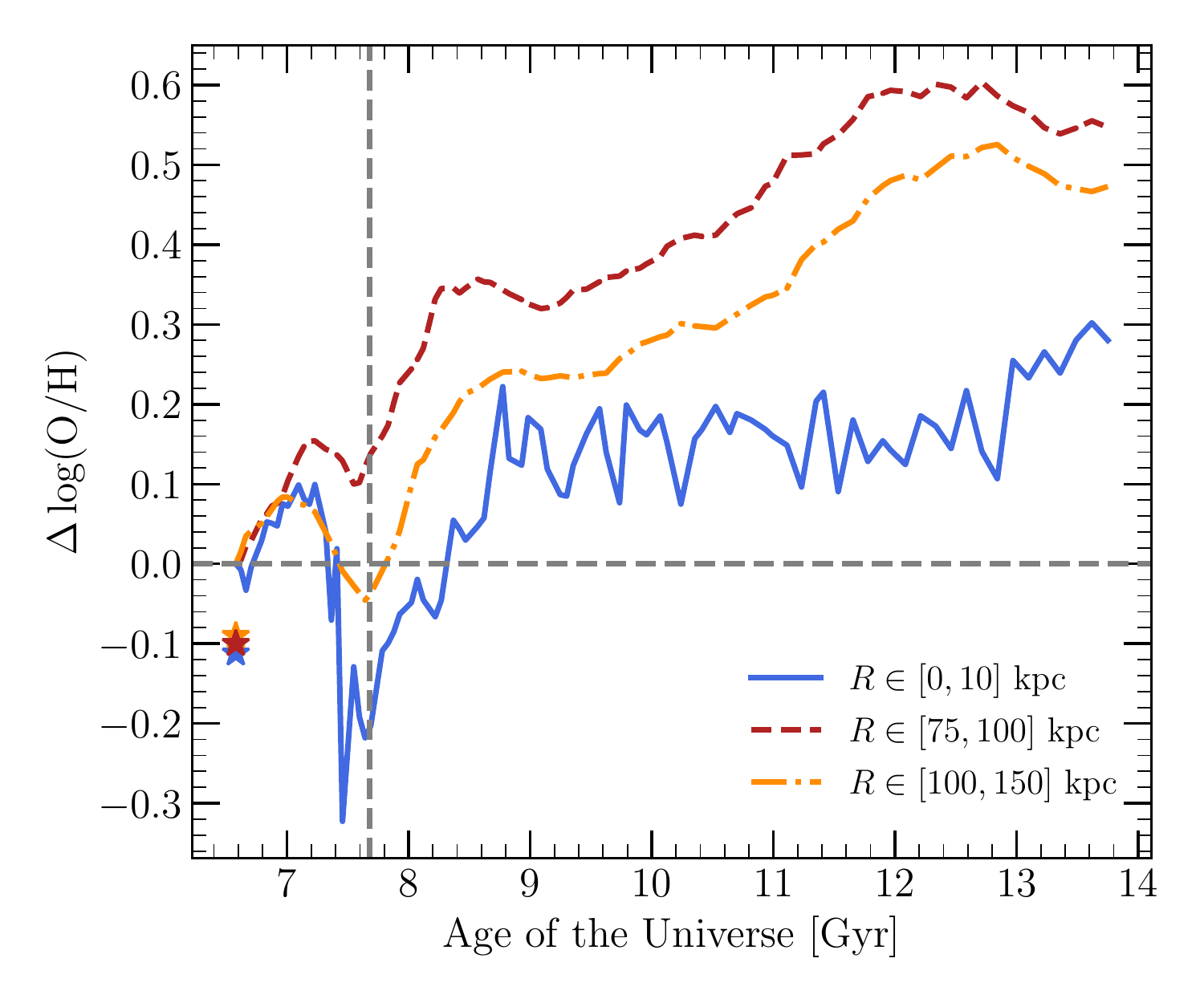}
\caption{The evolution of the median oxygen abundance (normalized to the pre-merger O/H) in different radial bins (coloured lines) during the merger. \rr{The metallicity of the secondary galaxy relative to the primary galaxy at the pre-merger epoch is marked by star symbols representing the different radial bins. The secondary galaxy is metal poor compared to the primary galaxy.} The vertical grey line marks the merger time. The merger increases the metallicity of the gas in the CGM. The long-lived enhancement propagates outwards; this is illustrated by the delayed response of the orange and brick-red lines. Known effects of mergers are also evident, i.e. the dilution effect in the central region of the galaxy due to metal poor gas being tidally funnelled into the centre during mergers.}
\label{fig:O/H_median_vs_t}
\end{figure}

\noindent
In addition to significant evolution in the gas density profile for merging galaxies, the gas metallicity profile can be impacted through metal enriched outflows that accompany the starburst phase of the merger sequence.  To explore changes in the gas metallicity profile, Figure \ref{fig:O/H_median_vs_t} shows the evolution of the median $\log$(O/H) in three radial bins. The vertical grey line represents the merger time. The blue line shows evident signs of dilution in the central regions ($R<10$ kpc) after first passage. The central dilution is caused by low metallicity gas being funnelled towards the centre of the galaxy due to merger-induced tidal torques, an effect which is well documented observationally \citep[e.g., ][]{Kewley2006, Ellison08, Scudder2012}. The central gas metallicity is replenished by stellar enrichment from the merger-induced star formation $\sim 700$ Myr after the merger. Meanwhile, the metallicity of the CGM, at $R>75$ kpc, monotonically increases after the merger. The CGM metallicity remains elevated by $\log$(O/H)$ \sim 0.2-0.3$ dex until $z=0$ (6 Gyr post-merger). Comparing the metallicity enhancement in the outer two radial bins, $R \in [75, 100]$ kpc and $R \in [100, 150]$ kpc, shows a delay in the metallicity enhancement which is indicative of out-flowing metal rich gas propagating to higher radii. The delay of $\sim 300$ Myr is consistent with the typical velocity of galactic winds (a few 100 km/s); for example, if we assume a wind with constant velocity $v \sim 100$ km/s travelling $\Delta R \sim 25$ kpc, the expected delay would be $\sim 244$ Myr. In the following section, we address in more detail the mechanisms responsible for enriching the CGM.

\begin{figure}
  \centering
  \includegraphics[width=\columnwidth]{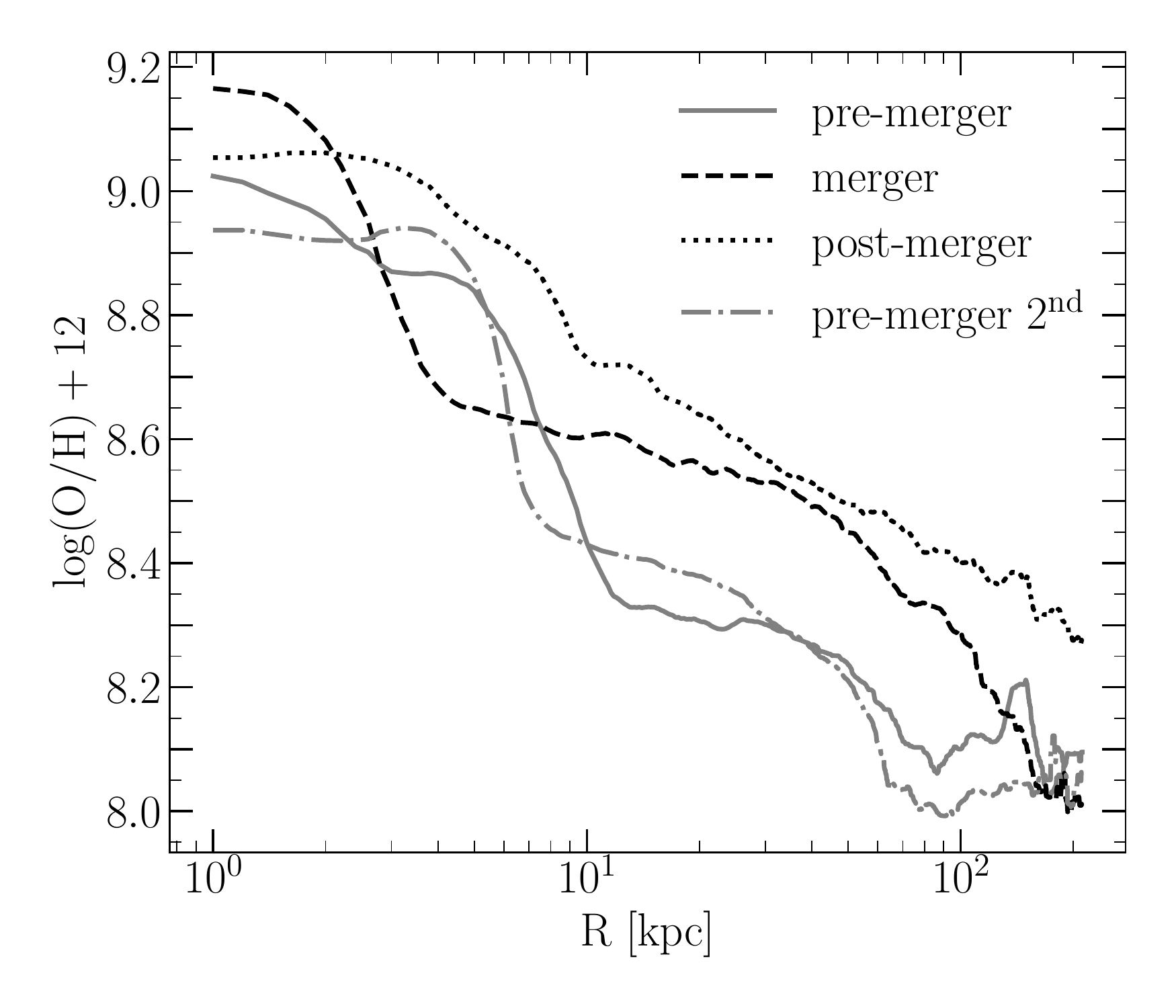}
\caption{The evolution of the radially-averaged, projected metallicity profile for three selected epochs: pre-merger (grey solid line), merger (black dashed line), and post-merger (black dotted line). \rr{The secondary galaxy's metallicity profile at the pre-merger epoch is shown by the dot--dashed grey line.} The merger produces a flatter metallicity profile by enriching the CGM metallicity.}
\label{fig:O/H_median_vs_r}
\end{figure}

The increase in the metallicity in the CGM is also evident as a flattening of the metallicity profile. Figure \ref{fig:O/H_median_vs_r} shows the metallicity profile of the galaxy's halo at three epochs: pre-merger (solid grey line), merger (dashed black line), and post-merger (dotted black line). After the merger, the CGM's metallicity increases, thus causing a flatter metallicity profile extending out to the halo. The flattening of the galactic disc's metallicity profile due to merger-driven inflows has been predicted by simulations \citep[e.g., ][]{Perez2006, Rupke2010_sim, Perez2011} and confirmed in observations \citep[e.g., ][]{Kewley2010, Rupke2010_obs}. Moreover, although a central metallicity dilution is visible when using a large aperture (see $R\in [0,10]$ kpc in Figure \ref{fig:O/H_median_vs_t}), the central 2 kpc region exhibits an enhancement in metallicity. The dilution effect is dependent on the initial metallicity profile of the pre-merger galactic discs \citep[e.g., ][]{Torrey2012}. A similar dependence of the dilution effect on the aperture has been reported in a recent spatially resolved analysis of the metallicity profiles of interacting galaxies, e.g., \citet{CALIFA_metallicity_2015} report an enhancement in the central SFRs of interacting galaxies accompanied by no variations in the central metallicity when compared to their isolated counterparts. The authors note that a central dilution in metallicity is seen when the aperture of the observation is large (similar to figure \ref{fig:O/H_median_vs_t}).

\textbf{The effect of the merger on the CGM metallicity:} Summarizing, the merger populates the CGM with metal rich gas therefore enhancing the total CGM metallicity by a factor of $\sim 3$ and producing a shallower halo metallicity profile.

%
\subsection{CGM ionization}
\label{sec:results/ionization}
\noindent
We have shown in the previous sub-sections a long-lasting increase in the covering fractions of hydrogen and metals, and an increase in CGM metallicity.  In this sub-section, we now compute the quantities that are directly measurable by observational studies of the CGM in absorption \cite[e.g., ][]{Tumlinson2011, Tumlinson2013, COS-Halos_metals,  Bordoloi2014, Peeples2014, Werk2014, Borthakur2015, COS-Gass, Werk2016, Prochaska2017}.  Specifically, we use the ionization calculations from {\sc cloudy} to predict the column densities of commonly observed species in absorption line studies. The ionization calculations in post-processing were done self-consistently with the physics models implemented in the simulation. As described in Section \ref{sec:methods/ionization}, we include radiation from an AGN and a time-varying UVB.

\begin{figure}
  \centering
  \includegraphics[width=\columnwidth]{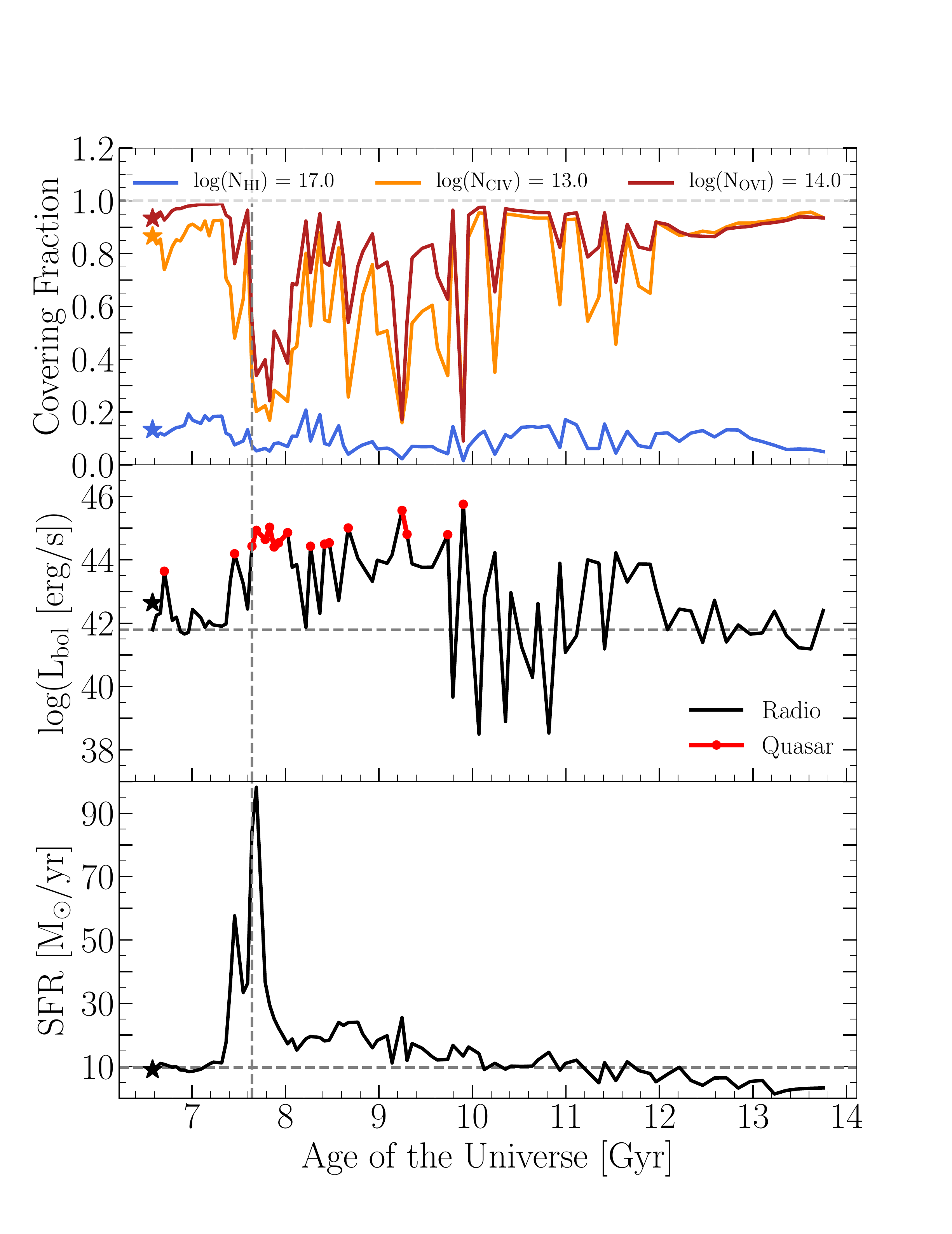}
  \caption{The evolution of the covering fractions of \ion{O}{vi}, \ion{C}{iv}, and \ion{H}{i} are shown in blue, orange, and brick-red in the top  panel, along with the AGN bolometric luminosity in the \rr{middle} panel (colours refer to the associated AGN feedback mode -- red: quasar mode, black: radio mode)\rr{, and the SFR in the bottom panel. The stars represent the properties of the secondary galaxy at the pre-merger time.} After the merger, the covering fraction of all ionization species decreases due to AGN radiative feedback decreasing the cooling rates of the CGM gas and increasing its internal energy. The stochastic fluctuations in AGN activity are matched by fluctuations in the ionization highlighting the effect of the AGN's hard ionizing radiation field on the ionization of the gas. Low ionization species (e.g., \ion{H}{i}) survive in self-shielding regions. \rr{Note that the peak SFR (merger-induced) is responsible for the outflows leading to the aforementioned increase in covering fractions and $R_{50}$.}}
\label{fig:ion_covfrac}
\end{figure}

The presence of a stochastically varying AGN manifests as stochastic variations in the ionization of various species. Figure \ref{fig:ion_covfrac} shows the evolution of the covering fractions of absorbers with column densities characteristic of the CGM for three species: \ion{O}{vi}, \ion{C}{iv}, and \ion{H}{i}. The species were chosen for demonstration purposes as their ionization energies span a large range. Consequently, these ions span gas that ranges in temperature from $T \sim 10^4$ K to $T \sim 10^{6}$ K. Before the merger ($t\sim 6.5$ Gyr), the galactic halo is mostly virialized and dominated by a hot gas traced by \ion{O}{vi}; the halo has a smooth profile with little visible substructure. After the merger ($t>7.6$ Gyr), the covering fractions of \ion{H}{i}, \ion{C}{iv}, and \ion{O}{vi} decline. The hard radiation field from the AGN ionizes even \ion{O}{vi} when radiating at a maximum ($\geq 10^{44}$ $\mathrm{erg \, s}^{-1}$). \rr{Additionally, the enhanced activity (AGN and SFR) during the merger is responsible for strong outflows which populate the CGM with cool gas and metals therefore increasing covering fractions and $R_{50}$; the peak in SFR coincides with the sharpest increase in $R_{50}$, covering fractions, and CGM metallicity.} When the AGN activity declines ($t>10$ Gyr) the CGM begins to cool and the covering fractions of the various species increase again. At this epoch, the CGM is populated by outflows from the galactic disc which manifest in a clumpy CGM rather than the halo's mostly smooth hot gas profile prior to the merger. Notice that the variation in the covering fractions coincides with AGN intensity peaks. This is indicative of the effectiveness of the AGN model at altering the ionization of the CGM. Coupled with the expectation from both simulations and observations that mergers can fuel AGNs, Figure \ref{fig:ion_covfrac} suggests that even modest enhancements in the AGN luminosity will lead to a dominance of more highly ionized species. This is consistent with \citet{oppenheimer_agn} who show that AGN radiation can be an effective ionization source for the CGM.

We caution the reader that the ionization calculations presented in this study may be prone to uncertainties and further corrections owing to the ionization equilibrium assumption. \citet{Segers2017} demonstrate that stochastic AGN behaviour can keep CGM gas (i.e. metals) out of ionization equilibrium even at large radii (i.e. $2 \times R_\mathrm{vir}$). Additionally, \citet{oppenheimer_agn} and \citet{Oppenheimer2017_OVI} illustrate the importance of non-equilibrium effects when modelling the ionization of the CGM. Simulations including non-ionization equilibrium \citep[e.g. ][]{Oppenheimer2017} predict CGM ionizations which are consistent with observed CGM properties. Therefore, assuming ionization equilibrium poses a principal limitation to the predictions of the work presented above. \rr{In order to understand the extent of the aforementioned limitation, we first examine the AGN variability time-scales. The AGN varies on time-scales $\gtrsim 45$ Myr which well exceed the recombination time-scales of the ions of interest, i.e. \ion{O}{vi}, \ion{C}{iv}, \ion{H}{i}, in the simulated CGM ($\lesssim 20$ Myr). Therefore, the simulated CGM reaches equilibrium on time-scales shorter than the variations in the underlying radiation field. Next, we investigate the presence of a proximity zone fossil (PZF) in the CGM. The simulated CGM shows no signs of a PZF which is consistent with studies showing that AGN lifetimes $\lesssim 1$ Myr are needed to produce a significant PZF \citep[i.e. ][]{Oppenheimer2017_OVI}. The absence of a PZF is evident in Figure \ref{fig:ion_covfrac} where the covering fractions of ions immediately trace changes in the AGN bolometric luminosity. Therefore, although the assumption of ionization equilibrium is a limitation of this work non-equilibrium effects appear to be negligible in the simulations we present. Furthermore}, we note that the ionization state of the merger remnant's CGM in this study is consistent with the results of the COS-Haloes survey \citep{Werk2014}; both the \ion{H}{i} and \ion{O}{vi} covering fractions for absorbers with $N_\mathrm{H} \geq 10^{17}$ cm$^{-2}$ and $N_\mathrm{O} \geq 10^{14}$ cm$^{-2}$, respectively, are in agreement with the covering fractions reported in COS-Haloes. \rr{The median redshift of the COS-Haloes sample corresponds to $t \sim 11$ Gyr ($\sim 3.5$ Gyr post-merger) where the simulated merger remnant is an isolated disc galaxy showing no visible signatures of the merger (e.g. asymmetries in the stellar or \ion{H}{i} profile) and the median AGN luminosity is $L_\text{AGN} \lesssim 10^{41}$ erg$/$s. The agreement with the COS-Haloes results may be suggestive of the insignificance of the non-equilibrium ionization effects in the CGM of the simulated galaxy in this study.}

\begin{figure}
  \centering
  \includegraphics[width=\columnwidth]{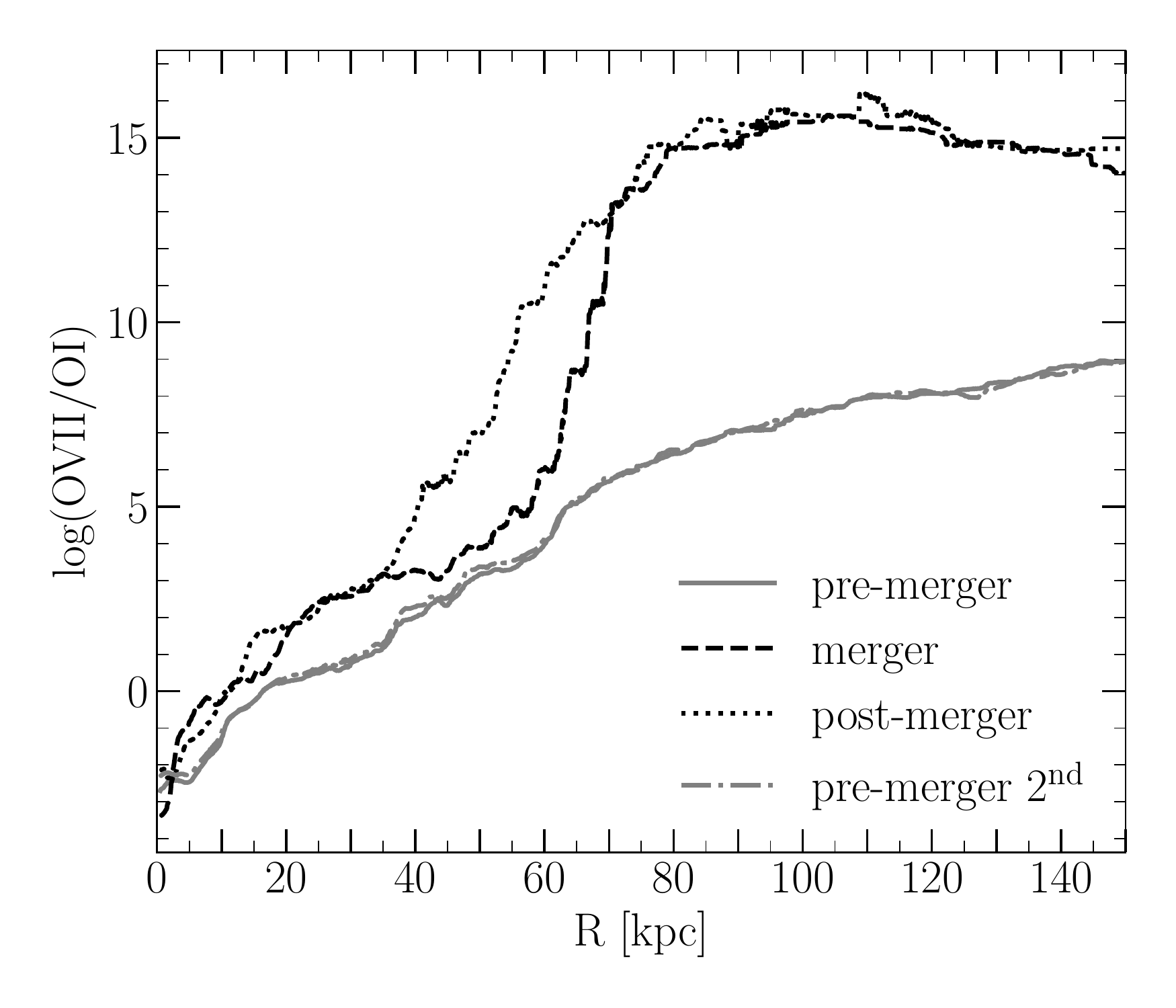}
\caption{The ionization profile of oxygen at three stages of the merger: pre-merger, merger, and post-merger. \rr{The secondary galaxy's ionization profile at the pre-merger time is shown by the dot--dashed grey line. The ionization states of the CGM of both merging galaxies show remarkable similarity.} After the merger the increase in the ionizing radiation field manifests as an increase in the gas ionization ratio. Low ionization species (i.e. \ion{O}{i}) only survive in regions where the gas is self-shielding (i.e. galactic disc). 1 Gyr after the merger, \ion{O}{vii} dominates the ionization profile as the CGM becomes more ionized.}
\label{fig:O_ionization}
\end{figure}

Figure \ref{fig:ion_covfrac} demonstrates that the presence of the AGN provides a hard radiation field such that even the \ion{O}{vi} covering fraction decreases. The ionized \ion{O}{vi} populates higher ionization states. Therefore, we use the (rarely observed) \ion{O}{vii} to trace the highly ionized gas by examining the gas ionization profile. Figure \ref{fig:O_ionization} shows the ionization profile, \ion{O}{vii} / \ion{O}{i}, at three evolutionary epochs during the merger. After the merger, the ionization of the gas increases at all radii, most notably in the CGM ($\mathrm{R} \in [60,150]$ kpc). This further supports the picture of AGN radiation increasing the ionization of the CGM gas (i.e. declining covering fraction of all species). We remind the reader that our ionization model does not account for contributions from the host galaxy's stars to the ionizing radiation field. 

In addition to the the AGN's hard ionizing radiation field, during galaxy mergers gas in the colliding galactic discs may be shock heated up to X-ray emitting temperatures \citep[i.e. ][]{Cox2006_agn}. \citet{Cox2006_agn} demonstrated that the hot X-ray metal enriched gas can then be launched into the galactic halo to populate the CGM especially in the presence of an AGN [see figures 1 and 3 from \citet{Cox2006_agn}].

\textbf{The effects of the merger on the CGM's ionization:} In summary, our simulations show that, in agreement with previous works, AGN fuelling can be enhanced during a merger. As a result, the CGM gas around merger systems experiences competing effects between (i) increases in CGM metallicity driven by metal rich outflows and (ii) decreases in the observable ionization fractions as AGN activity significantly increases photoionization. Therefore, despite the increase in CGM metallicity by a factor of $2-3$, the covering fraction of the majority of ions observed in rest-frame UV spectra may actually decrease during a merger, as a result of the dominance of the AGN's hard ionization field. To recover the increase covering fraction and $R_{50, \, \mathrm{X}}$ reported in Section \ref{sec:results/cover}, one would need to apply proper ionization corrections.

\begin{figure*}
  \centering
\includegraphics[width=\linewidth]{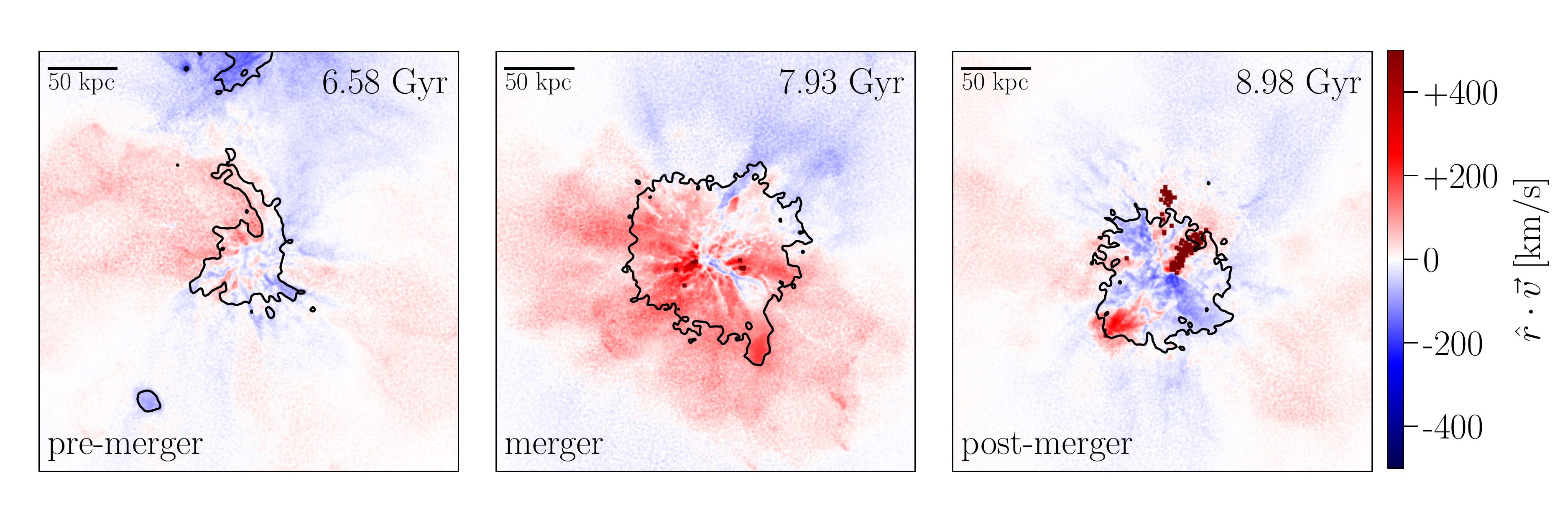}
\caption{Mass-weighted gas radial velocity projections ($\hat{r} \cdot \vec{v}$) in the galaxy's reference frame at three epochs: pre-merger, merger, and post-merger (left to right). The red colours indicate outflowing gas while the blue colours represent inflows. The black density contour shows the galaxy's gas distribution as shown in figure \ref{fig:projections_H-et-O}. After the merger, the velocity field becomes divergent as it is dominated by merger-driven outflows. These outflows are responsible for increasing the CGM metal content and therefore metal covering fractions.}
\label{fig:outflows}
\end{figure*}

\section{Discussion}
\label{sec:discussion}
\noindent

%
\subsection{Merger-induced outflows populating the CGM}
\label{sec:outflows}
\noindent

Galactic winds are observed to be ubiquitous in galaxies \citep[e.g., ][]{ Pettini2002, Adelberger2003, Adelberger2005, Martin2005, Rupke2005_starburst_outflow, Rupke2005_agn_outflow, Veilleux2005, Veilleux2013, McElroy2015, Zschaechner2016, outflow_sfr, JongHak_outflows}. Moreover, galactic winds are a common feature in simulations of galaxy evolution where stellar and AGN feedback prescriptions produce large scale outflows of metal-enriched gas \citep[e.g., ][]{Illustris,  fire, auriga_dm, Auriga}. While such feedback mechanisms are primarily introduced to regulate star formation in galaxies \citep[e.g., ][]{Oppenheimer_et_Dave2006, Puchwein_et_Springel2013}, they are also responsible for recycling the galactic metal content and enriching the CGM (as shown in this study). Simulations of galaxy evolution illustrate that galaxies with no/low feedback lack a sufficient amount of metals in their haloes \citep[e.g., ][]{Cox2006_agn,  Zahid_et_Torrey_2014, Stinson16}, where a significant fraction of CGM metals are carried by galactic outflows \citep[e.g., ][]{ Fumagalli2011, Muratov2017}.

In previous sections we have demonstrated that the merger increases the covering fractions of gas and metals in the CGM, the CGM metallicity, and CGM `size' (i.e. as explored with the $R_{50, \, \mathrm{X}}$ parameter introduced in this work). The enhanced metallicity in the CGM propagates outwards with time (see figure \ref{fig:O/H_median_vs_t}). There are two plausible mechanisms for transporting the gas into the CGM during a merger: tidal stripping and galactic winds/outflows. Whereas feedback driven outflows only directly impact the gas, tidal effects impact the stellar and gaseous components at a comparable level. Therefore, examining the spatial re-distribution of the stellar profile, one can better constrain the role of tidal torques in enriching the CGM gas. However, while inspecting the evolution of the stellar distribution throughout the merger, no significant transport of star particles is observed out to large radii which we attribute with the CGM ($R>50$ kpc). The absence of a tidally stripped stellar component propagating outwards disfavours the tidal stripping mechanism to be the leading mechanism in populating the CGM at radii $R>50$ kpc. 

On the other hand, a wind-enrichment scenario is supported by an examination of the gas velocity field within the simulation, shown in Figure \ref{fig:outflows}. The panels show mass-weighted projections ($300 \times 300$ kpc$^2$) of the gas radial velocity in the galaxy's reference frame at three stages of the merger (left to right). The direction and magnitude of the underlying gas velocity field is indicated by the colour map. Red colours indicate out-flowing gas, while blue colours show inflows. The galaxy's projected density profile is shown by the black contours. It is evident from Figure \ref{fig:outflows} that, post-merger, the velocity field becomes divergent consistent with merger-driven outflows. More evidence of powerful outflows can be seen in the strong decline in the covering fractions, $R_{50, \, \mathrm{H}}$, and $R_{50, \, \mathrm{O}}$ at $t>11$ Gyr which coincide with a strong AGN episode where the associated AGN feedback is returned in a series of radio bubbles (see Figure \ref{fig:ion_covfrac}; i.e. $11.6\text{ Gyr} \le t \le 11.9 \text{ Gyr}$). The radio-mode feedback is injected into the halo (at $R=50$ kpc) in the form of rising buoyant hot bubbles. This form of feedback can efficiently remove gas from the halo causing a decline in the covering fractions of denser clumps ($N_\mathrm{O} > 10^{17}$ cm$^{-2}$ and $N_\mathrm{H} > 10^{21}$ cm$^{-2}$). Meanwhile, the ambient hot dense medium probed by absorbers with $N_\mathrm{O} > 10^{16}$ cm$^{-2}$ is sustained. Note that the metallicity of the gas is not affected, however the covering fraction is.

The outflows induced during the merger are responsible for increasing the CGM ($50$ kpc $\leq R \leq 150$ kpc) metal mass by factors $3-4$ for up to 6 Gyr post-merger. This picture of outflows is consistent with current observational studies of the CGM which find a metal-enriched gas population out to impact parameters $\sim 150$ kpc \citep[e.g., ][]{Werk2014, Prochaska2017}. Other studies report enhanced low-ionization column densities (stronger Ly$\alpha$, \ion{Si}{ii}, and \ion{C}{iv}), and different dynamics (Ly$\alpha$ equivalent width and velocity offset from the host's systemic velocity) in the CGM of starburst/post-starburst galaxies \citep[i.e. ][]{COS-Burst_old2016, COS-Burst} when compared to normal star-forming galaxies; such differences can be explained by invoking galactic outflows from starburst regions within the hosts. Similar results are seen at high impact parameters in a sample of AGN hosts: The column densities of low-ionization species (e.g., \ion{H}{i}, \ion{Si}{ii}, \ion{Si}{iii}) are enhanced in the CGM of AGN hosts when compared to normal star-forming galaxies or passive galaxies matched in stellar mass \citep{COS-AGN_submitted}. Galactic outflows have also been shown to be the source of metals in the CGM in hydrodynamical simulations where the CGM gas of galaxies simulated without outflows remains relatively un-enriched \citep[e.g., ][]{Oppenheimer_et_Dave2006, Ford2013, Hummels2013, Suresh2015b}.

%

\begin{figure*}
  \centering
  \includegraphics[width=0.5\linewidth]{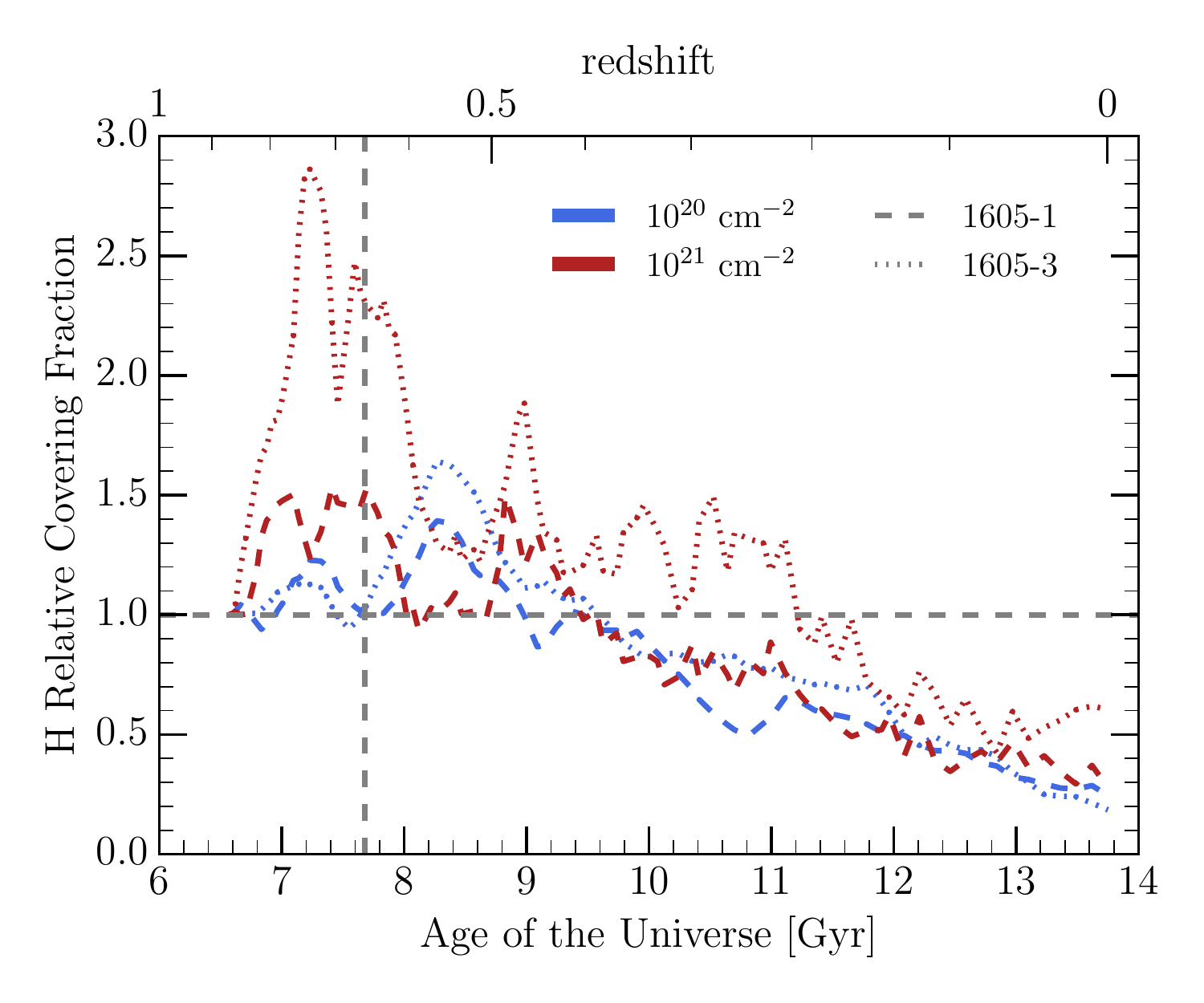}\includegraphics[width=0.5\linewidth]{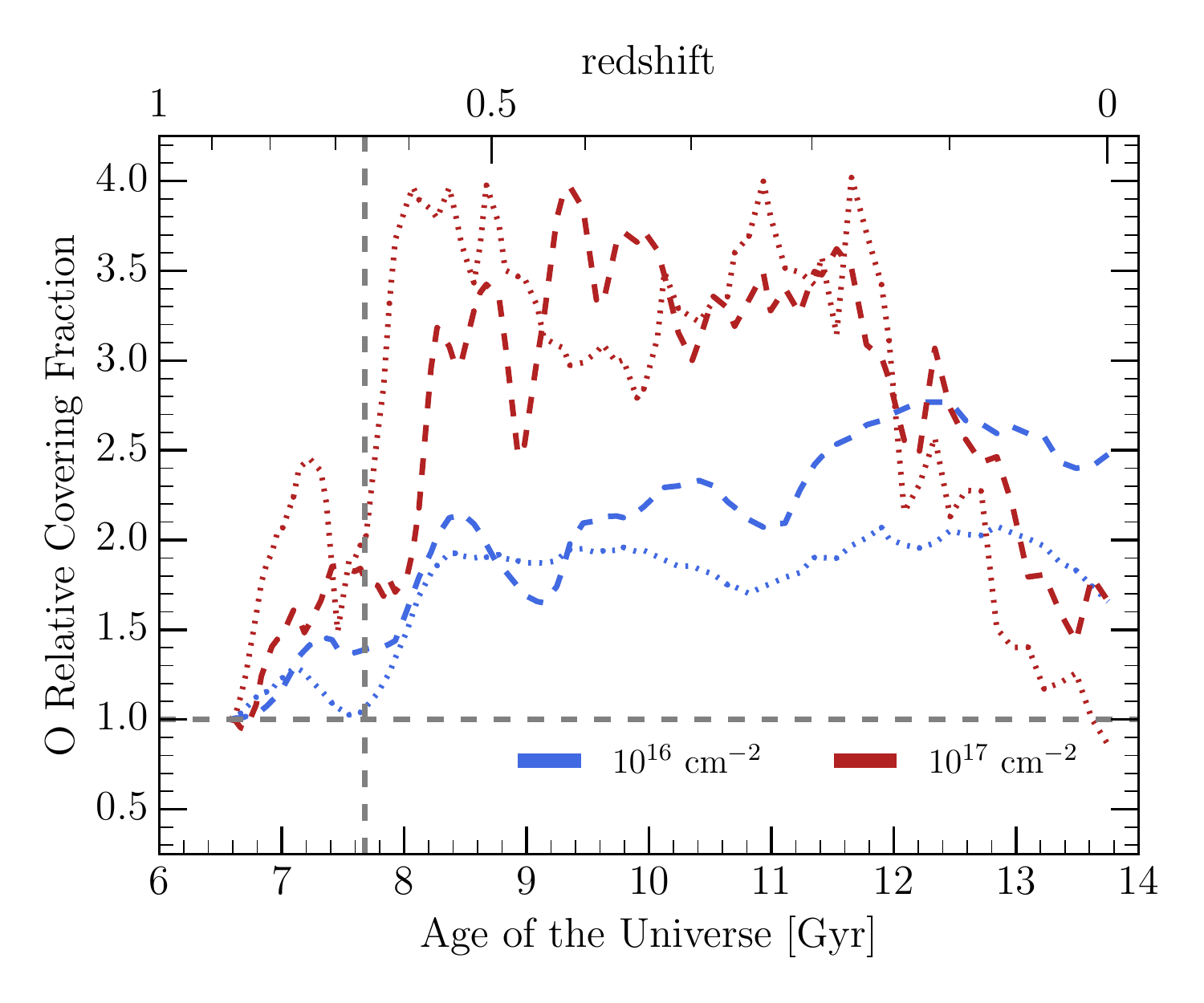}
\caption{The evolution of the column density of hydrogen (left) and oxygen (right) absorbers throughout the merger. The different colours indicate absorption thresholds, while the line styles represent the different resolutions -- 1605-1 ($\sim 1.4 \times$ higher than the original resolution of the Illustris simulation): dashed lines; 1605-3 (the fiducial resolution simulation used for analysis in this paper, $\sim 40 \times$ higher than Illustris): dotted lines. The vertical grey dashed line indicates the merger time in the high-resolution simulation. Changing the resolution by a factor of $\sim 40$ has a negligible quantitative effect on the relative change in CGM properties except at the highest gas column densities. Qualitatively, the simulated CGM covering fractions remain unchanged for different resolutions.}
\label{fig:resolution}
\end{figure*}

\subsection{Analysis limitations and the effects of resolution}
\label{sec:discussion/resolution_effect}
\noindent
The presented analysis focuses on studying the CGM of galaxies undergoing a major merger to understand the interplay between mergers, a major evolutionary path, and the CGM. We use cosmological zoom-in simulations which enable us to study the CGM in a self-consistent evolutionary context. It is, however, worth noting that the physical models and simulation resolution may potentially have a significant impact on the simulated CGM properties.

The physical models implemented in the simulations (i.e. feedback processes, chemical enrichment) may have remarkable impact on the predicted CGM metal and energy budget (i.e. abundances and ionization) especially in cases where the physical model does not convergence at varying resolutions. The convergence of the model has been demonstrated at resolutions comparable to the ones used in this study \citep[see ][]{Sparre2016a, Auriga}. Moreover, the galaxy formation model used in this study has been calibrated to reproduce physical properties of $z=0$ L$^*$ galaxies in cosmological zoom-in hydrodynamical simulations \citep{Marinacci2014,  Illustris, Auriga}. \citet{Auriga} simulated a sample of 30 L$^*$ galaxies and demonstrated that the galaxy formation model reproduces appropriate stellar masses, stellar disc sizes, rotation curves, SFRs, and metallicities as in the local Universe. We leave a detailed discussion of the effects of varying the physical models on the CGM to future work.

The simulation resolution may strongly limit the resolved CGM structure and gas properties \citep[i.e. density, metallicity, temperature; e.g., ][]{Nelson2016}. Nonetheless, resolution is a necessary sacrifice bearing in mind the computational expense involved in cosmological simulations. Simulating gas clouds on observationally motivated scales ($\sim$ pc scales) is not currently possible in cosmological simulations. For example, in order to achieve $\sim 1$ pc scale resolution in the CGM, with typical CGM densities $n_\mathrm{H} \lesssim 10^{-2}$ cm$^{-3}$, a baryonic mass resolution limit of $m_\mathrm{b} \lesssim 2.5\times 10^{-4}$ M$_\odot$ is required. Such an exorbitant mass resolution is well beyond the resolution limit of the current generation of galaxy evolution (even isolated simulations). However, small-scale, high-resolution simulations studying the evolution of single clouds in the CGM \citep[e.g., ][]{Heitsch2009, Armillotta2017} complement our approach by providing a sub-resolution understanding of cloud survival in the CGM.

In this section, we discuss the effect that the simulation resolution (mass and spatial resolution) can have on the results. We use the re-runs of \citet{Sparre2016a} of the same halo (1605) at different resolutions (1605-1 and 1605-3) to demonstrate the convergence of the CGM properties. The halo 1605-1 is simulated $\sim 1.4 \times$ the resolution of the Illustris project, while 1605-3 is the fiducial simulation with a $\sim 40 \times$  better mass resolution than Illustris. Figure \ref{fig:resolution} shows the evolution of the covering fractions (normalized to the pre-merger state) of hydrogen and oxygen throughout the merger. The colours represent column density absorption thresholds while the different line styles represent the two resolution runs (1605-1: dashed lines; 1605-3: dotted lines). Increasing the resolution has no qualitative effect on the relative change in the covering fraction induced by the merger except for the highest $N_\mathrm{H}$ absorbers. The global enhancement (and the strengths of the enhancement) in the covering fractions of metals in the CGM reported in previous sections are reproduced at low-resolution. At high column densities (i.e. $N_\mathrm{H} >10^{21}$ cm$^{-2}$), the contributing gas is dominated by high density gas cells. In the low resolution simulation, the high densities are not resolved as the dense substructure is dissolved into larger gas cells, therefore reducing the associated covering fractions. Similarly, increasing the resolution has little (qualitative) effect on the relative changes in the ionization of the CGM. The differences in the evolution of the CGM ionization (the strength and length of ionization) are dominated by the variations in the AGN accretion. At all resolutions the CGM's ionization directly follows the fluctuations in the BH's accretion rate. 

%
\subsection{An accounting of the CGM metal budget}
\label{sec:discussion/metals}
\begin{figure}
  \centering
  \includegraphics[width=\linewidth]{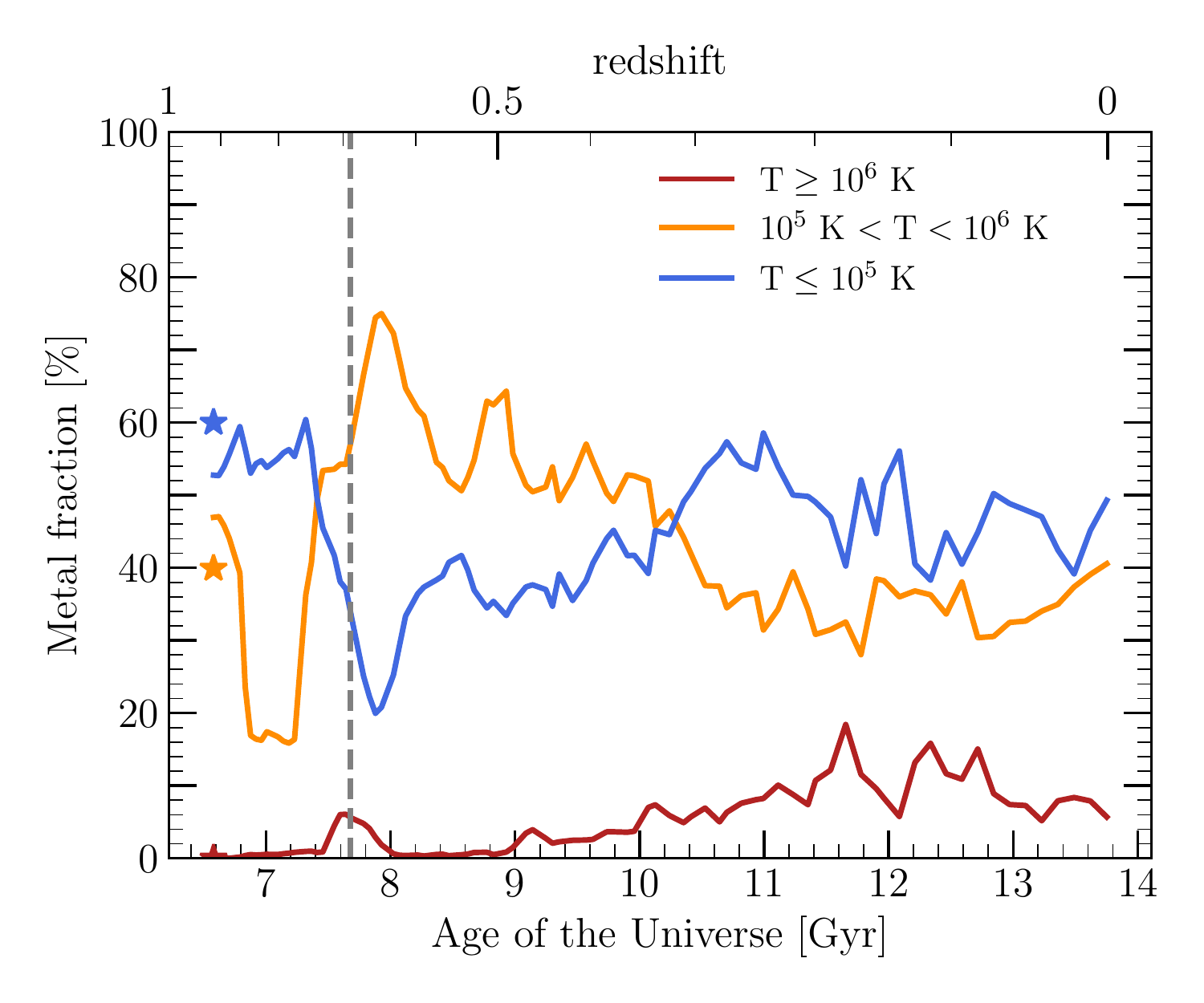}
\caption{An accounting of the CGM metal budget for the various physical phases (coloured lines \rr{show the evolution of the metal fraction by mass:} $M_{Z,\text{phase}} / M_{Z, \text{total}}$) around the merging halo ($50$ kpc $\leq \mathrm{R} \leq 150$ kpc). The vertical dashed grey line indicates the time of the merger. \rr{The stars mark the metal fractions, by mass, in the CGM of the secondary galaxy. Although the secondary galaxy and its CGM are metal poor when compared to the primary galaxy, both parent galaxies' CGM show remarkable similarity in their fractional distribution of metals.} After the merger, the induced stellar feedback enriches the gas while both stellar and AGN feedback increase the gas's internal energy. This is evident as an increase in the \rr{mass} fraction of metals in the hot ($\mathrm{T} \geq 10^6$ K; brick-red) and warm ($10^5$ K $< \mathrm{T} < 10^6$ K; orange) gas phases by a factor of $\sim 4$. While the hot metal-enriched gas cools within $\sim 1$ Gyr post-merger, the warm gas continues to account for $\gtrsim 50 \% $ of the \rr{halo's metal budget (by mass)} for $\sim 2.3$ Gyr post-merger.}
\label{fig:CGMmetals}
\end{figure}

\noindent
Galactic feedback can have a significant impact on the CGM gas: In addition to launching gas into the CGM, galactic feedback enriches the gas reservoir with metals (stellar feedback) and increases the internal energy of the gas \citep[stellar and AGN feedback; e.g., ][]{Cox2004, Cox2006_agn, Sinha2009}. It has been observationally inferred that sizeable amounts of gas and metals (equivalent to the gas within the ISM) reside in the extended gaseous halo surrounding galaxies \citep[e.g., ][]{Bordoloi2014, Peeples2014, Werk2014, Prochaska2017}. Therefore, tracking the CGM metal budget and accounting for the different gas phases constrains the effect of merger-induced feedback processes on the CGM. Figure \ref{fig:CGMmetals} shows a metal census in the CGM of the merging galaxy: $50$ kpc $\leq \mathrm{R} \leq 150$ kpc. After the merger, the metal \rr{mass} fraction in hot ($\mathrm{T} \geq 10^6$ K) and warm ($10^5$ K $< \mathrm{T} < 10^6$ K) gas increases by a factor of $\sim 4$. The hot gas cools within $< 1$ Gyr post-merger, while the warm gas continues to host $\gtrsim 50\%$ of the CGM's metals \rr{by mass}. The hot gas metal \rr{mass} fraction starts to increase again at $t > 9$ Gyr potentially due to: (1) metal-rich, fast outflows (see the right-hand panel of Figure \ref{fig:outflows}) populating the CGM, and (2) shock heating in the more massive merger remnant halo. The metal fraction hosted by the cool gas phase ($\mathrm{T} \leq 10^5$ K) increases after the merger ($t \sim 8$ Gyr) as more cool metal-enriched outflows are deposited at $\mathrm{R} > 50$ kpc. At $t\sim 10$ Gyr, the warm gas experiences significant cooling which coincides with a decrease in the mean AGN bolometric luminosity (see the lower panel of Figure \ref{fig:ion_covfrac}). This suggests that AGN feedback (and radiative contributions) plays a key role in maintaining a warm CGM component. Our result is consistent with the work of \citet{Oppenheimer2016} who showed that the covering fraction of \ion{O}{vi} (which probes temperatures characteristic of the warm gas shown in Figure \ref{fig:CGMmetals}) in simulations which do not include AGN feedback is well below the observed \ion{O}{vi} covering fraction. Therefore, the merger increases the metal \rr{mass} fraction in the warm/hot CGM gas for $> 3$ Gyr post-merger. 

\begin{figure}
  \centering
\includegraphics[width=\linewidth]{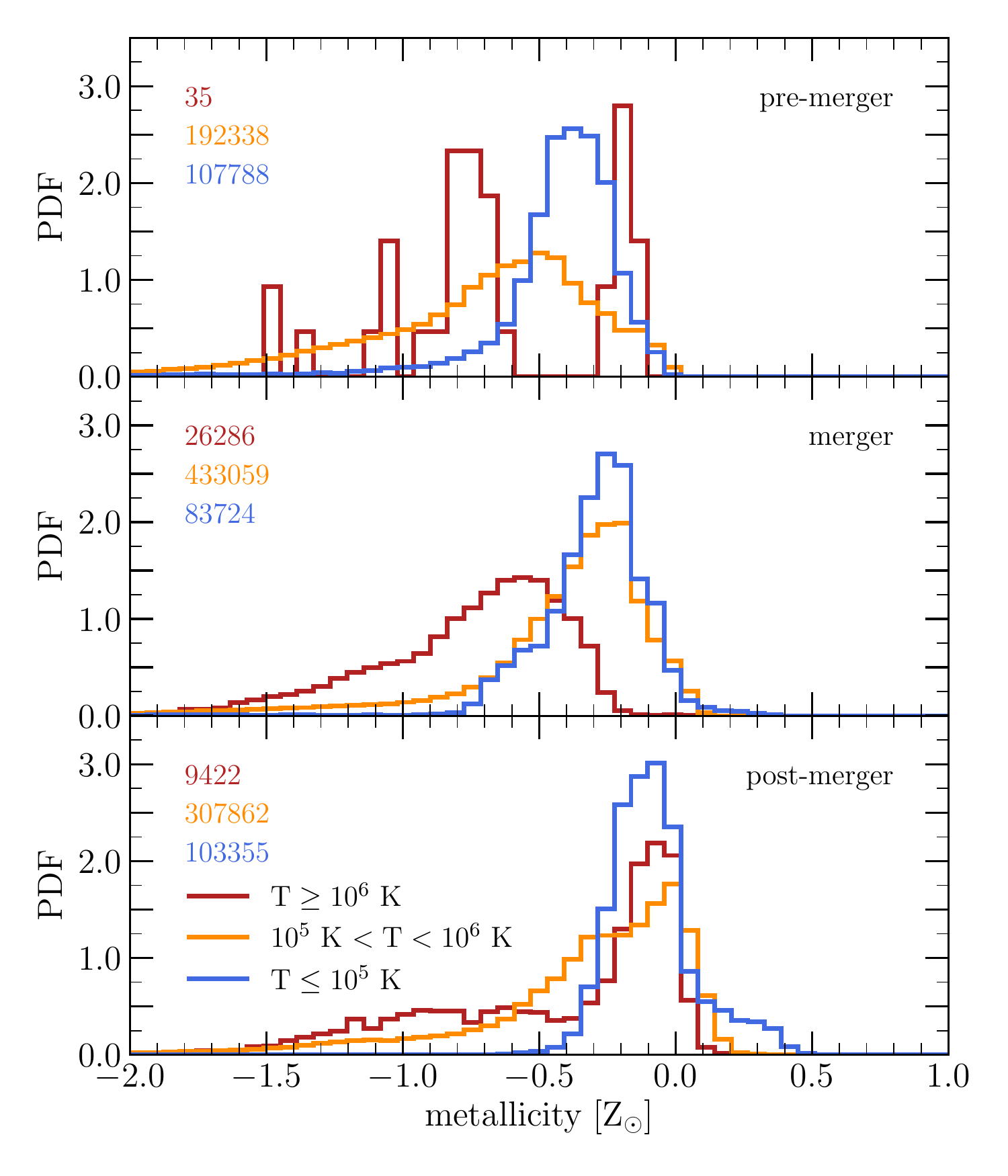}
\caption{A census of gas phase metallicities in the CGM of the merging halo ($50$ kpc $\leq \mathrm{R} \leq 150$ kpc). The coloured histograms represent different gas phases defined by their temperature. The number of gas cells contributing to each gas phase ($\propto M_\text{gas phase}$) is shown in the top right corner of each panel. After the merger, the median metallicity of all phases increases leading to a prominent super-solar metallicity gas component in all phases as the merger enriches the CGM with metals. Additionally, the budget of hot and warm gas ($\text{T}>10^5$ K) increases after the merger while a large fraction ($> 50 \%$) of that gas remains at low metallicities ($\log(Z/\text{Z}_\odot) < -0.23$) extending to $\log(Z/\text{Z}_\odot)< -1.5$.}
\label{fig:CGMphase_metallicity}
\end{figure}

The increase in the metal \rr{mass} fractions in the gas phases may be driven by two factors: (1) changes in the gas-mass in each phase, and/or (2) changes in the metallicity of the respective phases. Figure \ref{fig:CGMphase_metallicity} shows the probability density function of the gas-phase metallicity at three key epochs: pre-merger (top panel), merger (middle panel), and post-merger (lower panel). The coloured histograms represent the gas phases shown in Figure \ref{fig:CGMmetals}: $\text{T} \leq 10^5$ K (blue), $10^5 \text{ K} < \text{T} < 10^6 \text{ K}$ (orange), and $\text{T}\geq 10^6$ K (brick-red). The number of gas cells contributing to the individual gas phases is indicated in the top-right corner of each panel. It is evident that the merger drives an increase in the gas metallicity of all phases (notice the shift in the distributions' median to higher metallicities). Moreover, the merger induces the presence of a prominent super-solar metallicity gas component in all phases, although it is most evident in the cool gas phase ($\text{T} \leq 10^5$ K). The presence of high metallicity gas in the CGM is consistent with observations of the CGM \citep[e.g. ][]{Tumlinson2011, Prochaska2017}. For example, \citet{Prochaska2017} showed that $\sim 25 \%$ of the COS-Haloes sightlines show super-solar metallicities. Although the merger drives an increase in the global CGM gas metallicity, a warm/hot metal poor ($\log(Z/\text{Z}_\odot) < -0.25$) gas component persists after the merger.  Moreover, feedback processes excited during the merger increase the gas internal energy causing an enhancement in the amount of hot and warm gas during the merger. The heated gas then cools driving a decline in the amount of hot and warm gas in the post-merger epoch.

\section{Conclusions}
\label{sec:conclusions}
\noindent
This work's primary focus is to demonstrate the effect a major merger can have on the CGM of a galaxy. We use zoom-in cosmological hydrodynamical simulations \citep{Sparre2016a} to model the evolution of the CGM of a galaxy undergoing a major merger. We've selected the merging galaxies from the Illustris project such that the merger occurs at $z =0.66$ with a merger ratio of 1:1.16. The $z=0$ descendant is a Milky Way-like galaxy with stellar mass $M_\star = 10^{10.89}$ M$_\odot$ residing in a dark matter halo with total mass $M_{200} = 10^{12.00}$ M$_\odot$. We model gas ionization in post-processing using an ionization source which accounts for contributions from the host galaxy's AGN and a UVB consistent with the physics model adopted in the simulations \citep{ Marinacci2014, Illustris, Auriga}.

To quantify the effect of the major merger on the CGM chemical and ionization properties, we assess the change in the covering fractions of commonly observed ionic species, as well as total hydrogen and metals. Additionally, we define a characteristic CGM size ($R_{50, \, \mathrm{X}}$) to follow the spatial effect of the merger on the CGM gas/ionization profile.

We have demonstrated that:
\begin{enumerate}
\item{The major merger can increase the covering fractions of oxygen ($N_\mathrm{O} > 10^{16-17}$ cm$^{-2}$) by factors of $2-4$ for $>4$ Gyr post-merger. The total hydrogen covering fractions ($N_\mathrm{H} > 10^{20-21}$ cm$^{-2}$) can increase by factors of $1-1.25$ for $\sim 3$ Gyr post-merger. The increase in covering fraction is more evident for metals (e.g., oxygen) as the CGM gets enriched.}
\item{The newly populated CGM shows significant enhancement in its metal and baryonic budget, and its characteristic size ($R_{50,\, \mathrm{X}}$). The increasing size is more evident when probed using metal absorption which is consistent with metal rich outflows preferentially populating the CGM with metals. A factor of 1.3 increase in $R_{50, \, \mathrm{H}}(5 \times 10^{17}\, \mathrm{cm}^{-2}, \, t)$ persists for a $\sim 3$ Gyr post-merger. $R_{50, \, \mathrm{H}} (10^{20}\, \mathrm{cm}^{-2}, \, t > 11 \, \mathrm{Gyr})$ declines as the gas in the CGM diffuses below the density threshold. The presence of oxygen in the CGM extends to larger radii with a factoral increase in $R_{50, \, \mathrm{O}} (10^{16}\, \mathrm{cm}^{-2}, \, t)$ as high as $1.4\times$ persisting for $\sim 6$ Gyr.}
\item{The simulation predicts a $0.2-0.3$ dex increase in the CGM metallicity ($\log(\mathrm{O/H})$) that lasts for $\sim 6$ Gyr.}
\item{Merger-induced outflows populate the CGM with metals preferentially increasing the metal covering fraction and CGM metallicity.}
\item{The major merger triggers an AGN episode which, in addition to inducing outflows, returns significant amounts of energy to the surrounding halo causing an increase in the CGM ionization. Therefore, although the simulations predict an increase in the metallicity of the CGM, observationally we might expect a drop in the covering fraction or typical column densities of metal species in the rest-frame UV.}
\end{enumerate}

This work presented a case study which demonstrates the effect a major merger can have on the CGM of the galaxy. In future work we will further investigate the effect of the merger orbital properties on the CGM. Knowing that mergers with different orbital properties exhibit differences in their AGN/star formation activity (strength and length of activity) and  morphological asymmetries, one would expect the merger's orbit to affect the CGM as well. We will also investigate the effect of diverse merger histories on the CGM of $z=0$ galaxies in cosmological zoom-in simulations using a suite of Milky Way-like galaxies. This would be directly comparable to current observational surveys (i.e. COS-Haloes, COS-GASS).

\section*{Acknowledgments}
\rr{The authors thank Ben Oppenheimer for helpful discussions about non-equilibrium ionization effects in the CGM}, as well as Jess Werk and Volker Springel for fruitful discussions. \rr{The authors also thank the anonymous referee for a prompt response and helpful comments which made this manuscript better}. MHH acknowledges the receipt of a Vanier Canada Graduate Scholarship. MHH would also like to thank Volker Springel for supporting MHH's stay at the Heidelberg Institute for Theoretical Studies (HITS). MS acknowledges support from HITS and MIT. SLE acknowledges the receipt of an NSERC Discovery Grant. PT acknowledges support from NASA through Hubble Fellowship grant HST-HF2-51384.001-A, awarded by the STScI, which is operated by the Association of Universities for Research in Astronomy, Inc., for NASA, under contract NAS5-26555. MV acknowledges support through an MIT RSC award, the support of the Alfred P. Sloan Foundation, and support by NASA ATP grant NNX17AG29G.



\bibliographystyle{mnras}
\bibliography{references} 

\begin{thebibliography}{}
\makeatletter
\relax
\def\mn@urlcharsother{\let\do\@makeother \do\$\do\&\do\#\do\^\do\_\do\%\do\~}
\def\mn@doi{\begingroup\mn@urlcharsother \@ifnextchar [ {\mn@doi@}
  {\mn@doi@[]}}
\def\mn@doi@[#1]#2{\def\@tempa{#1}\ifx\@tempa\@empty \href
  {http://dx.doi.org/#2} {doi:#2}\else \href {http://dx.doi.org/#2} {#1}\fi
  \endgroup}
\def\mn@eprint#1#2{\mn@eprint@#1:#2::\@nil}
\def\mn@eprint@arXiv#1{\href {http://arxiv.org/abs/#1} {{\tt arXiv:#1}}}
\def\mn@eprint@dblp#1{\href {http://dblp.uni-trier.de/rec/bibtex/#1.xml}
  {dblp:#1}}
\def\mn@eprint@#1:#2:#3:#4\@nil{\def\@tempa {#1}\def\@tempb {#2}\def\@tempc
  {#3}\ifx \@tempc \@empty \let \@tempc \@tempb \let \@tempb \@tempa \fi \ifx
  \@tempb \@empty \def\@tempb {arXiv}\fi \@ifundefined
  {mn@eprint@\@tempb}{\@tempb:\@tempc}{\expandafter \expandafter \csname
  mn@eprint@\@tempb\endcsname \expandafter{\@tempc}}}

\bibitem[\protect\citeauthoryear{{Adelberger}, {Steidel}, {Shapley}  \&
  {Pettini}}{{Adelberger} et~al.}{2003}]{Adelberger2003}
{Adelberger} K.~L.,  {Steidel} C.~C.,  {Shapley} A.~E.,   {Pettini} M.,  2003,
  \mn@doi [\apj] {10.1086/345660}, \href
  {http://adsabs.harvard.edu/abs/2003ApJ...584...45A} {584, 45}

\bibitem[\protect\citeauthoryear{{Adelberger}, {Shapley}, {Steidel}, {Pettini},
  {Erb}  \& {Reddy}}{{Adelberger} et~al.}{2005}]{Adelberger2005}
{Adelberger} K.~L.,  {Shapley} A.~E.,  {Steidel} C.~C.,  {Pettini} M.,  {Erb}
  D.~K.,   {Reddy} N.~A.,  2005, \mn@doi [\apj] {10.1086/431753}, \href
  {http://adsabs.harvard.edu/abs/2005ApJ...629..636A} {629, 636}

\bibitem[\protect\citeauthoryear{{Armillotta}, {Fraternali}, {Werk},
  {Prochaska}  \& {Marinacci}}{{Armillotta} et~al.}{2017}]{Armillotta2017}
{Armillotta} L.,  {Fraternali} F.,  {Werk} J.~K.,  {Prochaska} J.~X.,
  {Marinacci} F.,  2017, \mn@doi [\mnras] {10.1093/mnras/stx1239}, \href
  {http://adsabs.harvard.edu/abs/2017MNRAS.470..114A} {470, 114}

\bibitem[\protect\citeauthoryear{{Barnes} \& {Hernquist}}{{Barnes} \&
  {Hernquist}}{1991}]{BH91}
{Barnes} J.~E.,  {Hernquist} L.~E.,  1991, \mn@doi [\apjl] {10.1086/185978},
  \href {http://adsabs.harvard.edu/abs/1991ApJ...370L..65B} {370, L65}

\bibitem[\protect\citeauthoryear{{Barnes} \& {Hernquist}}{{Barnes} \&
  {Hernquist}}{1996}]{BH96}
{Barnes} J.~E.,  {Hernquist} L.,  1996, \mn@doi [\apj] {10.1086/177957}, \href
  {http://adsabs.harvard.edu/abs/1996ApJ...471..115B} {471, 115}

\bibitem[\protect\citeauthoryear{{Barrera-Ballesteros}
  et~al.,}{{Barrera-Ballesteros} et~al.}{2015}]{CALIFA_metallicity_2015}
{Barrera-Ballesteros} J.~K.,  et~al., 2015, \mn@doi [\aap]
  {10.1051/0004-6361/201425397}, \href
  {http://adsabs.harvard.edu/abs/2015A%26A...579A..45B} {579, A45}

\bibitem[\protect\citeauthoryear{{Berg}, {Ellison}, {Tumlinson}, {Oppenheimer},
  {Horton}, {Bordoloi}  \& {Schaye}}{{Berg} et~al.}{2017}]{COS-AGN_submitted}
{Berg} T.~A.~M.,  {Ellison} S.~L.,  {Tumlinson} J.,  {Oppenheimer} B.,
  {Horton} R.,  {Bordoloi} R.,   {Schaye} J.,  2017, \mnras, submitted

\bibitem[\protect\citeauthoryear{{Bergeron}}{{Bergeron}}{1986}]{Bergeron1986}
{Bergeron} J.,  1986, \aap, \href
  {http://adsabs.harvard.edu/abs/1986A%26A...155L...8B} {155, L8}

\bibitem[\protect\citeauthoryear{{Bergeron} \& {Boiss{\'e}}}{{Bergeron} \&
  {Boiss{\'e}}}{1991}]{Bergeron_et_Boiss1991}
{Bergeron} J.,  {Boiss{\'e}} P.,  1991, \aap, \href
  {http://adsabs.harvard.edu/abs/1991A%26A...243..344B} {243, 344}

\bibitem[\protect\citeauthoryear{{Bird}, {Vogelsberger}, {Haehnelt}, {Sijacki},
  {Genel}, {Torrey}, {Springel}  \& {Hernquist}}{{Bird}
  et~al.}{2014}]{Bird2014}
{Bird} S.,  {Vogelsberger} M.,  {Haehnelt} M.,  {Sijacki} D.,  {Genel} S.,
  {Torrey} P.,  {Springel} V.,   {Hernquist} L.,  2014, \mn@doi [\mnras]
  {10.1093/mnras/stu1923}, \href
  {http://adsabs.harvard.edu/abs/2014MNRAS.445.2313B} {445, 2313}

\bibitem[\protect\citeauthoryear{{Bird}, {Haehnelt}, {Neeleman}, {Genel},
  {Vogelsberger}  \& {Hernquist}}{{Bird} et~al.}{2015}]{Bird2015}
{Bird} S.,  {Haehnelt} M.,  {Neeleman} M.,  {Genel} S.,  {Vogelsberger} M.,
  {Hernquist} L.,  2015, \mn@doi [\mnras] {10.1093/mnras/stu2542}, \href
  {http://adsabs.harvard.edu/abs/2015MNRAS.447.1834B} {447, 1834}

\bibitem[\protect\citeauthoryear{{Bird}, {Rubin}, {Suresh}  \&
  {Hernquist}}{{Bird} et~al.}{2016}]{Bird2016}
{Bird} S.,  {Rubin} K.~H.~R.,  {Suresh} J.,   {Hernquist} L.,  2016, \mn@doi
  [\mnras] {10.1093/mnras/stw1582}, \href
  {http://adsabs.harvard.edu/abs/2016MNRAS.462..307B} {462, 307}

\bibitem[\protect\citeauthoryear{{Bogd{\'a}n} et~al.,}{{Bogd{\'a}n}
  et~al.}{2015}]{Bogdan2015}
{Bogd{\'a}n} {\'A}.,  et~al., 2015, \mn@doi [\apj]
  {10.1088/0004-637X/804/1/72}, \href
  {http://adsabs.harvard.edu/abs/2015ApJ...804...72B} {804, 72}

\bibitem[\protect\citeauthoryear{{Bordoloi} et~al.,}{{Bordoloi}
  et~al.}{2014}]{Bordoloi2014}
{Bordoloi} R.,  et~al., 2014, \mn@doi [\apj] {10.1088/0004-637X/796/2/136},
  \href {http://adsabs.harvard.edu/abs/2014ApJ...796..136B} {796, 136}

\bibitem[\protect\citeauthoryear{{Bordoloi}, {Wagner}, {Heckman}  \&
  {Norman}}{{Bordoloi} et~al.}{2017}]{bordoloi_ionization}
{Bordoloi} R.,  {Wagner} A.~Y.,  {Heckman} T.~M.,   {Norman} C.~A.,  2017,
  \mn@doi [\apj] {10.3847/1538-4357/aa8e9c}, \href
  {http://adsabs.harvard.edu/abs/2017ApJ...848..122B} {848, 122}

\bibitem[\protect\citeauthoryear{{Borthakur}, {Heckman}, {Strickland}, {Wild}
  \& {Schiminovich}}{{Borthakur} et~al.}{2013}]{Borthakur2013}
{Borthakur} S.,  {Heckman} T.,  {Strickland} D.,  {Wild} V.,   {Schiminovich}
  D.,  2013, \mn@doi [\apj] {10.1088/0004-637X/768/1/18}, \href
  {http://adsabs.harvard.edu/abs/2013ApJ...768...18B} {768, 18}

\bibitem[\protect\citeauthoryear{{Borthakur} et~al.,}{{Borthakur}
  et~al.}{2015}]{Borthakur2015}
{Borthakur} S.,  et~al., 2015, \mn@doi [\apj] {10.1088/0004-637X/813/1/46},
  \href {http://adsabs.harvard.edu/abs/2015ApJ...813...46B} {813, 46}

\bibitem[\protect\citeauthoryear{{Borthakur} et~al.,}{{Borthakur}
  et~al.}{2016}]{COS-Gass}
{Borthakur} S.,  et~al., 2016, \mn@doi [\apj] {10.3847/1538-4357/833/2/259},
  \href {http://adsabs.harvard.edu/abs/2016ApJ...833..259B} {833, 259}

\bibitem[\protect\citeauthoryear{{Burchett} et~al.,}{{Burchett}
  et~al.}{2016}]{Burchett2016}
{Burchett} J.~N.,  et~al., 2016, \mn@doi [\apj] {10.3847/0004-637X/832/2/124},
  \href {http://adsabs.harvard.edu/abs/2016ApJ...832..124B} {832, 124}

\bibitem[\protect\citeauthoryear{{Casteels} et~al.,}{{Casteels}
  et~al.}{2014}]{Casteels2014}
{Casteels} K.~R.~V.,  et~al., 2014, \mn@doi [\mnras] {10.1093/mnras/stu1799},
  \href {http://adsabs.harvard.edu/abs/2014MNRAS.445.1157C} {445, 1157}

\bibitem[\protect\citeauthoryear{{Chabrier}}{{Chabrier}}{2003}]{Chabrier}
{Chabrier} G.,  2003, \mn@doi [\pasp] {10.1086/376392}, \href
  {http://adsabs.harvard.edu/abs/2003PASP..115..763C} {115, 763}

\bibitem[\protect\citeauthoryear{{Chen}, {Lanzetta}  \& {Webb}}{{Chen}
  et~al.}{2001}]{Chen2001}
{Chen} H.-W.,  {Lanzetta} K.~M.,   {Webb} J.~K.,  2001, \mn@doi [\apj]
  {10.1086/321537}, \href {http://adsabs.harvard.edu/abs/2001ApJ...556..158C}
  {556, 158}

\bibitem[\protect\citeauthoryear{{Chen}, {Helsby}, {Gauthier}, {Shectman},
  {Thompson}  \& {Tinker}}{{Chen} et~al.}{2010}]{Chen2010}
{Chen} H.-W.,  {Helsby} J.~E.,  {Gauthier} J.-R.,  {Shectman} S.~A.,
  {Thompson} I.~B.,   {Tinker} J.~L.,  2010, \mn@doi [\apj]
  {10.1088/0004-637X/714/2/1521}, \href
  {http://adsabs.harvard.edu/abs/2010ApJ...714.1521C} {714, 1521}

\bibitem[\protect\citeauthoryear{{Christensen}, {Dav{\'e}}, {Governato},
  {Pontzen}, {Brooks}, {Munshi}, {Quinn}  \& {Wadsley}}{{Christensen}
  et~al.}{2016}]{Christensen2016}
{Christensen} C.~R.,  {Dav{\'e}} R.,  {Governato} F.,  {Pontzen} A.,  {Brooks}
  A.,  {Munshi} F.,  {Quinn} T.,   {Wadsley} J.,  2016, \mn@doi [\apj]
  {10.3847/0004-637X/824/1/57}, \href
  {http://adsabs.harvard.edu/abs/2016ApJ...824...57C} {824, 57}

\bibitem[\protect\citeauthoryear{{Cooksey}, {Thom}, {Prochaska}  \&
  {Chen}}{{Cooksey} et~al.}{2010}]{Cooksey2010}
{Cooksey} K.~L.,  {Thom} C.,  {Prochaska} J.~X.,   {Chen} H.-W.,  2010, \mn@doi
  [\apj] {10.1088/0004-637X/708/1/868}, \href
  {http://adsabs.harvard.edu/abs/2010ApJ...708..868C} {708, 868}

\bibitem[\protect\citeauthoryear{{Corlies} \& {Schiminovich}}{{Corlies} \&
  {Schiminovich}}{2016}]{Corlies2016}
{Corlies} L.,  {Schiminovich} D.,  2016, \mn@doi [\apj]
  {10.3847/0004-637X/827/2/148}, \href
  {http://adsabs.harvard.edu/abs/2016ApJ...827..148C} {827, 148}

\bibitem[\protect\citeauthoryear{{Cox}, {Primack}, {Jonsson}  \&
  {Somerville}}{{Cox} et~al.}{2004}]{Cox2004}
{Cox} T.~J.,  {Primack} J.,  {Jonsson} P.,   {Somerville} R.~S.,  2004, \mn@doi
  [\apjl] {10.1086/421905}, \href
  {http://adsabs.harvard.edu/abs/2004ApJ...607L..87C} {607, L87}

\bibitem[\protect\citeauthoryear{{Cox}, {Jonsson}, {Primack}  \&
  {Somerville}}{{Cox} et~al.}{2006a}]{Cox2006}
{Cox} T.~J.,  {Jonsson} P.,  {Primack} J.~R.,   {Somerville} R.~S.,  2006a,
  \mn@doi [\mnras] {10.1111/j.1365-2966.2006.11107.x}, \href
  {http://adsabs.harvard.edu/abs/2006MNRAS.373.1013C} {373, 1013}

\bibitem[\protect\citeauthoryear{{Cox}, {Di Matteo}, {Hernquist}, {Hopkins},
  {Robertson}  \& {Springel}}{{Cox} et~al.}{2006b}]{Cox2006_agn}
{Cox} T.~J.,  {Di Matteo} T.,  {Hernquist} L.,  {Hopkins} P.~F.,  {Robertson}
  B.,   {Springel} V.,  2006b, \mn@doi [\apj] {10.1086/503284}, \href
  {http://adsabs.harvard.edu/abs/2006ApJ...643..692C} {643, 692}

\bibitem[\protect\citeauthoryear{{Cox}, {Jonsson}, {Somerville}, {Primack}  \&
  {Dekel}}{{Cox} et~al.}{2008}]{Cox2008}
{Cox} T.~J.,  {Jonsson} P.,  {Somerville} R.~S.,  {Primack} J.~R.,   {Dekel}
  A.,  2008, \mn@doi [\mnras] {10.1111/j.1365-2966.2007.12730.x}, \href
  {http://adsabs.harvard.edu/abs/2008MNRAS.384..386C} {384, 386}

\bibitem[\protect\citeauthoryear{{Dasyra} et~al.,}{{Dasyra}
  et~al.}{2006}]{Dasyra2006}
{Dasyra} K.~M.,  et~al., 2006, \mn@doi [\apj] {10.1086/499068}, \href
  {http://adsabs.harvard.edu/abs/2006ApJ...638..745D} {638, 745}

\bibitem[\protect\citeauthoryear{{Di Matteo}, {Springel}  \& {Hernquist}}{{Di
  Matteo} et~al.}{2005}]{DiMatteo2005}
{Di Matteo} T.,  {Springel} V.,   {Hernquist} L.,  2005, \mn@doi [\nat]
  {10.1038/nature03335}, \href
  {http://adsabs.harvard.edu/abs/2005Natur.433..604D} {433, 604}

\bibitem[\protect\citeauthoryear{{Di Matteo}, {Combes}, {Melchior}  \&
  {Semelin}}{{Di Matteo} et~al.}{2007}]{DiMatteo2007}
{Di Matteo} P.,  {Combes} F.,  {Melchior} A.-L.,   {Semelin} B.,  2007, \mn@doi
  [\aap] {10.1051/0004-6361:20066959}, \href
  {http://adsabs.harvard.edu/abs/2007A%26A...468...61D} {468, 61}

\bibitem[\protect\citeauthoryear{{Di Matteo}, {Bournaud}, {Martig}, {Combes},
  {Melchior}  \& {Semelin}}{{Di Matteo} et~al.}{2008}]{merger_properties}
{Di Matteo} P.,  {Bournaud} F.,  {Martig} M.,  {Combes} F.,  {Melchior} A.-L.,
   {Semelin} B.,  2008, \mn@doi [\aap] {10.1051/0004-6361:200809480}, \href
  {http://adsabs.harvard.edu/abs/2008A%26A...492...31D} {492, 31}

\bibitem[\protect\citeauthoryear{{Ellison}, {Patton}, {Simard}  \&
  {McConnachie}}{{Ellison} et~al.}{2008}]{Ellison08}
{Ellison} S.~L.,  {Patton} D.~R.,  {Simard} L.,   {McConnachie} A.~W.,  2008,
  \mn@doi [\aj] {10.1088/0004-6256/135/5/1877}, \href
  {http://adsabs.harvard.edu/abs/2008AJ....135.1877E} {135, 1877}

\bibitem[\protect\citeauthoryear{{Ellison}, {Patton}, {Mendel}  \&
  {Scudder}}{{Ellison} et~al.}{2011}]{Ellison2011}
{Ellison} S.~L.,  {Patton} D.~R.,  {Mendel} J.~T.,   {Scudder} J.~M.,  2011,
  \mn@doi [\mnras] {10.1111/j.1365-2966.2011.19624.x}, \href
  {http://adsabs.harvard.edu/abs/2011MNRAS.418.2043E} {418, 2043}

\bibitem[\protect\citeauthoryear{{Ellison}, {Mendel}, {Scudder}, {Patton}  \&
  {Palmer}}{{Ellison} et~al.}{2013}]{Ellison2013}
{Ellison} S.~L.,  {Mendel} J.~T.,  {Scudder} J.~M.,  {Patton} D.~R.,   {Palmer}
  M.~J.~D.,  2013, \mn@doi [\mnras] {10.1093/mnras/sts546}, \href
  {http://adsabs.harvard.edu/abs/2013MNRAS.430.3128E} {430, 3128}

\bibitem[\protect\citeauthoryear{{Farina}, {Falomo}, {Decarli}, {Treves}  \&
  {Kotilainen}}{{Farina} et~al.}{2013}]{Farina2013}
{Farina} E.~P.,  {Falomo} R.,  {Decarli} R.,  {Treves} A.,   {Kotilainen}
  J.~K.,  2013, \mn@doi [\mnras] {10.1093/mnras/sts410}, \href
  {http://adsabs.harvard.edu/abs/2013MNRAS.429.1267F} {429, 1267}

\bibitem[\protect\citeauthoryear{{Farina}, {Falomo}, {Scarpa}, {Decarli},
  {Treves}  \& {Kotilainen}}{{Farina} et~al.}{2014}]{Farina2014}
{Farina} E.~P.,  {Falomo} R.,  {Scarpa} R.,  {Decarli} R.,  {Treves} A.,
  {Kotilainen} J.~K.,  2014, \mn@doi [\mnras] {10.1093/mnras/stu585}, \href
  {http://adsabs.harvard.edu/abs/2014MNRAS.441..886F} {441, 886}

\bibitem[\protect\citeauthoryear{{Faucher-Gigu{\`e}re}, {Lidz}, {Zaldarriaga}
  \& {Hernquist}}{{Faucher-Gigu{\`e}re} et~al.}{2009}]{UVB}
{Faucher-Gigu{\`e}re} C.-A.,  {Lidz} A.,  {Zaldarriaga} M.,   {Hernquist} L.,
  2009, \mn@doi [\apj] {10.1088/0004-637X/703/2/1416}, \href
  {http://adsabs.harvard.edu/abs/2009ApJ...703.1416F} {703, 1416}

\bibitem[\protect\citeauthoryear{{Ferland} et~al.,}{{Ferland}
  et~al.}{2013}]{Cloudy}
{Ferland} G.~J.,  et~al., 2013, \rmxaa, \href
  {http://adsabs.harvard.edu/abs/2013RMxAA..49..137F} {49, 137}

\bibitem[\protect\citeauthoryear{{Ford}, {Oppenheimer}, {Dav{\'e}}, {Katz},
  {Kollmeier}  \& {Weinberg}}{{Ford} et~al.}{2013}]{Ford2013}
{Ford} A.~B.,  {Oppenheimer} B.~D.,  {Dav{\'e}} R.,  {Katz} N.,  {Kollmeier}
  J.~A.,   {Weinberg} D.~H.,  2013, \mn@doi [\mnras] {10.1093/mnras/stt393},
  \href {http://adsabs.harvard.edu/abs/2013MNRAS.432...89F} {432, 89}

\bibitem[\protect\citeauthoryear{{Ford}, {Dav{\'e}}, {Oppenheimer}, {Katz},
  {Kollmeier}, {Thompson}  \& {Weinberg}}{{Ford} et~al.}{2014}]{Ford2014}
{Ford} A.~B.,  {Dav{\'e}} R.,  {Oppenheimer} B.~D.,  {Katz} N.,  {Kollmeier}
  J.~A.,  {Thompson} R.,   {Weinberg} D.~H.,  2014, \mn@doi [\mnras]
  {10.1093/mnras/stu1418}, \href
  {http://adsabs.harvard.edu/abs/2014MNRAS.444.1260F} {444, 1260}

\bibitem[\protect\citeauthoryear{{Fumagalli}, {Prochaska}, {Kasen}, {Dekel},
  {Ceverino}  \& {Primack}}{{Fumagalli} et~al.}{2011}]{Fumagalli2011}
{Fumagalli} M.,  {Prochaska} J.~X.,  {Kasen} D.,  {Dekel} A.,  {Ceverino} D.,
  {Primack} J.~R.,  2011, \mn@doi [\mnras] {10.1111/j.1365-2966.2011.19599.x},
  \href {http://adsabs.harvard.edu/abs/2011MNRAS.418.1796F} {418, 1796}

\bibitem[\protect\citeauthoryear{{Genel} et~al.,}{{Genel}
  et~al.}{2014}]{IntroIllustris_G1}
{Genel} S.,  et~al., 2014, \mn@doi [\mnras] {10.1093/mnras/stu1654}, \href
  {http://adsabs.harvard.edu/abs/2014MNRAS.445..175G} {445, 175}

\bibitem[\protect\citeauthoryear{{Governato} et~al.,}{{Governato}
  et~al.}{2009}]{Governato2009}
{Governato} F.,  et~al., 2009, \mn@doi [\mnras]
  {10.1111/j.1365-2966.2009.15143.x}, \href
  {http://adsabs.harvard.edu/abs/2009MNRAS.398..312G} {398, 312}

\bibitem[\protect\citeauthoryear{{Grand} et~al.,}{{Grand}
  et~al.}{2017}]{Auriga}
{Grand} R.~J.~J.,  et~al., 2017, \mn@doi [\mnras] {10.1093/mnras/stx071}, \href
  {http://adsabs.harvard.edu/abs/2017MNRAS.467..179G} {467, 179}

\bibitem[\protect\citeauthoryear{{Hayward} \& {Hopkins}}{{Hayward} \&
  {Hopkins}}{2017}]{Hayward2017}
{Hayward} C.~C.,  {Hopkins} P.~F.,  2017, \mn@doi [\mnras]
  {10.1093/mnras/stw2888}, \href
  {http://adsabs.harvard.edu/abs/2017MNRAS.465.1682H} {465, 1682}

\bibitem[\protect\citeauthoryear{{Heckman} \& {Borthakur}}{{Heckman} \&
  {Borthakur}}{2016}]{COS-Burst_old2016}
{Heckman} T.~M.,  {Borthakur} S.,  2016, \mn@doi [\apj]
  {10.3847/0004-637X/822/1/9}, \href
  {http://adsabs.harvard.edu/abs/2016ApJ...822....9H} {822, 9}

\bibitem[\protect\citeauthoryear{{Heckman}, {Borthakur}, {Wild}, {Schiminovich}
   \& {Bordoloi}}{{Heckman} et~al.}{2017}]{COS-Burst}
{Heckman} T.,  {Borthakur} S.,  {Wild} V.,  {Schiminovich} D.,   {Bordoloi} R.,
   2017, \mn@doi [\apj] {10.3847/1538-4357/aa80dc}, \href
  {http://adsabs.harvard.edu/abs/2017ApJ...846..151H} {846, 151}

\bibitem[\protect\citeauthoryear{{Heitsch} \& {Putman}}{{Heitsch} \&
  {Putman}}{2009}]{Heitsch2009}
{Heitsch} F.,  {Putman} M.~E.,  2009, \mn@doi [\apj]
  {10.1088/0004-637X/698/2/1485}, \href
  {http://adsabs.harvard.edu/abs/2009ApJ...698.1485H} {698, 1485}

\bibitem[\protect\citeauthoryear{{Hern{\'a}ndez-Toledo}, {Avila-Reese},
  {Conselice}  \& {Puerari}}{{Hern{\'a}ndez-Toledo}
  et~al.}{2005}]{HernandezToledo2005}
{Hern{\'a}ndez-Toledo} H.~M.,  {Avila-Reese} V.,  {Conselice} C.~J.,
  {Puerari} I.,  2005, \mn@doi [\aj] {10.1086/427134}, \href
  {http://adsabs.harvard.edu/abs/2005AJ....129..682H} {129, 682}

\bibitem[\protect\citeauthoryear{{Hern{\'a}ndez-Toledo}, {Avila-Reese},
  {Salazar-Contreras}  \& {Conselice}}{{Hern{\'a}ndez-Toledo}
  et~al.}{2006}]{HernandezToledo2006}
{Hern{\'a}ndez-Toledo} H.~M.,  {Avila-Reese} V.,  {Salazar-Contreras} J.~R.,
  {Conselice} C.~J.,  2006, \mn@doi [\aj] {10.1086/504157}, \href
  {http://adsabs.harvard.edu/abs/2006AJ....132...71H} {132, 71}

\bibitem[\protect\citeauthoryear{{Hernquist}}{{Hernquist}}{1989}]{H89}
{Hernquist} L.,  1989, \mn@doi [\nat] {10.1038/340687a0}, \href
  {http://adsabs.harvard.edu/abs/1989Natur.340..687H} {340, 687}

\bibitem[\protect\citeauthoryear{{Hernquist} \& {Quinn}}{{Hernquist} \&
  {Quinn}}{1987}]{Hernquist-et-Quinn1987}
{Hernquist} L.,  {Quinn} P.~J.,  1987, \mn@doi [\apj] {10.1086/164844}, \href
  {http://adsabs.harvard.edu/abs/1987ApJ...312....1H} {312, 1}

\bibitem[\protect\citeauthoryear{{Hopkins} \& {Quataert}}{{Hopkins} \&
  {Quataert}}{2010}]{Hopkins_et_Quataert2010}
{Hopkins} P.~F.,  {Quataert} E.,  2010, \mn@doi [\mnras]
  {10.1111/j.1365-2966.2010.17064.x}, \href
  {http://adsabs.harvard.edu/abs/2010MNRAS.407.1529H} {407, 1529}

\bibitem[\protect\citeauthoryear{{Hopkins}, {Hernquist}, {Cox}  \& {Kere{\v
  s}}}{{Hopkins} et~al.}{2008a}]{Hopkins2008i}
{Hopkins} P.~F.,  {Hernquist} L.,  {Cox} T.~J.,   {Kere{\v s}} D.,  2008a,
  \mn@doi [\apjs] {10.1086/524362}, \href
  {http://adsabs.harvard.edu/abs/2008ApJS..175..356H} {175, 356}

\bibitem[\protect\citeauthoryear{{Hopkins}, {Cox}, {Kere{\v s}}  \&
  {Hernquist}}{{Hopkins} et~al.}{2008b}]{Hopkins2008ii}
{Hopkins} P.~F.,  {Cox} T.~J.,  {Kere{\v s}} D.,   {Hernquist} L.,  2008b,
  \mn@doi [\apjs] {10.1086/524363}, \href
  {http://adsabs.harvard.edu/abs/2008ApJS..175..390H} {175, 390}

\bibitem[\protect\citeauthoryear{{Hopkins}, {Kere{\v s}}, {O{\~n}orbe},
  {Faucher-Gigu{\`e}re}, {Quataert}, {Murray}  \& {Bullock}}{{Hopkins}
  et~al.}{2014}]{fire}
{Hopkins} P.~F.,  {Kere{\v s}} D.,  {O{\~n}orbe} J.,  {Faucher-Gigu{\`e}re}
  C.-A.,  {Quataert} E.,  {Murray} N.,   {Bullock} J.~S.,  2014, \mn@doi
  [\mnras] {10.1093/mnras/stu1738}, \href
  {http://adsabs.harvard.edu/abs/2014MNRAS.445..581H} {445, 581}

\bibitem[\protect\citeauthoryear{{Hummels}, {Bryan}, {Smith}  \&
  {Turk}}{{Hummels} et~al.}{2013}]{Hummels2013}
{Hummels} C.~B.,  {Bryan} G.~L.,  {Smith} B.~D.,   {Turk} M.~J.,  2013, \mn@doi
  [\mnras] {10.1093/mnras/sts702}, \href
  {http://adsabs.harvard.edu/abs/2013MNRAS.430.1548H} {430, 1548}

\bibitem[\protect\citeauthoryear{{Johnson}, {Chen}, {Mulchaey}, {Tripp},
  {Prochaska}  \& {Werk}}{{Johnson} et~al.}{2014}]{Johnson2014}
{Johnson} S.~D.,  {Chen} H.-W.,  {Mulchaey} J.~S.,  {Tripp} T.~M.,  {Prochaska}
  J.~X.,   {Werk} J.~K.,  2014, \mn@doi [\mnras] {10.1093/mnras/stt2409}, \href
  {http://adsabs.harvard.edu/abs/2014MNRAS.438.3039J} {438, 3039}

\bibitem[\protect\citeauthoryear{{Johnson}, {Chen}  \& {Mulchaey}}{{Johnson}
  et~al.}{2015a}]{Johnson2015}
{Johnson} S.~D.,  {Chen} H.-W.,   {Mulchaey} J.~S.,  2015a, \mn@doi [\mnras]
  {10.1093/mnras/stv553}, \href
  {http://adsabs.harvard.edu/abs/2015MNRAS.449.3263J} {449, 3263}

\bibitem[\protect\citeauthoryear{{Johnson}, {Chen}  \& {Mulchaey}}{{Johnson}
  et~al.}{2015b}]{Johnson2015QPQ}
{Johnson} S.~D.,  {Chen} H.-W.,   {Mulchaey} J.~S.,  2015b, \mn@doi [\mnras]
  {10.1093/mnras/stv1481}, \href
  {http://adsabs.harvard.edu/abs/2015MNRAS.452.2553J} {452, 2553}

\bibitem[\protect\citeauthoryear{{Keeney}, {Stocke}, {Danforth}  \&
  {Carilli}}{{Keeney} et~al.}{2011}]{Keeney2011}
{Keeney} B.~A.,  {Stocke} J.~T.,  {Danforth} C.~W.,   {Carilli} C.~L.,  2011,
  \mn@doi [\aj] {10.1088/0004-6256/141/2/66}, \href
  {http://adsabs.harvard.edu/abs/2011AJ....141...66K} {141, 66}

\bibitem[\protect\citeauthoryear{{Kewley}, {Geller}  \& {Barton}}{{Kewley}
  et~al.}{2006}]{Kewley2006}
{Kewley} L.~J.,  {Geller} M.~J.,   {Barton} E.~J.,  2006, \mn@doi [\aj]
  {10.1086/500295}, \href {http://adsabs.harvard.edu/abs/2006AJ....131.2004K}
  {131, 2004}

\bibitem[\protect\citeauthoryear{{Kewley}, {Rupke}, {Zahid}, {Geller}  \&
  {Barton}}{{Kewley} et~al.}{2010}]{Kewley2010}
{Kewley} L.~J.,  {Rupke} D.,  {Zahid} H.~J.,  {Geller} M.~J.,   {Barton} E.~J.,
   2010, \mn@doi [\apjl] {10.1088/2041-8205/721/1/L48}, \href
  {http://adsabs.harvard.edu/abs/2010ApJ...721L..48K} {721, L48}

\bibitem[\protect\citeauthoryear{{Knapen}, {Cisternas}  \&
  {Querejeta}}{{Knapen} et~al.}{2015}]{Knapen2015}
{Knapen} J.~H.,  {Cisternas} M.,   {Querejeta} M.,  2015, \mn@doi [\mnras]
  {10.1093/mnras/stv2135}, \href
  {http://adsabs.harvard.edu/abs/2015MNRAS.454.1742K} {454, 1742}

\bibitem[\protect\citeauthoryear{{Lacey} \& {Cole}}{{Lacey} \&
  {Cole}}{1993}]{LC1993}
{Lacey} C.,  {Cole} S.,  1993, \mnras, \href
  {http://adsabs.harvard.edu/abs/1993MNRAS.262..627L} {262, 627}

\bibitem[\protect\citeauthoryear{{Lambas}, {Tissera}, {Alonso}  \&
  {Coldwell}}{{Lambas} et~al.}{2003}]{Lambas2003}
{Lambas} D.~G.,  {Tissera} P.~B.,  {Alonso} M.~S.,   {Coldwell} G.,  2003,
  \mn@doi [\mnras] {10.1111/j.1365-2966.2003.07179.x}, \href
  {http://adsabs.harvard.edu/abs/2003MNRAS.346.1189L} {346, 1189}

\bibitem[\protect\citeauthoryear{{Lanzetta}, {Bowen}, {Tytler}  \&
  {Webb}}{{Lanzetta} et~al.}{1995}]{Lanzetta1995}
{Lanzetta} K.~M.,  {Bowen} D.~V.,  {Tytler} D.,   {Webb} J.~K.,  1995, \mn@doi
  [\apj] {10.1086/175459}, \href
  {http://adsabs.harvard.edu/abs/1995ApJ...442..538L} {442, 538}

\bibitem[\protect\citeauthoryear{{Lehner}, {O'Meara}, {Fox}, {Howk},
  {Prochaska}, {Burns}  \& {Armstrong}}{{Lehner} et~al.}{2014}]{Lehner2014}
{Lehner} N.,  {O'Meara} J.~M.,  {Fox} A.~J.,  {Howk} J.~C.,  {Prochaska} J.~X.,
   {Burns} V.,   {Armstrong} A.~A.,  2014, \mn@doi [\apj]
  {10.1088/0004-637X/788/2/119}, \href
  {http://adsabs.harvard.edu/abs/2014ApJ...788..119L} {788, 119}

\bibitem[\protect\citeauthoryear{{Lehner}, {Howk}  \& {Wakker}}{{Lehner}
  et~al.}{2015}]{Lehner2015}
{Lehner} N.,  {Howk} J.~C.,   {Wakker} B.~P.,  2015, \mn@doi [\apj]
  {10.1088/0004-637X/804/2/79}, \href
  {http://adsabs.harvard.edu/abs/2015ApJ...804...79L} {804, 79}

\bibitem[\protect\citeauthoryear{{Lelli}, {Verheijen}  \& {Fraternali}}{{Lelli}
  et~al.}{2014a}]{Lelli2014i}
{Lelli} F.,  {Verheijen} M.,   {Fraternali} F.,  2014a, \mn@doi [\mnras]
  {10.1093/mnras/stu1804}, \href
  {http://adsabs.harvard.edu/abs/2014MNRAS.445.1694L} {445, 1694}

\bibitem[\protect\citeauthoryear{{Lelli}, {Verheijen}  \& {Fraternali}}{{Lelli}
  et~al.}{2014b}]{Lelli2014ii}
{Lelli} F.,  {Verheijen} M.,   {Fraternali} F.,  2014b, \mn@doi [\aap]
  {10.1051/0004-6361/201322657}, \href
  {http://adsabs.harvard.edu/abs/2014A%26A...566A..71L} {566, A71}

\bibitem[\protect\citeauthoryear{{Liang} \& {Chen}}{{Liang} \&
  {Chen}}{2014}]{Liang-et-Chen2014}
{Liang} C.~J.,  {Chen} H.-W.,  2014, \mn@doi [\mnras] {10.1093/mnras/stu1901},
  \href {http://adsabs.harvard.edu/abs/2014MNRAS.445.2061L} {445, 2061}

\bibitem[\protect\citeauthoryear{{Lotz}, {Jonsson}, {Cox}  \& {Primack}}{{Lotz}
  et~al.}{2008}]{Lotz2008}
{Lotz} J.~M.,  {Jonsson} P.,  {Cox} T.~J.,   {Primack} J.~R.,  2008, \mn@doi
  [\mnras] {10.1111/j.1365-2966.2008.14004.x}, \href
  {http://adsabs.harvard.edu/abs/2008MNRAS.391.1137L} {391, 1137}

\bibitem[\protect\citeauthoryear{{Lotz}, {Jonsson}, {Cox}  \& {Primack}}{{Lotz}
  et~al.}{2010a}]{Lotz2010massratio}
{Lotz} J.~M.,  {Jonsson} P.,  {Cox} T.~J.,   {Primack} J.~R.,  2010a, \mn@doi
  [\mnras] {10.1111/j.1365-2966.2010.16268.x}, \href
  {http://adsabs.harvard.edu/abs/2010MNRAS.404..575L} {404, 575}

\bibitem[\protect\citeauthoryear{{Lotz}, {Jonsson}, {Cox}  \& {Primack}}{{Lotz}
  et~al.}{2010b}]{Lotz2010gasfraction}
{Lotz} J.~M.,  {Jonsson} P.,  {Cox} T.~J.,   {Primack} J.~R.,  2010b, \mn@doi
  [\mnras] {10.1111/j.1365-2966.2010.16269.x}, \href
  {http://adsabs.harvard.edu/abs/2010MNRAS.404..590L} {404, 590}

\bibitem[\protect\citeauthoryear{{Maiolino} et~al.,}{{Maiolino}
  et~al.}{2017}]{outflow_sfr}
{Maiolino} R.,  et~al., 2017, \mn@doi [\nat] {10.1038/nature21677}, \href
  {http://adsabs.harvard.edu/abs/2017Natur.544..202M} {544, 202}

\bibitem[\protect\citeauthoryear{{Marinacci}, {Pakmor}  \&
  {Springel}}{{Marinacci} et~al.}{2014a}]{Marinacci2014}
{Marinacci} F.,  {Pakmor} R.,   {Springel} V.,  2014a, \mn@doi [\mnras]
  {10.1093/mnras/stt2003}, \href
  {http://adsabs.harvard.edu/abs/2014MNRAS.437.1750M} {437, 1750}

\bibitem[\protect\citeauthoryear{{Marinacci}, {Pakmor}, {Springel}  \&
  {Simpson}}{{Marinacci} et~al.}{2014b}]{Marinacci2014feedback}
{Marinacci} F.,  {Pakmor} R.,  {Springel} V.,   {Simpson} C.~M.,  2014b,
  \mn@doi [\mnras] {10.1093/mnras/stu1136}, \href
  {http://adsabs.harvard.edu/abs/2014MNRAS.442.3745M} {442, 3745}

\bibitem[\protect\citeauthoryear{{Marinacci}, {Grand}, {Pakmor}, {Springel},
  {G{\'o}mez}, {Frenk}  \& {White}}{{Marinacci} et~al.}{2017}]{Marinacci2017}
{Marinacci} F.,  {Grand} R.~J.~J.,  {Pakmor} R.,  {Springel} V.,  {G{\'o}mez}
  F.~A.,  {Frenk} C.~S.,   {White} S.~D.~M.,  2017, \mn@doi [\mnras]
  {10.1093/mnras/stw3366}, \href
  {http://adsabs.harvard.edu/abs/2017MNRAS.466.3859M} {466, 3859}

\bibitem[\protect\citeauthoryear{{Martin}}{{Martin}}{2005}]{Martin2005}
{Martin} C.~L.,  2005, \mn@doi [\apj] {10.1086/427277}, \href
  {http://adsabs.harvard.edu/abs/2005ApJ...621..227M} {621, 227}

\bibitem[\protect\citeauthoryear{{McElroy}, {Croom}, {Pracy}, {Sharp}, {Ho}  \&
  {Medling}}{{McElroy} et~al.}{2015}]{McElroy2015}
{McElroy} R.,  {Croom} S.~M.,  {Pracy} M.,  {Sharp} R.,  {Ho} I.-T.,
  {Medling} A.~M.,  2015, \mn@doi [\mnras] {10.1093/mnras/stu2224}, \href
  {http://adsabs.harvard.edu/abs/2015MNRAS.446.2186M} {446, 2186}

\bibitem[\protect\citeauthoryear{{Mihos} \& {Hernquist}}{{Mihos} \&
  {Hernquist}}{1996}]{Mihos-et-Hernquist1996}
{Mihos} J.~C.,  {Hernquist} L.,  1996, \mn@doi [\apj] {10.1086/177353}, \href
  {http://adsabs.harvard.edu/abs/1996ApJ...464..641M} {464, 641}

\bibitem[\protect\citeauthoryear{{Montuori}, {Di Matteo}, {Lehnert}, {Combes}
  \& {Semelin}}{{Montuori} et~al.}{2010}]{Montouri2010}
{Montuori} M.,  {Di Matteo} P.,  {Lehnert} M.~D.,  {Combes} F.,   {Semelin} B.,
   2010, \mn@doi [\aap] {10.1051/0004-6361/201014304}, \href
  {http://adsabs.harvard.edu/abs/2010A%26A...518A..56M} {518, A56}

\bibitem[\protect\citeauthoryear{{Moreno}, {Torrey}, {Ellison}, {Patton},
  {Bluck}, {Bansal}  \& {Hernquist}}{{Moreno} et~al.}{2015}]{Moreno2015}
{Moreno} J.,  {Torrey} P.,  {Ellison} S.~L.,  {Patton} D.~R.,  {Bluck}
  A.~F.~L.,  {Bansal} G.,   {Hernquist} L.,  2015, \mn@doi [\mnras]
  {10.1093/mnras/stv094}, \href
  {http://adsabs.harvard.edu/abs/2015MNRAS.448.1107M} {448, 1107}

\bibitem[\protect\citeauthoryear{{Muratov}, {Kere{\v s}},
  {Faucher-Gigu{\`e}re}, {Hopkins}, {Quataert}  \& {Murray}}{{Muratov}
  et~al.}{2015}]{Muratov2015}
{Muratov} A.~L.,  {Kere{\v s}} D.,  {Faucher-Gigu{\`e}re} C.-A.,  {Hopkins}
  P.~F.,  {Quataert} E.,   {Murray} N.,  2015, \mn@doi [\mnras]
  {10.1093/mnras/stv2126}, \href
  {http://adsabs.harvard.edu/abs/2015MNRAS.454.2691M} {454, 2691}

\bibitem[\protect\citeauthoryear{{Muratov} et~al.,}{{Muratov}
  et~al.}{2017}]{Muratov2017}
{Muratov} A.~L.,  et~al., 2017, \mn@doi [\mnras] {10.1093/mnras/stx667}, \href
  {http://adsabs.harvard.edu/abs/2017MNRAS.468.4170M} {468, 4170}

\bibitem[\protect\citeauthoryear{{Nelson}, {Genel}, {Pillepich},
  {Vogelsberger}, {Springel}  \& {Hernquist}}{{Nelson}
  et~al.}{2016}]{Nelson2016}
{Nelson} D.,  {Genel} S.,  {Pillepich} A.,  {Vogelsberger} M.,  {Springel} V.,
   {Hernquist} L.,  2016, \mn@doi [\mnras] {10.1093/mnras/stw1191}, \href
  {http://adsabs.harvard.edu/abs/2016MNRAS.460.2881N} {460, 2881}

\bibitem[\protect\citeauthoryear{{Nulsen} \& {Fabian}}{{Nulsen} \&
  {Fabian}}{2000}]{Nulsen_et_Fabian2000}
{Nulsen} P.~E.~J.,  {Fabian} A.~C.,  2000, \mn@doi [\mnras]
  {10.1046/j.1365-8711.2000.03038.x}, \href
  {http://adsabs.harvard.edu/abs/2000MNRAS.311..346N} {311, 346}

\bibitem[\protect\citeauthoryear{{Oppenheimer} \& {Dav{\'e}}}{{Oppenheimer} \&
  {Dav{\'e}}}{2006}]{Oppenheimer_et_Dave2006}
{Oppenheimer} B.~D.,  {Dav{\'e}} R.,  2006, \mn@doi [\mnras]
  {10.1111/j.1365-2966.2006.10989.x}, \href
  {http://adsabs.harvard.edu/abs/2006MNRAS.373.1265O} {373, 1265}

\bibitem[\protect\citeauthoryear{{Oppenheimer} \& {Schaye}}{{Oppenheimer} \&
  {Schaye}}{2013}]{oppenheimer_agn}
{Oppenheimer} B.~D.,  {Schaye} J.,  2013, \mn@doi [\mnras]
  {10.1093/mnras/stt1150}, \href
  {http://adsabs.harvard.edu/abs/2013MNRAS.434.1063O} {434, 1063}

\bibitem[\protect\citeauthoryear{{Oppenheimer} et~al.,}{{Oppenheimer}
  et~al.}{2016}]{Oppenheimer2016}
{Oppenheimer} B.~D.,  et~al., 2016, \mn@doi [\mnras] {10.1093/mnras/stw1066},
  \href {http://adsabs.harvard.edu/abs/2016MNRAS.460.2157O} {460, 2157}

\bibitem[\protect\citeauthoryear{{Oppenheimer}, {Segers}, {Schaye}, {Richings}
  \& {Crain}}{{Oppenheimer} et~al.}{2017a}]{Oppenheimer2017_OVI}
{Oppenheimer} B.~D.,  {Segers} M.,  {Schaye} J.,  {Richings} A.~J.,   {Crain}
  R.~A.,  2017a, preprint, \href
  {http://adsabs.harvard.edu/abs/2017arXiv170507897O} {} (\mn@eprint {arXiv}
  {1705.07897})

\bibitem[\protect\citeauthoryear{{Oppenheimer}, {Schaye}, {Crain}, {Werk}  \&
  {Richings}}{{Oppenheimer} et~al.}{2017b}]{Oppenheimer2017}
{Oppenheimer} B.~D.,  {Schaye} J.,  {Crain} R.~A.,  {Werk} J.~K.,   {Richings}
  A.~J.,  2017b, preprint, \href
  {http://adsabs.harvard.edu/abs/2017arXiv170907577O} {} (\mn@eprint {arXiv}
  {1709.07577})

\bibitem[\protect\citeauthoryear{{Pakmor}, {Springel}, {Bauer}, {Mocz},
  {Munoz}, {Ohlmann}, {Schaal}  \& {Zhu}}{{Pakmor}
  et~al.}{2016}]{Arepo_revised}
{Pakmor} R.,  {Springel} V.,  {Bauer} A.,  {Mocz} P.,  {Munoz} D.~J.,
  {Ohlmann} S.~T.,  {Schaal} K.,   {Zhu} C.,  2016, \mn@doi [\mnras]
  {10.1093/mnras/stv2380}, \href
  {http://adsabs.harvard.edu/abs/2016MNRAS.455.1134P} {455, 1134}

\bibitem[\protect\citeauthoryear{{Patton}, {Ellison}, {Simard}, {McConnachie}
  \& {Mendel}}{{Patton} et~al.}{2011}]{Patton2011}
{Patton} D.~R.,  {Ellison} S.~L.,  {Simard} L.,  {McConnachie} A.~W.,
  {Mendel} J.~T.,  2011, \mn@doi [\mnras] {10.1111/j.1365-2966.2010.17932.x},
  \href {http://adsabs.harvard.edu/abs/2011MNRAS.412..591P} {412, 591}

\bibitem[\protect\citeauthoryear{{Patton}, {Torrey}, {Ellison}, {Mendel}  \&
  {Scudder}}{{Patton} et~al.}{2013}]{Patton2013}
{Patton} D.~R.,  {Torrey} P.,  {Ellison} S.~L.,  {Mendel} J.~T.,   {Scudder}
  J.~M.,  2013, \mn@doi [\mnras] {10.1093/mnrasl/slt058}, \href
  {http://adsabs.harvard.edu/abs/2013MNRAS.433L..59P} {433, L59}

\bibitem[\protect\citeauthoryear{{Patton}, {Qamar}, {Ellison}, {Bluck},
  {Simard}, {Mendel}, {Moreno}  \& {Torrey}}{{Patton}
  et~al.}{2016}]{Patton2016}
{Patton} D.~R.,  {Qamar} F.~D.,  {Ellison} S.~L.,  {Bluck} A.~F.~L.,  {Simard}
  L.,  {Mendel} J.~T.,  {Moreno} J.,   {Torrey} P.,  2016, \mn@doi [\mnras]
  {10.1093/mnras/stw1494}, \href
  {http://adsabs.harvard.edu/abs/2016MNRAS.461.2589P} {461, 2589}

\bibitem[\protect\citeauthoryear{{Peeples}, {Werk}, {Tumlinson}, {Oppenheimer},
  {Prochaska}, {Katz}  \& {Weinberg}}{{Peeples} et~al.}{2014}]{Peeples2014}
{Peeples} M.~S.,  {Werk} J.~K.,  {Tumlinson} J.,  {Oppenheimer} B.~D.,
  {Prochaska} J.~X.,  {Katz} N.,   {Weinberg} D.~H.,  2014, \mn@doi [\apj]
  {10.1088/0004-637X/786/1/54}, \href
  {http://adsabs.harvard.edu/abs/2014ApJ...786...54P} {786, 54}

\bibitem[\protect\citeauthoryear{{Perez}, {Tissera}, {Scannapieco}, {Lambas}
  \& {de Rossi}}{{Perez} et~al.}{2006}]{Perez2006}
{Perez} M.~J.,  {Tissera} P.~B.,  {Scannapieco} C.,  {Lambas} D.~G.,   {de
  Rossi} M.~E.,  2006, \mn@doi [\aap] {10.1051/0004-6361:20054761}, \href
  {http://adsabs.harvard.edu/abs/2006A%26A...459..361P} {459, 361}

\bibitem[\protect\citeauthoryear{{Perez}, {Michel-Dansac}  \&
  {Tissera}}{{Perez} et~al.}{2011}]{Perez2011}
{Perez} J.,  {Michel-Dansac} L.,   {Tissera} P.~B.,  2011, \mn@doi [\mnras]
  {10.1111/j.1365-2966.2011.19300.x}, \href
  {http://adsabs.harvard.edu/abs/2011MNRAS.417..580P} {417, 580}

\bibitem[\protect\citeauthoryear{{Pettini}, {Rix}, {Steidel}, {Adelberger},
  {Hunt}  \& {Shapley}}{{Pettini} et~al.}{2002}]{Pettini2002}
{Pettini} M.,  {Rix} S.~A.,  {Steidel} C.~C.,  {Adelberger} K.~L.,  {Hunt}
  M.~P.,   {Shapley} A.~E.,  2002, \mn@doi [\apj] {10.1086/339355}, \href
  {http://adsabs.harvard.edu/abs/2002ApJ...569..742P} {569, 742}

\bibitem[\protect\citeauthoryear{{Pointon}, {Nielsen}, {Kacprzak}, {Muzahid},
  {Churchill}  \& {Charlton}}{{Pointon} et~al.}{2017}]{Pointon2017}
{Pointon} S.~K.,  {Nielsen} N.~M.,  {Kacprzak} G.~G.,  {Muzahid} S.,
  {Churchill} C.~W.,   {Charlton} J.~C.,  2017, \mn@doi [\apj]
  {10.3847/1538-4357/aa7743}, \href
  {http://adsabs.harvard.edu/abs/2017ApJ...844...23P} {844, 23}

\bibitem[\protect\citeauthoryear{{Pop}, {Pillepich}, {Amorisco}  \&
  {Hernquist}}{{Pop} et~al.}{2017}]{Pop2017}
{Pop} A.-R.,  {Pillepich} A.,  {Amorisco} N.~C.,   {Hernquist} L.,  2017,
  preprint, \href {http://adsabs.harvard.edu/abs/2017arXiv170606102P} {}
  (\mn@eprint {arXiv} {1706.06102})

\bibitem[\protect\citeauthoryear{{Prochaska}, {Weiner}, {Chen}, {Mulchaey}  \&
  {Cooksey}}{{Prochaska} et~al.}{2011}]{Prochaska2011}
{Prochaska} J.~X.,  {Weiner} B.,  {Chen} H.-W.,  {Mulchaey} J.,   {Cooksey} K.,
   2011, \mn@doi [\apj] {10.1088/0004-637X/740/2/91}, \href
  {http://adsabs.harvard.edu/abs/2011ApJ...740...91P} {740, 91}

\bibitem[\protect\citeauthoryear{{Prochaska} et~al.,}{{Prochaska}
  et~al.}{2017}]{Prochaska2017}
{Prochaska} J.~X.,  et~al., 2017, \mn@doi [\apj] {10.3847/1538-4357/aa6007},
  \href {http://adsabs.harvard.edu/abs/2017ApJ...837..169P} {837, 169}

\bibitem[\protect\citeauthoryear{{Puchwein} \& {Springel}}{{Puchwein} \&
  {Springel}}{2013}]{Puchwein_et_Springel2013}
{Puchwein} E.,  {Springel} V.,  2013, \mn@doi [\mnras] {10.1093/mnras/sts243},
  \href {http://adsabs.harvard.edu/abs/2013MNRAS.428.2966P} {428, 2966}

\bibitem[\protect\citeauthoryear{{Putman}, {Peek}  \& {Joung}}{{Putman}
  et~al.}{2012}]{Putman2012}
{Putman} M.~E.,  {Peek} J.~E.~G.,   {Joung} M.~R.,  2012, \mn@doi [\araa]
  {10.1146/annurev-astro-081811-125612}, \href
  {http://adsabs.harvard.edu/abs/2012ARA%26A..50..491P} {50, 491}

\bibitem[\protect\citeauthoryear{{Rahmati}, {Pawlik}, {Rai{\v c}evic}  \&
  {Schaye}}{{Rahmati} et~al.}{2013}]{Rahmati2013}
{Rahmati} A.,  {Pawlik} A.~H.,  {Rai{\v c}evic} M.,   {Schaye} J.,  2013,
  \mn@doi [\mnras] {10.1093/mnras/stt066}, \href
  {http://adsabs.harvard.edu/abs/2013MNRAS.430.2427R} {430, 2427}

\bibitem[\protect\citeauthoryear{{Rodriguez-Gomez} et~al.,}{{Rodriguez-Gomez}
  et~al.}{2017}]{Rodriguez-Gomez2017}
{Rodriguez-Gomez} V.,  et~al., 2017, \mn@doi [\mnras] {10.1093/mnras/stx305},
  \href {http://adsabs.harvard.edu/abs/2017MNRAS.467.3083R} {467, 3083}

\bibitem[\protect\citeauthoryear{{Rupke}, {Veilleux}  \& {Sanders}}{{Rupke}
  et~al.}{2005a}]{Rupke2005_starburst_outflow}
{Rupke} D.~S.,  {Veilleux} S.,   {Sanders} D.~B.,  2005a, \mn@doi [\apjs]
  {10.1086/432889}, \href {http://adsabs.harvard.edu/abs/2005ApJS..160..115R}
  {160, 115}

\bibitem[\protect\citeauthoryear{{Rupke}, {Veilleux}  \& {Sanders}}{{Rupke}
  et~al.}{2005b}]{Rupke2005_agn_outflow}
{Rupke} D.~S.,  {Veilleux} S.,   {Sanders} D.~B.,  2005b, \mn@doi [\apj]
  {10.1086/444451}, \href {http://adsabs.harvard.edu/abs/2005ApJ...632..751R}
  {632, 751}

\bibitem[\protect\citeauthoryear{{Rupke}, {Kewley}  \& {Barnes}}{{Rupke}
  et~al.}{2010a}]{Rupke2010_sim}
{Rupke} D.~S.~N.,  {Kewley} L.~J.,   {Barnes} J.~E.,  2010a, \mn@doi [\apjl]
  {10.1088/2041-8205/710/2/L156}, \href
  {http://adsabs.harvard.edu/abs/2010ApJ...710L.156R} {710, L156}

\bibitem[\protect\citeauthoryear{{Rupke}, {Kewley}  \& {Chien}}{{Rupke}
  et~al.}{2010b}]{Rupke2010_obs}
{Rupke} D.~S.~N.,  {Kewley} L.~J.,   {Chien} L.-H.,  2010b, \mn@doi [\apj]
  {10.1088/0004-637X/723/2/1255}, \href
  {http://adsabs.harvard.edu/abs/2010ApJ...723.1255R} {723, 1255}

\bibitem[\protect\citeauthoryear{{Satyapal}, {Ellison}, {McAlpine}, {Hickox},
  {Patton}  \& {Mendel}}{{Satyapal} et~al.}{2014}]{Satyapal2014}
{Satyapal} S.,  {Ellison} S.~L.,  {McAlpine} W.,  {Hickox} R.~C.,  {Patton}
  D.~R.,   {Mendel} J.~T.,  2014, \mn@doi [\mnras] {10.1093/mnras/stu650},
  \href {http://adsabs.harvard.edu/abs/2014MNRAS.441.1297S} {441, 1297}

\bibitem[\protect\citeauthoryear{{Schaye} et~al.,}{{Schaye}
  et~al.}{2015}]{auriga_dm}
{Schaye} J.,  et~al., 2015, \mn@doi [\mnras] {10.1093/mnras/stu2058}, \href
  {http://adsabs.harvard.edu/abs/2015MNRAS.446..521S} {446, 521}

\bibitem[\protect\citeauthoryear{{Scott} \& {Kaviraj}}{{Scott} \&
  {Kaviraj}}{2014}]{Scott_et_Kaviraj2014}
{Scott} C.,  {Kaviraj} S.,  2014, \mn@doi [\mnras] {10.1093/mnras/stt2014},
  \href {http://adsabs.harvard.edu/abs/2014MNRAS.437.2137S} {437, 2137}

\bibitem[\protect\citeauthoryear{{Scott} et~al.,}{{Scott}
  et~al.}{2014}]{HIasymm}
{Scott} T.~C.,  et~al., 2014, \mn@doi [\aap] {10.1051/0004-6361/201423701},
  \href {http://adsabs.harvard.edu/abs/2014A%26A...567A..56S} {567, A56}

\bibitem[\protect\citeauthoryear{{Scudder}, {Ellison}, {Torrey}, {Patton}  \&
  {Mendel}}{{Scudder} et~al.}{2012}]{Scudder2012}
{Scudder} J.~M.,  {Ellison} S.~L.,  {Torrey} P.,  {Patton} D.~R.,   {Mendel}
  J.~T.,  2012, \mn@doi [\mnras] {10.1111/j.1365-2966.2012.21749.x}, \href
  {http://adsabs.harvard.edu/abs/2012MNRAS.426..549S} {426, 549}

\bibitem[\protect\citeauthoryear{{Segers}, {Oppenheimer}, {Schaye}  \&
  {Richings}}{{Segers} et~al.}{2017}]{Segers2017}
{Segers} M.~C.,  {Oppenheimer} B.~D.,  {Schaye} J.,   {Richings} A.~J.,  2017,
  \mn@doi [\mnras] {10.1093/mnras/stx1633}, \href
  {http://adsabs.harvard.edu/abs/2017MNRAS.471.1026S} {471, 1026}

\bibitem[\protect\citeauthoryear{{Shen}, {Madau}, {Aguirre}, {Guedes}, {Mayer}
  \& {Wadsley}}{{Shen} et~al.}{2012}]{Shen2012}
{Shen} S.,  {Madau} P.,  {Aguirre} A.,  {Guedes} J.,  {Mayer} L.,   {Wadsley}
  J.,  2012, \mn@doi [\apj] {10.1088/0004-637X/760/1/50}, \href
  {http://adsabs.harvard.edu/abs/2012ApJ...760...50S} {760, 50}

\bibitem[\protect\citeauthoryear{{Shen}, {Madau}, {Guedes}, {Mayer},
  {Prochaska}  \& {Wadsley}}{{Shen} et~al.}{2013}]{Shen2013}
{Shen} S.,  {Madau} P.,  {Guedes} J.,  {Mayer} L.,  {Prochaska} J.~X.,
  {Wadsley} J.,  2013, \mn@doi [\apj] {10.1088/0004-637X/765/2/89}, \href
  {http://adsabs.harvard.edu/abs/2013ApJ...765...89S} {765, 89}

\bibitem[\protect\citeauthoryear{{Sijacki}, {Vogelsberger}, {Genel},
  {Springel}, {Torrey}, {Snyder}, {Nelson}  \& {Hernquist}}{{Sijacki}
  et~al.}{2015}]{IntroIllustris_S1}
{Sijacki} D.,  {Vogelsberger} M.,  {Genel} S.,  {Springel} V.,  {Torrey} P.,
  {Snyder} G.~F.,  {Nelson} D.,   {Hernquist} L.,  2015, \mn@doi [\mnras]
  {10.1093/mnras/stv1340}, \href
  {http://adsabs.harvard.edu/abs/2015MNRAS.452..575S} {452, 575}

\bibitem[\protect\citeauthoryear{{Sinha} \& {Holley-Bockelmann}}{{Sinha} \&
  {Holley-Bockelmann}}{2009}]{Sinha2009}
{Sinha} M.,  {Holley-Bockelmann} K.,  2009, \mn@doi [\mnras]
  {10.1111/j.1365-2966.2009.14955.x}, \href
  {http://adsabs.harvard.edu/abs/2009MNRAS.397..190S} {397, 190}

\bibitem[\protect\citeauthoryear{{Sparre} \& {Springel}}{{Sparre} \&
  {Springel}}{2016}]{Sparre2016a}
{Sparre} M.,  {Springel} V.,  2016, \mn@doi [\mnras] {10.1093/mnras/stw1793},
  \href {http://adsabs.harvard.edu/abs/2016MNRAS.462.2418S} {462, 2418}

\bibitem[\protect\citeauthoryear{{Sparre} \& {Springel}}{{Sparre} \&
  {Springel}}{2017}]{Sparre2017}
{Sparre} M.,  {Springel} V.,  2017, \mn@doi [\mnras] {10.1093/mnras/stx1516},
  \href {http://adsabs.harvard.edu/abs/2017MNRAS.470.3946S} {470, 3946}

\bibitem[\protect\citeauthoryear{{Springel}}{{Springel}}{2010}]{Arepo}
{Springel} V.,  2010, \mn@doi [\mnras] {10.1111/j.1365-2966.2009.15715.x},
  \href {http://adsabs.harvard.edu/abs/2010MNRAS.401..791S} {401, 791}

\bibitem[\protect\citeauthoryear{{Springel} \& {Hernquist}}{{Springel} \&
  {Hernquist}}{2003}]{ISM}
{Springel} V.,  {Hernquist} L.,  2003, \mn@doi [\mnras]
  {10.1046/j.1365-8711.2003.06206.x}, \href
  {http://adsabs.harvard.edu/abs/2003MNRAS.339..289S} {339, 289}

\bibitem[\protect\citeauthoryear{{Springel} \& {Hernquist}}{{Springel} \&
  {Hernquist}}{2005}]{SH2005}
{Springel} V.,  {Hernquist} L.,  2005, \mn@doi [\apjl] {10.1086/429486}, \href
  {http://adsabs.harvard.edu/abs/2005ApJ...622L...9S} {622, L9}

\bibitem[\protect\citeauthoryear{{Springel}, {White}, {Tormen}  \&
  {Kauffmann}}{{Springel} et~al.}{2001}]{Subfind}
{Springel} V.,  {White} S.~D.~M.,  {Tormen} G.,   {Kauffmann} G.,  2001,
  \mn@doi [\mnras] {10.1046/j.1365-8711.2001.04912.x}, \href
  {http://adsabs.harvard.edu/abs/2001MNRAS.328..726S} {328, 726}

\bibitem[\protect\citeauthoryear{{Stern}, {Hennawi}, {Prochaska}  \&
  {Werk}}{{Stern} et~al.}{2016}]{Stern2016}
{Stern} J.,  {Hennawi} J.~F.,  {Prochaska} J.~X.,   {Werk} J.~K.,  2016,
  \mn@doi [\apj] {10.3847/0004-637X/830/2/87}, \href
  {http://adsabs.harvard.edu/abs/2016ApJ...830...87S} {830, 87}

\bibitem[\protect\citeauthoryear{{Stinson} et~al.,}{{Stinson}
  et~al.}{2012a}]{Stinson2012}
{Stinson} G.~S.,  et~al., 2012a, \mn@doi [\mnras]
  {10.1111/j.1365-2966.2012.21522.x}, \href
  {http://adsabs.harvard.edu/abs/2012MNRAS.425.1270S} {425, 1270}

\bibitem[\protect\citeauthoryear{{Stinson} et~al.,}{{Stinson}
  et~al.}{2012b}]{Stinson16}
{Stinson} G.~S.,  et~al., 2012b, \mn@doi [\mnras]
  {10.1111/j.1365-2966.2012.21522.x}, \href
  {http://adsabs.harvard.edu/abs/2012MNRAS.425.1270S} {425, 1270}

\bibitem[\protect\citeauthoryear{{Strickland} \& {Heckman}}{{Strickland} \&
  {Heckman}}{2009}]{Strickland2009}
{Strickland} D.~K.,  {Heckman} T.~M.,  2009, \mn@doi [\apj]
  {10.1088/0004-637X/697/2/2030}, \href
  {http://adsabs.harvard.edu/abs/2009ApJ...697.2030S} {697, 2030}

\bibitem[\protect\citeauthoryear{{Suresh}, {Bird}, {Vogelsberger}, {Genel},
  {Torrey}, {Sijacki}, {Springel}  \& {Hernquist}}{{Suresh}
  et~al.}{2015}]{Suresh2015b}
{Suresh} J.,  {Bird} S.,  {Vogelsberger} M.,  {Genel} S.,  {Torrey} P.,
  {Sijacki} D.,  {Springel} V.,   {Hernquist} L.,  2015, \mn@doi [\mnras]
  {10.1093/mnras/stu2762}, \href
  {http://adsabs.harvard.edu/abs/2015MNRAS.448..895S} {448, 895}

\bibitem[\protect\citeauthoryear{{Suresh}, {Rubin}, {Kannan}, {Werk},
  {Hernquist}  \& {Vogelsberger}}{{Suresh} et~al.}{2017}]{Suresh2017}
{Suresh} J.,  {Rubin} K.~H.~R.,  {Kannan} R.,  {Werk} J.~K.,  {Hernquist} L.,
  {Vogelsberger} M.,  2017, \mn@doi [\mnras] {10.1093/mnras/stw2499}, \href
  {http://adsabs.harvard.edu/abs/2017MNRAS.465.2966S} {465, 2966}

\bibitem[\protect\citeauthoryear{{Thom} et~al.,}{{Thom}
  et~al.}{2012}]{Thom2012}
{Thom} C.,  et~al., 2012, \mn@doi [\apjl] {10.1088/2041-8205/758/2/L41}, \href
  {http://adsabs.harvard.edu/abs/2012ApJ...758L..41T} {758, L41}

\bibitem[\protect\citeauthoryear{{Torrey}, {Cox}, {Kewley}  \&
  {Hernquist}}{{Torrey} et~al.}{2012}]{Torrey2012}
{Torrey} P.,  {Cox} T.~J.,  {Kewley} L.,   {Hernquist} L.,  2012, \mn@doi
  [\apj] {10.1088/0004-637X/746/1/108}, \href
  {http://adsabs.harvard.edu/abs/2012ApJ...746..108T} {746, 108}

\bibitem[\protect\citeauthoryear{{Tripp}, {Savage}  \& {Jenkins}}{{Tripp}
  et~al.}{2000}]{Tripp2000}
{Tripp} T.~M.,  {Savage} B.~D.,   {Jenkins} E.~B.,  2000, \mn@doi [\apjl]
  {10.1086/312644}, \href {http://adsabs.harvard.edu/abs/2000ApJ...534L...1T}
  {534, L1}

\bibitem[\protect\citeauthoryear{{Tumlinson} et~al.,}{{Tumlinson}
  et~al.}{2011}]{Tumlinson2011}
{Tumlinson} J.,  et~al., 2011, \mn@doi [Science] {10.1126/science.1209840},
  \href {http://adsabs.harvard.edu/abs/2011Sci...334..948T} {334, 948}

\bibitem[\protect\citeauthoryear{{Tumlinson} et~al.,}{{Tumlinson}
  et~al.}{2013}]{Tumlinson2013}
{Tumlinson} J.,  et~al., 2013, \mn@doi [\apj] {10.1088/0004-637X/777/1/59},
  \href {http://adsabs.harvard.edu/abs/2013ApJ...777...59T} {777, 59}

\bibitem[\protect\citeauthoryear{{Tumlinson}, {Peeples}  \& {Werk}}{{Tumlinson}
  et~al.}{2017}]{Tumlinson2017}
{Tumlinson} J.,  {Peeples} M.~S.,   {Werk} J.~K.,  2017, \mn@doi [\araa]
  {10.1146/annurev-astro-091916-055240}, \href
  {http://adsabs.harvard.edu/abs/2017ARA%26A..55..389T} {55, 389}

\bibitem[\protect\citeauthoryear{{Turk}, {Smith}, {Oishi}, {Skory}, {Skillman},
  {Abel}  \& {Norman}}{{Turk} et~al.}{2011}]{yt}
{Turk} M.~J.,  {Smith} B.~D.,  {Oishi} J.~S.,  {Skory} S.,  {Skillman} S.~W.,
  {Abel} T.,   {Norman} M.~L.,  2011, \mn@doi [The Astrophysical Journal
  Supplement Series] {10.1088/0067-0049/192/1/9}, \href
  {http://adsabs.harvard.edu/abs/2011ApJS..192....9T} {192, 9}

\bibitem[\protect\citeauthoryear{{Veilleux}, {Cecil}  \&
  {Bland-Hawthorn}}{{Veilleux} et~al.}{2005}]{Veilleux2005}
{Veilleux} S.,  {Cecil} G.,   {Bland-Hawthorn} J.,  2005, \mn@doi [\araa]
  {10.1146/annurev.astro.43.072103.150610}, \href
  {http://adsabs.harvard.edu/abs/2005ARA%26A..43..769V} {43, 769}

\bibitem[\protect\citeauthoryear{{Veilleux} et~al.,}{{Veilleux}
  et~al.}{2013}]{Veilleux2013}
{Veilleux} S.,  et~al., 2013, \mn@doi [\apj] {10.1088/0004-637X/776/1/27},
  \href {http://adsabs.harvard.edu/abs/2013ApJ...776...27V} {776, 27}

\bibitem[\protect\citeauthoryear{{Vogelsberger}, {Genel}, {Sijacki}, {Torrey},
  {Springel}  \& {Hernquist}}{{Vogelsberger} et~al.}{2013}]{Illustris}
{Vogelsberger} M.,  {Genel} S.,  {Sijacki} D.,  {Torrey} P.,  {Springel} V.,
  {Hernquist} L.,  2013, \mn@doi [\mnras] {10.1093/mnras/stt1789}, \href
  {http://adsabs.harvard.edu/abs/2013MNRAS.436.3031V} {436, 3031}

\bibitem[\protect\citeauthoryear{{Vogelsberger} et~al.,}{{Vogelsberger}
  et~al.}{2014a}]{IntroIllustris_V1}
{Vogelsberger} M.,  et~al., 2014a, \mn@doi [\mnras] {10.1093/mnras/stu1536},
  \href {http://adsabs.harvard.edu/abs/2014MNRAS.444.1518V} {444, 1518}

\bibitem[\protect\citeauthoryear{{Vogelsberger} et~al.,}{{Vogelsberger}
  et~al.}{2014b}]{IntroIllustris_V2}
{Vogelsberger} M.,  et~al., 2014b, \mn@doi [\nat] {10.1038/nature13316}, \href
  {http://adsabs.harvard.edu/abs/2014Natur.509..177V} {509, 177}

\bibitem[\protect\citeauthoryear{{Werk}, {Prochaska}, {Thom}, {Tumlinson},
  {Tripp}, {O'Meara}  \& {Peeples}}{{Werk} et~al.}{2013}]{COS-Halos_metals}
{Werk} J.~K.,  {Prochaska} J.~X.,  {Thom} C.,  {Tumlinson} J.,  {Tripp} T.~M.,
  {O'Meara} J.~M.,   {Peeples} M.~S.,  2013, \mn@doi [\apjs]
  {10.1088/0067-0049/204/2/17}, \href
  {http://adsabs.harvard.edu/abs/2013ApJS..204...17W} {204, 17}

\bibitem[\protect\citeauthoryear{{Werk} et~al.,}{{Werk}
  et~al.}{2014}]{Werk2014}
{Werk} J.~K.,  et~al., 2014, \mn@doi [\apj] {10.1088/0004-637X/792/1/8}, \href
  {http://adsabs.harvard.edu/abs/2014ApJ...792....8W} {792, 8}

\bibitem[\protect\citeauthoryear{{Werk} et~al.,}{{Werk}
  et~al.}{2016}]{Werk2016}
{Werk} J.~K.,  et~al., 2016, \mn@doi [\apj] {10.3847/1538-4357/833/1/54}, \href
  {http://adsabs.harvard.edu/abs/2016ApJ...833...54W} {833, 54}

\bibitem[\protect\citeauthoryear{{Weston}, {McIntosh}, {Brodwin}, {Mann},
  {Cooper}, {McConnell}  \& {Nielsen}}{{Weston} et~al.}{2017}]{Weston2017}
{Weston} M.~E.,  {McIntosh} D.~H.,  {Brodwin} M.,  {Mann} J.,  {Cooper} A.,
  {McConnell} A.,   {Nielsen} J.~L.,  2017, \mn@doi [\mnras]
  {10.1093/mnras/stw2620}, \href
  {http://adsabs.harvard.edu/abs/2017MNRAS.464.3882W} {464, 3882}

\bibitem[\protect\citeauthoryear{{Wiersma}, {Schaye}, {Theuns}, {Dalla Vecchia}
   \& {Tornatore}}{{Wiersma} et~al.}{2009}]{sim_chem_evol_3}
{Wiersma} R.~P.~C.,  {Schaye} J.,  {Theuns} T.,  {Dalla Vecchia} C.,
  {Tornatore} L.,  2009, \mn@doi [\mnras] {10.1111/j.1365-2966.2009.15331.x},
  \href {http://adsabs.harvard.edu/abs/2009MNRAS.399..574W} {399, 574}

\bibitem[\protect\citeauthoryear{{Woo}, {Son}  \& {Bae}}{{Woo}
  et~al.}{2017}]{JongHak_outflows}
{Woo} J.-H.,  {Son} D.,   {Bae} H.-J.,  2017, \mn@doi [\apj]
  {10.3847/1538-4357/aa6894}, \href
  {http://adsabs.harvard.edu/abs/2017ApJ...839..120W} {839, 120}

\bibitem[\protect\citeauthoryear{{Zahid}, {Torrey}, {Vogelsberger},
  {Hernquist}, {Kewley}  \& {Dav{\'e}}}{{Zahid}
  et~al.}{2014}]{Zahid_et_Torrey_2014}
{Zahid} H.~J.,  {Torrey} P.,  {Vogelsberger} M.,  {Hernquist} L.,  {Kewley} L.,
    {Dav{\'e}} R.,  2014, \mn@doi [\apss] {10.1007/s10509-013-1666-0}, \href
  {http://adsabs.harvard.edu/abs/2014Ap%26SS.349..873Z} {349, 873}

\bibitem[\protect\citeauthoryear{{Zschaechner} et~al.,}{{Zschaechner}
  et~al.}{2016}]{Zschaechner2016}
{Zschaechner} L.~K.,  et~al., 2016, \mn@doi [\apj]
  {10.3847/0004-637X/832/2/142}, \href
  {http://adsabs.harvard.edu/abs/2016ApJ...832..142Z} {832, 142}

\bibitem[\protect\citeauthoryear{{van de Voort}, {Schaye}, {Altay}  \&
  {Theuns}}{{van de Voort} et~al.}{2012}]{vandeVoort2012}
{van de Voort} F.,  {Schaye} J.,  {Altay} G.,   {Theuns} T.,  2012, \mn@doi
  [\mnras] {10.1111/j.1365-2966.2012.20487.x}, \href
  {http://adsabs.harvard.edu/abs/2012MNRAS.421.2809V} {421, 2809}

\makeatother
\end{thebibliography}



\bsp	
\label{lastpage}
\end{document}